\documentclass[twocolumn]{emulateapj}  
\usepackage{epstopdf}
\usepackage{ulem}
\usepackage{amssymb}
\usepackage{times}
\usepackage{graphicx}
\usepackage[usenames,dvipsnames]{pstricks}
\usepackage{psfrag}
\usepackage{lipsum}
\usepackage{natbib}
\usepackage{textcase}

\usepackage[colorlinks=true,linkcolor=blue,citecolor=blue]{hyperref}
\def\araa{ARA\&A}
\def\apj{ApJ}
\def\apjl{ApJ}
\def\aap{A\&A}
\def\mnras{MNRAS}
\def\prd{Phys.~Rev.~D}
\def\prl{Phys.~Rev.~Lett.}
\def\nat{Nature}

\newcommand{\be}{\begin{equation}}
\newcommand{\ee}{\end{equation}}
\newcommand{\bary}{\begin{eqnarray}}
\newcommand{\eary}{\end{eqnarray}}

\begin{document}
\title{Afterglow light curves of the non-relativistic ejecta mass\\ in a stratified circumstellar medium} 
%
\author{N. Fraija\altaffilmark{1$\dagger$},  B. Betancourt Kamenetskaia\altaffilmark{1,2,3},  M.G. Dainotti\altaffilmark{4,5,6},  R. Barniol Duran\altaffilmark{7},  A. G\'alvan G\'amez\altaffilmark{1},   S.~ Dichiara\altaffilmark{8,9}  and  Pedreira, A. C. Caligula do E. S.\altaffilmark{1}}   
\affil{$^1$Instituto de Astronom\' ia, Universidad Nacional Aut\'onoma de M\'exico, Circuito Exterior, C.U., A. Postal 70-264, 04510 M\'exico City, M\'exico}
\affil{$^2$LMU Physics Department, Ludwig Maxmillians University,Theresienstr. 37, 80333 Munich, Germany}
\affil{$^3$TUM Physics Department, Technical University of Munich, James-Franck-Str, 85748 Garching, Germany}
\affil{$^4$Physics Department, Stanford University, 382 Via Pueblo Mall, Stanford, USA}
\affil{$^5$Space Science Institute, Boulder, Colorado}
\affil{$^6$Obserwatorium Astronomiczne, Uniwersytet Jagiello\'nski, ul. Orla 171, 31-501  Krak\'ow, Poland.}
\affil{$^{7}$Department of Physics and Astronomy, California State University, Sacramento, 6000 J Street, Sacramento, CA 95819-6041, USA}
\affil{$^8$ Department of Astronomy, University of Maryland, College Park, MD 20742-4111, USA}
\affil{$^{9}$ Astrophysics Science Division, NASA Goddard Space Flight Center, 8800 Greenbelt Rd, Greenbelt, MD 20771, USA}

\email{$\dagger$ nifraija@astro.unam.mx}
%
%
\begin{abstract}
We present the afterglow light curves produced by the deceleration of the non-relativistic ejecta mass in a stratified circumstellar medium with a density profile $n(r)\propto r^{-k}$ with $k=0$, $1$, $1.5$, $2$ and $2.5$. Once the ejecta mass is launched with equivalent kinetic energy parametrized by $E(>\beta)\propto \beta^{-\alpha}$ (where beta is the ejecta velocity) and propagates into the surrounding circumstellar medium, it first moves with constant velocity (the free-coasting phase), and later it decelerates (the Sedov-Taylor expansion). We present the predicted synchrotron and synchrotron-self Compton light curves during the free-coasting phase, and the subsequent Sedov-Taylor expansion. In particular cases, we show the corresponding light curves generated by the deceleration of several ejecta masses with different velocities launched during the coalescence of binary compact objects and the core-collapse of dying massive stars which will contribute at distinct timescales, frequencies, and intensities. We find that before the radiation escapes from the kilonova (KN) and produces a peak in the light curve, the KN afterglow ejecta is very weak and the corresponding ejecta velocity moves as $\propto t^{-\frac{2}{\alpha+1}}$. Finally, using the multi-wavelength observations and upper limits collected by a large campaign of orbiting satellites and ground telescopes, we constrain the parameter space of both the KN afterglow in GW170817 and the possibly generated KN afterglow in S190814bv. Further observations on timescales of years post-merger are needed to derive tighter constraints.
\end{abstract}
\section{Introduction}
Long-duration gamma-ray bursts \citep[lGRBs; $T_{90}\gtrsim 2\,{\rm s}$;][]{1993ApJ...413L.101K} are connected with the core-collapse (CC) of dying massive stars \citep{1993ApJ...405..273W, 1998Natur.395..670G} that lead to supernovae \citep[SNe;][]{1999Natur.401..453B, 2006ARA&A..44..507W}.   Short-duration GRBs (sGRBs; $T_{90}\lesssim 2\,{\rm s}$) are associated with the coalescence of binary compact objects (NS-NS or BH-NS)\footnote{NS corresponds to neutron star and BH to black hole.} that lead to kilonovae\footnote{A fairly isotropic thermal transient powered by the radioactive decay of rapid neutron capture process nuclei and isotopes} \citep[KNe;][]{1998ApJ...507L..59L, 2005ApJ...634.1202R, 2010MNRAS.406.2650M, 2013ApJ...774...25K, 2017LRR....20....3M}. The dying massive stars and the coalescence of binary compact objects are believed to launch considerable amounts of materials with broad ranges of velocities. In the framework of the coalescence of NS-NS mergers,  non-relativistic ejecta masses with velocities in the range $0.03\lesssim \beta\lesssim 0.8$ (expressed, hereafter,  in units of the speed of light) such as the dynamical ejecta,  the cocoon material, the shock breakout material and the wind ejecta are launched \citep[e.g., see][]{2009ApJ...690.1681D, 2014MNRAS.441.3444M, 2015MNRAS.446..750F,2014MNRAS.437L...6K, 2015MNRAS.446.1115M, 2014ApJ...784L..28N, 2014ApJ...788L...8M,2017ApJ...848L...6L, 2018PhRvL.120x1103L, 2011ApJ...738L..32G, 2013ApJ...778L..16H,2013ApJ...773...78B, 2014ApJ...789L..39W}. In the framework of CC-SNe (depending on the type of SN association), several ejecta masses with non-relativistic velocities less than $\beta\lesssim0.3$ have been reported \citep[e.g., see][]{2020ApJ...892..153M, 2020arXiv200405941I, 2020NatAs.tmp...78N, 2019Natur.565..324I, 2017hsn..book..195G,2008MNRAS.383.1485V, 1998Natur.395..663K,1999Natur.401..453B, 2006ARA&A..44..507W}.  While the inferred mass of the ejecta for the first GRB/SN association \citep[GRB 980425/SN1998bw;][]{1998Natur.395..663K} was $10^{-5}\,M_{\rm \odot}$ with a velocity of $\beta\sim 0.2-0.3$, the inferred mass for the first GRB/KN association \citep[GRB 170817A/AT2017gfo;][]{2017Sci...358.1556C} was $10^{-4} - 10^{-2}\,M_{\odot}$, ejected with a velocity of $\beta\sim 0.1- 0.3$ \citep{2017Natur.551...64A, 2017ApJ...848L..17C, 2017ApJ...848L..18N, 2019LRR....23....1M}.\\
\cite{1982ApJ...258..790C} studied the interaction of an adiabatic flow in a circumstellar density profile for type II SNe of the form $n(r)\propto r^{-2}$. Such a power law has since been the usual convention and it has been used in following papers for different types of SNe, such as those by \cite{2004MNRAS.354L..13K}, \cite{1984ApJ...285L..63C}, \cite{2006ApJ...651.1005S} and \cite{1996ApJ...472..257B}, amongst others.  Nevertheless, in a more recent study, \cite{2012ApJ...747..118M} showed that the spectral diversity of type II luminous SNe may be explained by the diversity in the density slope of the surrounding dense wind. To this effect, they proposed a wind density structure with a profile $\propto r^{-\rm k}$ and noticed that the ratio of the diffusion timescale in the optically thick region of the wind and the shock propagation timescale after the shock breakout have a strong dependence on the stratification parameter ${\rm k}$ which led to differences in the SN spectral evolution.\\
The interaction of the ejecta mass  with the surrounding circumburst medium in the non-relativistic regime has been proposed to describe the multi-wavelength afterglow observations in timescales from days to several years \citep[e.g., see][]{1997MNRAS.288L..51W, 1999ApJ...519L.155D, 1999MNRAS.309..513H, 2000ApJ...538..187L, 2003MNRAS.341..263H, 2013ApJ...778..107S, 2015MNRAS.454.1711B}. Several authors, \citep[e.g see ][]{2014MNRAS.437.1821M, 2016ApJ...831..141F, 2020ApJ...890..102L, 2020arXiv200607434S}, considered the material launched during the coalescence of binary compact objects and computed the synchrotron emission in the radio bands. The authors assumed a free-coasting phase, and the subsequent Sedov-Taylor expansion.  \cite{2001ApJ...551..946T} considered that the  kinetic energy of the shock wave could be described in terms of a power-law (PL) velocity distribution. Henceforth, several authors have considered that the material launched during the coalescence of binary compact objects and the CC-SNe may be described by a PL velocity distribution  \citep[e.g., see][]{2013ApJ...773...78B, 2014MNRAS.437L...6K, 2015MNRAS.450.1430H, 2015MNRAS.448..417B, 2017LRR....20....3M, 2018Natur.554..207M, 2019ApJ...871..200F, 2019MNRAS.487.3914K, 2019LRR....23....1M, 2012ApJ...750...68L, 2013ApJ...778...63H, 2013ApJ...778...18M, 2014ApJ...797..107M}.    In the context of the coalescence of a NS binary, for instance,  \cite{2019ApJ...871..200F}  and  \cite{2019MNRAS.487.3914K} assumed a PL distribution  for the energy of the shock breakout  (the  outermost matter) and the kilonova material in GW170817, and calculated the non-thermal emission generated by the interaction of these materials with the uniform-density medium.  In the case of dying massive stars, for example, \cite{2012ApJ...750...68L} presented a set of numerical simulations of stellar explosions induced by outflows and compared the results  with observational properties of type Ibc SNe  as a function of  the equivalent kinetic energy and velocity of the ejecta.\\ 
In this paper, we present, based on analytic arguments, a theoretical model that predicts the multi-wavelength afterglow emission generated by the deceleration of the non-relativistic ejecta mass in a circumstellar medium with profile $n(r)\propto r^{-k}$. Once the ejecta mass propagates into the circumstellar medium, it first moves with a constant velocity, and later, when the swept up quantity of material is similar to the ejected mass, it begins to decelerate.  During these stages,  electrons are accelerated in the forward shocks, and are cooled down by synchrotron and synchrotron-self Compton (SSC) processes. We present the predicted synchrotron and SSC light curves for ${\rm k}=0$, $1$, $1.5$, $2$ and $2.5$ that cover several ejecta masses launched during the coalescence of binary compact objects and the CC-SNe. Particularly, with the multi-wavelength observations and upper limits collected by a large campaign of orbiting satellites and ground telescopes,   we constrain the parameter space of both the KN afterglow in GW170817 and the possibly generated KN afterglow produced by the coalescence of a BH-NS system in S190814bv.   The paper is organized as follows: In Section 2, we introduce the theoretical model that predicts the multi-wavelength afterglow emission generated by the deceleration of the non-relativistic ejecta mass.  In Section 3, we show the predicted synchrotron and SSC light curves for a density profile with ${\rm k}=0$, $1$, $1.5$, $2$ and $2.5$. In Section 4, we show the  synchrotron and SSC light curves from material launched during the coalescence of a NS binary.   In section 5, we discuss the non-relativistic ejected mass around the KN emission. In Section 6, we apply our model to two particular cases, to GW170817 and S190814bv and finally in Section 7, we present the discussion and summary.  We adopt the convention $Q_{\rm x}=\frac{Q}{10^{\rm x}}$ in c.g.s. units and assume  for the cosmological constants a spatially flat universe $\Lambda$CDM model with  $H_0=69.6\,{\rm km\,s^{-1}\,Mpc^{-1}}$, $\Omega_{\rm M}=0.286$ and  $\Omega_\Lambda=0.714$ \citep{2016A&A...594A..13P}. Prime and unprimed quantities are used for the comoving and observer frames, respectively.
\section{Afterglow light curves of a non-relativistic ejecta mass}
In this section, we show the dynamics of the forward shocks given by the deceleration of the ejecta mass in a density profile $n(r)=A_{\rm k}\,r^{-\rm k}$ with $A_{\rm k}$ the density parameter and $0\leq {\rm k} < 3$. We use analytical arguments in order to compute the synchrotron and SSC light curves expected in the non-relativistic regime.   The synchrotron and SSC radiation is  derived in  the  fully  adiabatic  regime during the free-coasting and the deceleration phase  assuming an electron distribution described by $dN/d\gamma_{\rm e}\propto \gamma^{-p}_{\rm e}$, for $\gamma_{\rm e}\geq \gamma_{\rm m}$, where $\gamma_{\rm m}$ is the Lorentz factors of the lowest-energy electrons and $p$ is the index of the electron distribution.   We only consider the dynamics of the forward shocks instead of reverse shocks because electrons accelerated in the  reverse shocks generate short-lived emissions and we are interested in describing emissions  extended  in timescales from days to years.\\ 
Irrespective of the progenitor,  non-relativistic materials  with a wide range of masses and velocities are launched into the circumstellar medium. After a time ${\rm t}$ following the coalescence or collapse,  a generic ejecta mass moves  with a constant velocity $\beta$ and expands with a mean radius $r \approx \beta t$ \citep[e.g., see][]{1998Natur.395..663K, 2019LRR....23....1M}.    The initial expansion phase is not affected by the circumstellar medium, but once the swept up quantity of material is as large as the ejected mass, the ejecta begins to be decelerated \citep{2014MNRAS.439..744R}. During the deceleration phase, numerical simulations indicate that the velocity of matter in the front the ejecta is faster than the one that moves in the back, so that the ejecta acquires a velocity structure \citep[e.g., see][]{2014MNRAS.439..744R, 2001ApJ...551..946T}.
%
\subsection{The free-coasting phase}
During the initial expansion, the ejecta mass  is not affected by the circumstellar medium \citep{2014MNRAS.439..744R}, so the velocity  is constant $\beta \propto t^0$ and the radius evolves as $r \propto  (1+z)^{-1}t$.
\subsubsection{Synchrotron emission}
During the coasting phase,  the post-shock magnetic field evolves as, $B'\propto \epsilon^\frac12_{\rm B}\,A^\frac12_{\rm k}$ $\beta^{\frac{2-k}{2}}\,t^{-\frac{k}{2}}$ where $\epsilon_{\rm B}$ is  the microphysical parameter associated to the magnetic density.  The Lorentz factors of the lowest-energy electrons and of the higher energy electrons, which are efficiently cooled by synchrotron emission are
{\small
\bary\label{gamma_c}
\gamma_{\rm m}&=&  \gamma^0_{\rm m}\,  g(p) \epsilon_{\rm e,-1}\,\beta_{-0.5}^2\,\cr
\gamma_{\rm c}&=& \gamma^0_{\rm c}\,\left(\frac{1+z}{1.022}\right)^{1-k}(1+Y)^{-1}\,\epsilon^{-1}_{\rm B,-2}\,A^{-1}_{\rm k} \beta_{-0.5}^{k-2}\,t_6^{k-1}\,,
\eary
}
where $Y$ is the Compton parameter \citep{2001ApJ...548..787S}, $g(p)=\frac{p-2}{p-1}$, $\epsilon_{\rm e}$ is the microphysical parameter associated to the electron density and $z$ is the redshift of a generic observer located at 100 Mpc.   Given the evolution of the synchrotron frequencies as a function of the electron Lorentz factors  $\nu^{\rm syn}_{\rm i}\propto \gamma^2_{\rm i}$ for ${\rm i}={\rm m}$ and ${\rm c}$, using eq. (\ref{gamma_c}) the corresponding synchrotron break frequencies can be written as  
{\small
\bary\label{nu_syn}
\nu^{\rm syn}_{\rm m}&=&\nu^{\rm syn,0}_{\rm m}\,g^2(p) \left(\frac{1+z}{1.022}\right)^{\frac{k-2}{2}}\epsilon_{\rm e,-1}^2\,\epsilon_{\rm B,-1}^\frac12 A^\frac{1}{2}_{\rm k} \beta_{-0.5}^\frac{10-k}{2}\,t_6^{-\frac{k}{2}}\cr
\nu^{\rm syn}_{\rm c}&=& \nu^{\rm syn,0}_{\rm c} \left(\frac{1+z}{1.022}\right)^{\frac{2-3k}{2}} (1+Y)^{-2}\,  \epsilon_{\rm B,-2}^{-\frac32}\, A^{-\frac32}_{\rm k} \beta_{-0.5}^{\frac{3k-6}{2}}\,t_6^{\frac{3k-4}{2}}.\,\,\,\,\,\,\,
\eary
}
In the self-absorption regime, the synchrotron break frequencies are  \citep{2000ApJ...543...66P,2015ApJ...810..160G}
{\small
\bary\label{nu_syn_a}
\nu^{\rm syn}_{\rm a,1}&=& \nu^{\rm syn,0}_{\rm a,1} \left(\frac{1+z}{1.022}\right)^{\frac{4(k-2)}{5}}g(p)^{-1}\, \epsilon_{\rm e,-1}^{-1}\, \epsilon_{\rm B,-2}^{\frac15}\, A^{\frac45}_{\rm k} \beta_{-0.5}^{-\frac{4k+5}{5}}\,t_6^{\frac{3-4k}{5}}\,,\cr
\nu^{\rm syn}_{\rm a,2}&=& \nu^{\rm syn,0}_{\rm a,2} \left(\frac{1+z}{1.022}\right)^{\frac{(k-2)(p+6)}{2(p+4)}}g(p)^{\frac{2(p-1)}{p+4}}\, \epsilon_{\rm e,-1}^{\frac{2(p-1)}{p+4}}\, \epsilon_{\rm B,-2}^{\frac{p+2}{2(p+4)}}\cr
&&\hspace{3cm} \times\,  A^{\frac{p+6}{2(p+4)}}_{\rm k}\beta_{-0.5}^{ \frac{10p-kp-6k}{2(p+4)} }\,t_6^{\frac{4-kp-6k}{2(p+4)}},\cr
\nu^{\rm syn}_{\rm a,3}&=& \nu^{\rm syn,0}_{\rm a,3} \left(\frac{1+z}{1.022}\right)^{\frac{9k-13}{5}}\,(1+Y)\, \epsilon_{\rm B,-2}^{\frac65}\, A^{\frac95}_{\rm k} \beta_{-0.5}^{\frac{15-9k}{5}}\,t_6^{\frac{8-9k}{5}}\,,
\eary
}
where  $\nu^{\rm syn}_{\rm a,l}$   for ${\rm l}$=1, 2 and 3 are defined in the range of {\small $\nu^{\rm syn}_{\rm a,1}\leq \nu^{\rm syn}_{\rm m} \leq \nu^{\rm syn}_{\rm c}$},  {\small $\nu^{\rm syn}_{\rm m} \leq\nu^{\rm syn}_{\rm a,2}\leq \nu^{\rm syn}_{\rm c}$ and  {\small $\nu^{\rm syn}_{\rm a,3}\leq \nu^{\rm syn}_{\rm c} \leq  \nu^{\rm syn}_{\rm m}$}, respectively.    Taking into account that the peak spectral power evolves as $P_{\rm \nu, max} \propto (1+z)^{\frac{k-2}{2}}\, \epsilon^\frac12_{\rm B}\,A^\frac12_{\rm k}$ $\beta^{\frac{2-k}{2}}\,t^{-\frac{k}{2}}$  and that the number of swept-up electrons in the post-shock is $N_{\rm e}\propto (1+z)^{k-3}A_{\rm k}\beta^{3-k}\, t^{3-k}$,  the spectral peak flux density is given by
{\small
\be\label{f_syn}
F^{\rm syn}_{\rm \nu,max}= F^{\rm syn,0}_{\rm \nu,max} \left(\frac{1+z}{1.022}\right)^{\frac{3k-4}{2}}\, \epsilon^{\frac12}_{\rm B,-2}\, d_{\rm z,26.5}^{-2}\, A^{\frac32}_{\rm k}\, \beta_{-0.5}^{\frac{8-3k}{2}}\,t_6^{\frac{3(2-k)}{2}}\,,
\ee
}
where {\small $d_{\rm z}=(1+z)\frac{c}{H_0}\int^z_0\,\frac{d\tilde{z}}{\sqrt{\Omega_{\rm M}(1+\tilde{z})^3+\Omega_\Lambda}}$}  \citep{1972gcpa.book.....W} is the luminosity distance with $c$ the speed of light.
The terms $\gamma^0_{\rm m}$, $\gamma^0_{\rm c}$, $\nu^{\rm syn,0}_{\rm m}$, $\nu^{\rm syn,0}_{\rm c}$ and $F^{\rm syn,0}_{\rm \nu,max}$ are given in Table \ref{table1} for ${\rm k}=0$, $1$, $1.5$, $2$ and $2.5$.\\  
Using the synchrotron break frequencies (eqs. \ref{nu_syn} and \ref{nu_syn_a}) and the spectral peak flux density (eq. \ref{f_syn}),  the synchrotron light curves in the fast- and the slow-cooling regime evolve as
{\small
\begin{eqnarray}
\label{fc_coast}
F^{\rm syn}_{\rm \nu}\propto \cases{ 
t^{1+k}\, \nu^{2},\hspace{1.8cm} \nu<\nu^{\rm syn}_{\rm a,3}, \cr
t^{\frac{11-6k}{3}}\, \nu^{\frac13},\hspace{1.4cm} \nu^{\rm syn}_{\rm a,3}<\nu<\nu^{\rm syn}_{\rm c}, \cr
t^{\frac{8-3k} {4}}\, \nu^{-\frac{1}{2}},\hspace{1.25cm} \nu^{\rm syn}_{\rm c}<\nu<\nu^{\rm syn}_{\rm m},\,\,\,\,\, \cr
t^{\frac{8-k(p+2)} {4}}\,\nu^{-\frac{p}{2}},\,\,\,\,\hspace{0.5cm}   \nu^{\rm syn}_{\rm m}<\nu\,, \cr
}
\end{eqnarray}
}
{\small
\begin{eqnarray}
\label{sc_coast}
F^{\rm syn}_{\rm \nu}\propto \cases{ 
t^{2}\, \nu^{2},\hspace{2.45cm} \nu<\nu^{\rm syn}_{\rm a,1}, \cr
t^{\frac{9-4k}{3}}\, \nu^{\frac13},\hspace{1.8cm} \nu^{\rm syn}_{\rm a,1} <\nu<\nu^{\rm syn}_{\rm m}, \cr
t^{\frac{12-k(p+5)} {4}}\, \nu^{-\frac{p-1}{2}},\hspace{0.6cm} \nu^{\rm syn}_{\rm m}<\nu<\nu^{\rm syn}_{\rm c},\,\,\,\,\, \cr
t^{\frac{8-k(p+2)} {4}}\,\nu^{-\frac{p}{2}},\,\,\,\,\hspace{0.8cm}   \nu^{\rm syn}_{\rm c}<\nu\,, \cr
}
\end{eqnarray}
}
and 
{\small
\begin{eqnarray}
\label{sc_coast}
F^{\rm syn}_{\rm \nu}\propto \cases{ 
t^{2}\, \nu^{2},\hspace{2.3cm} \nu<\nu^{\rm syn}_{\rm m}, \cr
t^{\frac{8+k}{4}}\, \nu^{\frac52},\hspace{1.8cm} \nu^{\rm syn}_{\rm m} < \nu<\nu^{\rm syn}_{\rm a,2}, \cr
t^{\frac{12-k(p+5)} {4}}\, \nu^{-\frac{p-1}{2}},\hspace{0.6cm} \nu^{\rm syn}_{\rm a,2}<\nu<\nu^{\rm syn}_{\rm c},\,\,\,\,\, \cr
t^{\frac{8-k(p+2)} {4}}\,\nu^{-\frac{p}{2}},\,\,\,\,\hspace{0.8cm}   \nu^{\rm syn}_{\rm c}<\nu\,, \cr
}
\end{eqnarray}
}
respectively.      It is worth noting that the synchrotron light curves in the fast-cooling regime are derived for completeness, since they are not relevant for the problems investigated here.
\subsubsection{SSC emission}
The electron distribution accelerated during the forward shock  up-scatters synchrotron photons, yielding the SSC spectrum which is characterized by the break frequencies $\nu^{\rm ssc}_{\rm ij}= 4 \gamma^2_{\rm i} \nu^{\rm syn}_{\rm j} x_0$ with ${\rm i,\,j=a}$, ${\rm m}$  and ${\rm c}$ where $x_0$ is a parameter that assures energy conservation \citep[e.g., see][]{2001ApJ...548..787S, 2001ApJ...559..110Z, 2013MNRAS.435.2520G}.   The  SSC break frequencies, the corresponding spectral peak flux density and the light curves in the fast- and slow-cooling regime are shown in appendix.
\subsection{The deceleration phase}
During the deceleration phase, the ejecta acquires a velocity structure,  the velocity of matter in the front of the ejecta is faster than the one that moves in the back \citep{2000ApJ...535L..33S}.  \cite{2001ApJ...551..946T} studied  the acceleration of the ejecta mass with relativistic and sub-relativistic velocities. They found that the equivalent kinetic energy in the non- and ultra-relativistic limit  can be expressed as a PL velocity distribution given by {\small $E_{\rm k} (\geq \beta) \propto \beta^{-5.2}$ for   $\beta\ll 1$ and  $E_{\rm k} (\geq \beta\Gamma) \propto \left( \beta\Gamma \right)^{-1.1}$} for $\beta\Gamma\gg 1$ (with $\Gamma=\sqrt{1/1-\beta^2}$), respectively.\footnote{The polytropic index $n_p=3$ is used.}    Here,  we consider the non-relativistic regime, so the equivalent kinetic energy distribution given by {\small $E_{\rm k} (\geq \beta)= \tilde{E}\,\beta^{-\alpha}$} with  $\tilde{E}$ the fiducial energy and the values $3 \leq \alpha \leq 5.2$ are used.  We adopt this range of values motivated by the numerical simulations presented in \cite{2001ApJ...551..946T}.\\
During this phase the dynamics of the non-relativistic ejecta mass is described by the Sedov-Taylor solution.   Then,  the velocity and the blast wave radius can be written as 
{\small
\be\label{beta_dec}
\beta=\beta^0 \,\left(\frac{1+z}{1.022}\right)^{\frac{3-k}{\alpha+5-k}}\,A^{-\frac{1}{\alpha+5-k}}_{\rm k}\,\tilde{E}_{51}^{\frac{1}{\alpha+5-k}}\, t_7^{\frac{k-3}{\alpha+5-k}}\,,
\ee
}
and
{\small
\be\label{R_dec}
r=r^0\,\left(\frac{1+z}{1.022}\right)^{-\frac{\alpha+2}{\alpha+5-k}}\,A^{-\frac{1}{\alpha+5-k}}_{\rm k}\, \tilde{E}_{51}^{\frac{1}{\alpha+5-k}}\,t_7^{\frac{\alpha+2}{\alpha+5-k}}\,,
\ee
}
respectively, where  the fiducial energy ($\tilde{E}$) can be estimated by calculating the term $\propto \beta^2 r^3$.   From eq. (\ref{beta_dec}) it can be noticed that the deceleration time evolves as
{\small
\be\label{t_dec}
t_{\rm dec}= t^0_{\rm dec} \left(\frac{1+z}{1.022}\right)\,A^{\frac{1}{k-3}}_{\rm k}\,\tilde{E}_{51}^{\frac{1}{3-k}}\, \beta_{-0.5}^{\frac{\alpha+5-k}{k-3}}\,.
\ee
}
The terms $r^0$ and $t^0_{\rm dec}$ are given in Table \ref{table1} for ${\rm k}=0$, $1$, $1.5$, $2$ and $2.5$.  It is worth noting that for a uniform-density medium (k=0) and $\alpha=0$,  the velocity and blast wave radius derived in \cite{2013ApJ...778..107S} are recovered.
\subsubsection{Synchrotron emission}
During the deceleration phase,  the post-shock magnetic field evolves as, $B'\propto \,t^{-\frac{6+k\alpha}{2(\alpha+5-k)}}$.   The Lorentz factors of the lowest-energy electrons and of the higher energy electrons, which are efficiently cooled by synchrotron emission are
{\small
\bary\label{gamma_dec}
\gamma_{\rm m}&=&\gamma^0_{\rm m}\,\left(\frac{1+z}{1.022}\right)^{\frac{2(3-k)}{\alpha+5-k}}\,g(p)\, \epsilon_{\rm e,-1}\, A^{-\frac{2}{\alpha+5-k}}_{\rm k}\,\tilde{E}_{51}^{\frac{2}{\alpha+5-k}}\, t_7^{\frac{2(k-3)}{\alpha+5-k}}\cr
\gamma_{\rm c}&=&\gamma^0_{\rm c}\left(\frac{1+z}{1.022}\right)^{-\frac{k+1+\alpha(k-1)}{\alpha+5-k}}\, (1+Y)^{-1} \epsilon^{-1}_{\rm B,-2}\, A^{-\frac{\alpha+3}{\alpha+5-k}}_{\rm k}\,\cr
&& \hspace{3.7cm} \times\, \tilde{E}_{51}^{\frac{k-2}{\alpha+5-k}}\, t_7^{\frac{k+1+\alpha(k-1)}{\alpha+5-k}}\,.
\eary
}

The corresponding synchrotron break frequencies are given by
{\small
\bary\label{nu_syn_de}
\nu^{\rm syn}_{\rm m}&=&\nu^{\rm syn,0}_{\rm m}\,\left(\frac{1+z}{1.022}\right)^{\frac{20+k(\alpha-6)-2\alpha }{2(\alpha+5-k)}}\,g(p)^{2} \epsilon^2_{\rm e,-1}\,\epsilon^\frac12_{\rm B,-2}\,  A^{\frac{\alpha-5}{2(\alpha+5-k)}}_{\rm k}\,\cr
&&\hspace{3.5cm}\,\times\, \tilde{E}_{51}^{\frac{10-k}{2(\alpha+5-k)}}\,t_7^{-\frac{30 + k(\alpha-8)}{2(\alpha+5-k)}}\cr
\nu^{\rm syn}_{\rm c}&=&\nu^{\rm syn,0}_{\rm c}\,\left(\frac{1+z}{1.022}\right)^{-\frac{8-2\alpha + k(3\alpha+2)}{2(\alpha +5-k)}}\, \epsilon^{-\frac32}_{\rm B,-2}\, (1+Y)^{-2} \,\cr
&&\hspace{1.cm}\,\times\, A^{-\frac{3(\alpha+3)}{2(\alpha+5-k)}}_{\rm k}\tilde{E}_{51}^{\frac{3(k-2)}{2(\alpha+5-k)}}\,t_7^{\frac{k(4+3\alpha) - 4\alpha-2}{2(\alpha+5-k)}}\,.
\eary
}

In the self-absorption regime, the synchrotron break frequencies are 
{\small
\bary
\nu^{\rm syn}_{\rm a,1}&=& \nu^{\rm syn,0}_{\rm a,1} \left(\frac{1+z}{1.022}\right)^{m_{\rm 11}} g(p)^{-1} \epsilon_{\rm e,-1}^{-1} \epsilon_{\rm B,-2}^{\frac15}\, A^{\frac{25+4\alpha}{5(\alpha+5-k)}}_{\rm k}\cr
&& \hspace{4.3cm} \times\,\tilde{E}_{51}^{-\frac{4k+5}{5(\alpha+5-k)}}  t_6^{m_{\rm 21}}\,\cr
\nu^{\rm syn}_{\rm a,2}&=& \nu^{\rm syn,0}_{\rm a,2} \left(\frac{1+z}{1.022}\right)^{m_{\rm 12}} g(p)^{\frac{2(p-1)}{p+4}} \epsilon_{\rm e,-1}^{\frac{2(p-1)}{p+4}}\, A^{\frac{\alpha p+6\alpha-5p+30}{2(p+4)(\alpha+5-k)}}_{\rm k} \cr
&&\hspace{3.1cm} \times\, \epsilon_{\rm B,-2}^{\frac{p+2}{2(p+4)}}\, \tilde{E}_{51}^{\frac{10p-kp-6k}{2(p+4)(\alpha+5-k)}}\,t_6^{m_{\rm 22}},\cr
\nu^{\rm syn}_{\rm a, 3}&=& \nu^{\rm syn,0}_{\rm a,3} \left(\frac{1+z}{1.022}\right)^{m_{\rm 13}}\,(1+Y)\, \epsilon_{\rm B,-2}^{\frac65}\, A^{\frac{3(3\alpha+10)}{ 5(\alpha+5-k)}}_{\rm k}\cr
&& \hspace{4.3cm} \times\, \tilde{E}_{51}^{\frac{15-9k}{5(\alpha+5-k)}} t_6^{m_{\rm 23}},
\eary
}
where the PL indices $m_{\rm ij}$ for $i,\,j=$1, 2 and 3 are explicitly shown in the appendix.\\
Taking into account that the peak spectral power evolves as $P_{\rm \nu, max} \propto\,t^{-\frac{6+k\alpha}{2(\alpha+5-k)}}$ and the number of swept-up electrons in the post-shock develops as $N_{\rm e}\propto  t^\frac{6-2k+\alpha(3-k)}{\alpha+5-k}$,  the spectral peak flux density becomes
{\small
\bary\label{f_syn_de}
F^{\rm syn}_{\rm \nu,max}&=&F^{\rm syn,0}_{\rm \nu,max}\,\left(\frac{1+z}{1.022}\right)^{\frac{4+2k-4\alpha+3k\alpha}{2(\alpha+5-k)}}\, \epsilon^{\frac12}_{\rm B,-2}\, d_{\rm z,26.5}^{-2}\, A^{\frac{3\alpha+7}{2(\alpha+5-k)}}_{\rm k}\,\cr
&&\hspace{3.cm}\,\times\,   \tilde{E}_{51}^{\frac{8-3k}{2(\alpha+5-k)}}\,t_7^{\frac{6-4k + 6\alpha-3k\alpha}{2(\alpha+5-k)}}.\,\,\,\,\,
\eary
}
The terms $\gamma^0_{\rm m}$, $\gamma^0_{\rm c}$, $\nu^{\rm syn,0}_{\rm m}$, $\nu^{\rm syn,0}_{\rm c}$ and $F^{\rm syn,0}_{\rm \nu,max}$ are given in Table \ref{table1} for ${\rm k}=0$, $1$, $1.5$, $2$ and $2.5$.\\  
Using the synchrotron break frequencies (eq.~\ref{nu_syn_de}) and the spectral peak flux density (eq.~\ref{f_syn_de}),  the synchrotron light curves in the fast- and the slow-cooling regime are
{\small
\begin{eqnarray}
\label{fc_dec}
F^{\rm syn}_{\rm \nu}\propto \cases{ 
t^{\frac{ 5+k+ \alpha + k\alpha}{\alpha+5-k}}\, \nu^{2},\hspace{3.cm} \nu<\nu^{\rm syn}_{\rm a,3}, \cr
t^{\frac{ 10-8k+11\alpha-6k \alpha}{3(\alpha+5-k)}}\, \nu^{\frac13},\hspace{2.3cm}  \nu^{\rm syn}_{\rm a,3}  < \nu<\nu^{\rm syn}_{\rm c}, \cr
t^{\frac{10-4k+8\alpha-3k \alpha}{4(\alpha+5-k)}}\, \nu^{-\frac{1}{2}},\hspace{2.2cm} \nu^{\rm syn}_{\rm c}<\nu<\nu^{\rm syn}_{\rm m},\,\,\,\,\, \cr
t^{-\frac{30p+kp(\alpha-8) -8(\alpha+5)+2k(\alpha+6) }{4(\alpha+5-k)}}\,\nu^{-\frac{p}{2}},\,\,\hspace{0.1cm}   \nu^{\rm syn}_{\rm m}<\nu\,, \cr
}
\end{eqnarray}
}
{\small
\begin{eqnarray}
\label{sc_dec}
F^{\rm syn}_{\rm \nu}\propto \cases{ 
t^{\frac{2(\alpha+k-1)}{\alpha+5-k}}\, \nu^{2},\hspace{3.3cm}  \nu<\nu^{\rm syn}_{\rm a,1}, \cr
t^{\frac{24+9\alpha-2k(2\alpha+5)}{3(\alpha+5-k)}}\, \nu^{\frac13},\hspace{2.4cm}  \nu^{\rm syn}_{\rm a,1}< \nu<\nu^{\rm syn}_{\rm m}, \cr
t^{-\frac{6(5p-2\alpha-7) +k(16+p(\alpha-8)+5\alpha)}{4(\alpha+5-k)}}\, \nu^{-\frac{p-1}{2}},\hspace{0.01cm} \nu^{\rm syn}_{\rm m}<\nu<\nu^{\rm syn}_{\rm c},\,\,\,\,\, \cr
t^{-\frac{30p+kp(\alpha-8) -8(\alpha+5)+2k(\alpha+6) }{4(\alpha+5-k)}}\,\nu^{-\frac{p}{2}},\,\,\,\,\hspace{0.05cm}   \nu^{\rm syn}_{\rm c}<\nu\,, \cr
}
\end{eqnarray}
}
and
{\small
\begin{eqnarray}
\label{sc_dec}
F^{\rm syn}_{\rm \nu}\propto \cases{
t^{\frac{2(\alpha+k -1)}{\alpha+5-k}}\, \nu^{2},\hspace{3.4cm} \nu<\nu^{\rm syn}_{\rm m}, \cr
t^{\frac{22+\alpha(k+8)}{4(\alpha+5-k)}}\, \nu^{\frac52},\hspace{3.1cm}  \nu^{\rm syn}_{\rm m} <  \nu<\nu^{\rm syn}_{\rm a,2}, \cr
t^{-\frac{6(5p-2\alpha-7) +k(16+p(\alpha-8)+5\alpha)}{4(\alpha+5-k)}}\, \nu^{-\frac{p-1}{2}},\hspace{0.01cm} \nu^{\rm syn}_{\rm a,2}<\nu<\nu^{\rm syn}_{\rm c},\,\,\,\,\, \cr
t^{-\frac{30p+kp(\alpha-8) -8(\alpha+5)+2k(\alpha+6) }{4(\alpha+5-k)}}\,\nu^{-\frac{p}{2}},\,\,\,\,\hspace{0.05cm}   \nu^{\rm syn}_{\rm c}<\nu\,, \cr
}
\end{eqnarray}
}
respectively.

\subsubsection{SSC emission}
The SSC break frequencies,  the corresponding spectral peak flux density and the light curves in the fast- and slow-cooling regime are shown in Appendix.
\section{Analysis of the multi-wavelength light curves}
\subsection{Analysis of synchrotron light curves}
Figures \ref{k_0} - \ref{k_25} show the predicted synchrotron light curves produced by the deceleration of the non-relativistic ejecta in a circumstellar medium described by a density profile $A_{\rm k} r^{-\rm k}$ with ${\rm k}=0$, $1$, $1.5$, $2$ and $2.5$, respectively. Panels from top to bottom correspond to the electromagnetic bands in radio  at 6 GHz, optical at 1 eV and X-rays at 1 keV for $\tilde{E}=10^{51}\,{\rm erg}$, $\epsilon_{\rm B}=10^{-2}$, $\epsilon_{\rm e}=10^{-1}$ and $d_z=100\,{\rm Mpc}$.  The left-hand panels show the light curves for $p=2.6$ with $\alpha=3$, $4$ and $5$, and the right-hand panels show the light curves for $\alpha=3$ with $p=2.2$, $2.8$  and $3.4$.  All the figures lie in the slow-cooling regime and exhibit a deceleration timescale from several months to a few years due to the set of parameter values considered.   Similar timescales have been observed in the light curves from radio to hard X-rays in some SNe \citep[e.g., SN2014C and SN2016aps;][respectively]{2017ApJ...835..140M, 2020NatAs.tmp...78N}.  If we had chosen another set of parameters such as  a fiducial energy $\tilde{E}\simeq 10^{52}\,{\rm erg}$, a uniform-density medium $A_0\simeq 10\,{\rm cm^{-3}}$ and  equipartition parameters $\epsilon_{\rm e}\simeq 0.5$ and $\epsilon_{\rm B}\simeq 0.1$,  the synchrotron light curves would lie in the fast-cooling regime.   Consequently,  using  a set  of parameters such as  $\tilde{E}\simeq 10^{47}\,{\rm erg}$,  $A_2\simeq 3\times 10^{35}\,{\rm cm^{-1}}$ and $\beta=0.5$ for ${\rm k}=2$, and $\tilde{E}\sim 10^{47}\,{\rm erg}$, $A_0\simeq 1\,{\rm cm^{-3}}$ and $\beta=0.5$ for ${\rm k}=0$,  deceleration timescales from hours to months would have been obtained. This is also the case, for instance, where the ejecta mass is decelerated in a very dense medium  \citep[for discussion,  see][]{2011ApJ...729L...6C, 2020NatAs.tmp...78N}.\\
The density profile of $n(r)=A_{\rm k}r^{-\rm k}$ with ${\rm k}=0$, $1$, $1.5$, $2$ and $2.5$ for the  circumstellar medium that covers short and long GRB progenitors is used.  While  the uniform-density  medium (${\rm k}=0$) is expected to be connected with binary compact objects and CC-SNe, the stratified medium ($1\leq {\rm k} \leq 2.5$) is only expected to be associated to dying massive stars with different mass-loss evolution. For instance,   \cite{2013ApJ...776..120Y} and \cite{2013ApJ...774...13L} studied the dynamics of synchrotron external-shock emission in a sample of 19 and 146 GRBs and found that they were successfully described when the outflow was decelerated in an external environment with $0.4\leq {\rm k} \leq 1. 4$ and ${\rm k} \approx 1$, respectively.    \cite{2020arXiv200405941I} presented multi-wavelength observations of the nearby SN 2020bvc. The authors found that the X-ray observations were consistent with the afterglow emission generated by an off-axis jet with viewing angle of $23^\circ$ when it was decelerated in a circumburst medium  with a density profile with ${\rm k}=1.5$.\\
Figure  \ref{k_0} exhibits that during the coasting phase, the flux increases gradually, and during the deceleration phase, depending on the values of $\alpha$ and $p$, a flattening  or decrease in the light curve is expected.   Figures \ref{k_10} - \ref{k_25} show that for a density profile with $1< {\rm k} \leq 2.5$  the rebrightening in the light curves is not so evident. Therefore, a flattening or rebrightening at  timescales from months to years  in the light curves together with GW detection would be associated with the deceleration of a non-relativistic ejecta launched during the coalescence of binary compact objects or a CC-SN.   Even a flattening or rebrightening at  timescales of days could be detected with extreme values of circumstellar density or fiducial energy.    Additionally, it can be concluded that an observed flux that decreases early would be associated with the deceleration of  a non-relativistic ejecta launched by dying massive stars with different mass-loss evolution (the mass loss rate $\dot{M}$ and/or the wind velocity $v_{\rm w}$) at the end of their life \citep{2005ApJ...631..435R,  2006A&A...460..105V}.\\ 
The synchrotron light curve in the slow-cooling regime as a function of the density parameter during the deceleration phase is given by 
{
\small
\begin{eqnarray}
\label{sc_dec}
F^{\rm syn}_{\rm \nu}\propto \cases{
A_{\rm k}^{-\frac{4}{\alpha+5-k}},\hspace{1.4cm} \nu<\nu^{\rm syn}_{\rm a,1}, \cr 
A_{\rm k}^{\frac{4\alpha+13}{3(\alpha+5-k)}},\hspace{1.4cm} \nu^{\rm syn}_{\rm a,1}<\nu^{\rm syn}_{\rm m}, \cr
A_{\rm k}^{\frac{19+p(\alpha-5)+5\alpha}{4(\alpha+5-k)}},\hspace{0.8cm} \nu^{\rm syn}_{\rm m}<\nu<\nu^{\rm syn}_{\rm c},\,\,\,\,\, \cr
A_{\rm k}^{\frac{p(\alpha-5)+2(\alpha+5)}{4(\alpha+5-k)}},\,\,\,\,\hspace{0.55cm}   \nu^{\rm syn}_{\rm c}<\nu\,. \cr
}
\end{eqnarray}
}

The predicted flux is less sensitive to the density parameter for higher frequencies than for the lower ones (e.g., the radio light curve is more sensitive to variations of the density parameter than the X-ray light curve).   It can also be noticed that the predicted flux is more sensitive to the density parameter for larger values of $k$ and $\alpha$.   It is worth noting  that a transition phase from stellar-wind (k=2) to uniform-density (k=0) medium would be more evident in lower-frequency fluxes (e.g., this transition is more noticeable in radio or optical than in X-ray bands).  A similar signature was useful to describe the type of the progenitor, the mass-loss evolution, the afterglow emission by the deceleration of the relativistic jet,  and also to estimate the transition radius at $\sim (0.1 - 1)$ pc of  some bursts \cite[e.g. see GRB 050319, 080109A, 160625B and 190114C;][]{2007ApJ...664L...5K, 2009MNRAS.400.1829J, 2017ApJ...848...15F, 2019ApJ...879L..26F}.\\  
 Figures \ref{k_0} - \ref{k_25} show the synchrotron light curves for distinct values of $p$ and $\alpha$, and the same values of the microphysical parameters $\epsilon_{\rm B}=10^{-2}$ and  $\epsilon_{\rm e}=10^{-1}$.  However, any variation of the microphysical parameters will increase or decrease the predicted fluxes. For example, the synchrotron light curve in the slow-cooling regime as a function of these parameters evolves as  {\small $F^{\rm syn}_{\nu}:\propto   \epsilon^{-\frac23}_{\rm e}\, \epsilon^{\frac13}_{\rm B}$}  for {\small $\nu <\nu^{\rm syn}_{\rm m}$},  {\small $\propto  \epsilon^{p-1}_{\rm e}\, \epsilon^{\frac{p+1}{4}}_{\rm B}$} for {\small $\nu^{\rm syn}_{\rm m}< \nu < \nu^{\rm syn}_{\rm c}$} and {\small $\propto \epsilon^{p-1}_{\rm e}\, \epsilon^{\frac{p-2}{4}}_{\rm B}$} for {\small $\nu^{\rm syn}_{\rm c}< \nu$}.  The expected flux is more sensitive to the parameter $\epsilon_{\rm B}$ for lower frequencies than for higher ones, and the parameter $\epsilon_{\rm e}$ for higher frequencies than for  lower ones.\\
The synchrotron spectral breaks during the non-relativistic regime evolve as  $\nu^{\rm syn}_{\rm m}\propto t^{-\frac{k}{2}}$ and $\nu^{\rm syn}_{\rm c}\propto t^{\frac{3k-4}{2}}$ in the coasting phase, and $\nu^{\rm syn}_{\rm m}\propto t^{-\frac{30 + k(\alpha-8)}{2(\alpha+5-k)}}$  and  $\nu^{\rm syn}_{\rm c}\propto t^{\frac{k(4+3\alpha) - 4\alpha-2}{2(\alpha+5-k)}}$ in the deceleration phase.  For instance, these breaks evolve as  $\nu^{\rm syn}_{\rm m}\propto t^0$ and $\nu^{\rm syn}_{\rm c}\propto t^{-2}$ in the coasting phase and $\nu^{\rm syn}_{\rm m}\propto t^{-3}$ and $\nu^{\rm syn}_{\rm c}\propto t^{-\frac15}$ in the deceleration phase for a uniform-density medium, and   $\nu^{\rm syn}_{\rm m}\propto t^{-1}$ and $\nu^{\rm syn}_{\rm c}\propto t$ in the coasting phase and $\nu^{\rm syn}_{\rm m}\propto t^{-\frac73}$ and $\nu^{\rm syn}_{\rm c}\propto t$ in the deceleration phase for a stellar-wind medium.   Here, we present a valuable tool to pinpoint the emission from the non-relativistic ejecta as previously done by \cite{1999ApJ...524L..47G} in the case of the relativistic regime. \cite{1999ApJ...524L..47G} analyzed the prompt gamma-ray emission in the BATSE\footnote{Burst and Transient Source Experiment} detected burst GRB 980923. The light curve exhibited a main prompt episode lasting $\sim 40\,{\rm s}$ followed by a smooth emission tail that lasted $\sim 400\,{\rm s}$. The authors found that the spectrum in the smooth tail evolved as the synchrotron cooling break  $\propto t^{-0.52\pm0.12}$, concluding that the afterglow began during the prompt gamma-ray emission. Afterward, spectra analysis of GRB tails were done in order to identify early afterglows \citep[e.g., see][]{2005ApJ...635L.133B, 2006MNRAS.369..311Y}.\\

\subsection{Analysis of SSC light curves}
Figure \ref{ssc_all_k} shows the predicted SSC light curves at 100 keV (upper), 10 GeV (medium) and 100 GeV (lower) for the deceleration of the non-relativistic ejecta mass in a circumstellar medium with a density profile $A_{\rm k}r^{-\rm k}$ with ${\rm k}=0$, $1$, $1.5$, $2$ and $2.5$, respectively.  The left-hand panels show the light curves for $p=2.4$ and $\alpha=3$, and the right-hand panels for $p=2.8$ and $\alpha=5$.  In order to obtain the SSC light curves, we use the same typical values that are used for the synchrotron light curves.   The effect of the extragalactic background light (EBL) absorption modelled in \cite{2017A&A...603A..34F} is used.\\
\\
The purple solid line (${\rm k}=0$) in Figure  \ref{ssc_all_k} shows that a flattening or rebrightening in the light curve is expected, but not in a density profile with ${\rm k}>1$. This feature at timescales from months to years  together with GW detection would be associated with the deceleration of a non-relativistic ejecta launched during the coalescence of a binary compact object or a CC-SN. On the contrary, it is expected that an observed flux that decreases early  would be associated with the deceleration of a non-relativistic ejecta launched in the collapse of massive stars with different mass-loss evolution at the end of their life \citep{2005ApJ...631..435R,  2006A&A...460..105V}.\\ 
The SSC light curve in the slow-cooling regime as a function of the density parameter during the deceleration phase can be written as 
{
\small
\begin{eqnarray}
\label{sc_dec}
F^{\rm ssc}_{\rm \nu}\propto \cases{ 
A_{\rm k}^{\frac{9(\alpha+5)}{5(\alpha+5-k)}},\hspace{4.3cm} \nu<\nu^{\rm ssc}_{\rm a,1}, \cr
A_{\rm k}^{\frac{29 + 7\alpha}{3(\alpha+5-k)}},\hspace{4.3cm} \nu^{\rm ssc}_{\rm a,1}<\nu^{\rm ssc}_{\rm m}, \cr
A_{\rm k}^{\frac{43+p(\alpha-13)+9\alpha}{4(\alpha+5-k)}}\left(C_{1}+C_{2} {\rm ln} A_{\rm k}^{-\frac{7(\alpha+3)}{2(\alpha+5-k)}}\right),\hspace{0.0cm} \nu^{\rm ssc}_{\rm m}<\nu<\nu^{\rm ssc}_{\rm c},\,\,\,\,\, \cr
A_{\rm k}^{\frac{22-13p+2\alpha+\alpha p}{4(\alpha+5-k)}}\left(C_{3}+C_{4} {\rm ln} A_{\rm k}^{\frac{7(\alpha+3)}{2(\alpha+5-k)}}\right),\,\,\,\,\hspace{0.1cm}   \nu^{\rm ssc}_{\rm c}<\nu\,. \cr
}
\end{eqnarray}
}
The parameters $C_{\rm s}$ do not evolve with the density parameter.  Considering that the contribution from the logarithm function is small, this light curve has a similar behavior to the synchrotron light curves. 
\\
Inverse Compton (IC) scattering between an electron distribution accelerated  during the shock wave and the photon field from SNe has been explored in order to explain the X-ray emission observed in timescales of days after the explosion \citep{2004ApJ...605..823B, 2006ApJ...651..381C, 2006ApJ...641.1029C, 2013ApJ...778...18M}. \cite{2004ApJ...605..823B} studied the X-ray and radio emission from SN 2002ap.  The authors proposed that IC scattering could explain the X-ray excess observed at late times. They computed that the IC spectrum six days after the explosion peaks at energies of some MeV.   \cite{2006ApJ...651..381C} found that although the IC emission could reproduce the  X-ray flux observed in SN2002ap for a spectral index of $p=3$, this process is not promising for other SNe unless the electron energy density is much larger than the magnetic energy density. On the other hand,   \cite{2012ApJ...751..134M} introduced an analytic formalism for the IC scattering in the X-ray energy range and in the SN scenario with compact progenitors.  They showed that the IC luminosity evolves as $\propto t^{1.29-0.58p}$ for a uniform-density medium and $\propto t^{-(0.24p+0.64)}$ for a wind medium.  In this paper, we propose that  in addition to the external IC scattering process, the SSC mechanism could be present.  Figure  \ref{ssc_all_k} shows that in a density profile with $k>1$, the SSC flux could be detected in timescales of days with the evolution given by eqs. (\ref{fc_ssc_t}) and (\ref{sc_ssc_t}).   For instance, the SSC fluxes in the slow-cooling regime evolve as  {\small $\propto t^{\frac{4\alpha+14}{\alpha+5}}$} for {\small $\nu<\nu^{\rm ssc}_{\rm m}$},   {\small $\propto t^\frac{37 - 27p +8\alpha}{2(\alpha+5)}$} for {\small $\nu^{\rm ssc}_{\rm m}<\nu<\nu^{\rm ssc}_{\rm c}$} and  {\small $\propto t^\frac{38 - 27p +4\alpha}{2(\alpha+5)}$} for  {\small $\nu^{\rm ssc}_{\rm c}<\nu$} for a uniform-density medium,  and as {\small $\propto t^{\frac{2(1-\alpha)}{3(\alpha+3)}}$} for {\small $\nu<\nu^{\rm ssc}_{\rm m}$},   {\small $\propto t^\frac{5-\alpha-p(\alpha+11)}{2(\alpha+3)}$} for {\small $\nu^{\rm ssc}_{\rm m}<\nu<\nu^{\rm ssc}_{\rm c}$} and  {\small $\propto t^\frac{14 - 11p +\alpha(2-p)}{2(\alpha+3)}$} for  {\small $\nu^{\rm ssc}_{\rm c}<\nu$} for a stellar-wind medium, which have a different evolution  to the IC scattering model derived in \cite{2012ApJ...751..134M}.\\
\section{The non-relativistic masses ejected from the coalescence of a NS binary}
The coalescence of NS mergers is widely accepted  to launch significant masses with different velocities which will contribute at distinct timescales, frequencies and intensities. Once these ejecta masses move into the circumstellar medium,  the initial expansion phase is not affected until the swept up quantity of material is equal to the ejected masses. In this moment, the ejecta masses start to be decelerated. The non-relativistic masses ejected from the coalescence of the NS merger are the dynamical ejecta,  the cocoon material, the shock breakout material and the wind ejecta. As follows we give a brief introduction of each of these ejecta masses.

\subsection{The dynamical ejecta}
The dynamical ejecta is formed during the coalescence  due to hydrodynamical and gravitational interactions \citep{1994ApJ...431..742D, 1997A&A...319..122R, 1999A&A...341..499R}. Based on numerical simulations, the ejecta mass liberated, the kinetic energy and the velocities lie in the ranges of  $10^{-4}\lesssim M_{\rm ej}\lesssim10^{-2}\,{\rm M_{\odot}}$,  $10^{49}\lesssim E\lesssim10^{51}\,{\rm erg}$ and $0.1\lesssim\beta\Gamma\lesssim0.3$, respectively \citep[e.g., see][]{2011ApJ...738L..32G, 2013ApJ...778L..16H,2013ApJ...773...78B, 2013MNRAS.430.2121P, 2014ApJ...789L..39W, 2014MNRAS.439..757G}.
\subsection{The cocoon material}
As the GRB jet makes its way through the neutrino-driven or magnetically driven wind (previously ejected during the coalescence of the NS merger), it will deposit energy around its way.   The energy deposited laterally  will form a cocoon which may have an energy comparable to the  electromagnetic emission created by the jet.    \cite{2014ApJ...788L...8M} studied the necessary conditions for a cocoon production as a function of the jet luminosity.  The authors found that when the jet has a low or a high luminosity, a weak cocoon emission is expected.  \cite{2014ApJ...784L..28N} numerically showed that when a low-luminosity jet is considered, a hot cocoon confining the jet  is formed.  As soon as the relativistic jet reaches the shock-breakout material, the cocoon  breakouts and expands along the jet's axis.  Beyond the breakout material, the external pressure decreases abruptly, so the cocoon can accelerate and expand relativistically until it becomes transparent.  Accelerated material from the cocoon fireball will continue moving in the jet's axis.  It is worth mentioning that the significance of the cocoon also depends on the delay time between the merger and jet launching \citep{2019ApJ...877L..40G}.   The ejecta mass liberated in the cocoon, the kinetic energy and the velocities lie in the ranges of  $10^{-6}\lesssim M_{\rm ej}\lesssim10^{-4}\,{\rm M_{\odot}}$,  $10^{47}\lesssim E\lesssim10^{50.5}\,{\rm erg}$ and $0.2\lesssim\beta\Gamma\lesssim10$, respectively  \cite[e.g., see][]{2014ApJ...784L..28N, 2014ApJ...788L...8M,2017ApJ...848L...6L, 2018PhRvL.120x1103L, 2017ApJ...834...28N,2018MNRAS.473..576G}.
\subsection{The shock breakout material}
The shock breakout material  properties depend on the mass, radius and velocity of the merger remnant.  Immediately  after the coalescence occurs, a shock formed at the interface between the two NSs  is initially  ejected from the NS core to the NS crust at sub-relativistic velocities \citep[$\beta_{\rm in}\simeq 0.25$   e.g., see][]{2014MNRAS.437L...6K, 2015MNRAS.446.1115M}.   Once the shock reaches half of the escape velocity, a fraction of the thermal energy is converted into kinetic energy and  it can leave the  merger \citep[for details see][]{2014MNRAS.437L...6K, 2019ApJ...871..200F}.  Numerical simulations indicate that the ejecta mass, the kinetic energy and the velocities lie in the ranges of $10^{-6}\lesssim M_{\rm ej}\lesssim10^{-4}\,{\rm M_{\odot}}$,  $10^{47}\lesssim E\lesssim10^{50.5}\,{\rm erg}$ and $\beta\Gamma \gtrsim0.8$, respectively  \citep[e.g., see][]{2014MNRAS.437L...6K, 2015MNRAS.446.1115M}.\\
\subsection{The disk wind ejecta}
The coalescence of the NS binary will finish in a tidal disruption, resulting in the formation of an accretion disk around the central remnant.  The accretion disk will have a mass  in the range of $10^{-3}\lesssim M_{\rm ej}\lesssim 0.3\,{\rm M_{\odot}}$
\citep{2006PhRvD..73f4027S, 2013ApJ...778L..16H}, and due to a large source of thermal neutrinos \citep{1999ApJ...518..356P}, it could generate an outflow driven by neutrino heating similar to neutrino driven proto-NS winds in CC-SNe \citep{2008ApJ...679L.117S, 2008MNRAS.385.1455M}.   It  represents a significant source of ejecta mass that might even dominate over other ejecta masses as suggested by \cite{2017PhRvL.119w1102S}.  The ejecta mass, the kinetic energy and the velocities lie in the ranges of  $10^{-4}\lesssim M_{\rm ej}\lesssim10^{-1.3}\,{\rm M_{\odot}}$,  $10^{47}\lesssim E\lesssim10^{50}\,{\rm erg}$ and $0.03 \lesssim \beta\Gamma \lesssim0.1$, respectively  \citep[e.g., see][]{2009ApJ...690.1681D, 2014MNRAS.441.3444M, 2015MNRAS.446..750F}.\\
\subsection{Analysis of the multi-wavelength light curves}\label{analysis_sgrbs}
Figure \ref{lc_ns} shows the synchrotron and SSC light curves generated by the non-relativistic masses  ejected from the coalescence of a NS binary such as  the dynamical ejecta,  the cocoon material, the shock breakout material and the wind ejecta.\footnote{The pair of values $(\tilde{E}=10^{50}\,{\rm erg} ; \beta=0.2)$ is used for  the dynamical ejecta, $(10^{48}\,{\rm erg}; 0.3)$ for the cocoon material, $(10^{48.5}\,{\rm erg}; 0.8)$ for the shock breakout material and $(10^{50}\,{\rm erg}; 0.07)$ for  the wind. All light curves are shown  for $n=1\,{\rm cm^{-3}}$, $\alpha=3$, $\epsilon_{\rm B}=10^{-2}$, $\epsilon_{\rm e}=10^{-1}$, $p=3.2$ and $d_z=100\,{\rm Mpc}$.}    The electromagnetic emission from the non-relativistic masses is shown as a bump at a timescale of $\geq 10^3$ days.   In addition, we consider the synchrotron and SSC emission from an on-axis and off-axis relativistic jet with viewing angles of $\theta=15^\circ$, $30^\circ$ and $60^\circ$.\footnote{The values of $\tilde{E}=10^{50}\,{\rm erg}$,  $\epsilon_{\rm B}=10^{-2}$, $\epsilon_{\rm e}=10^{-1}$,  $n=10^{-2}\,{\rm cm^{-3}}$,       $p=2.2$  and $d_z=100\,{\rm Mpc}$  are used for the relativistic jet.}  The light curves correspondent to the off-axis jet  are plotted in accordance with the afterglow model introduced in  \cite{2019ApJ...884...71F}.  The left-hand panels show the synchrotron light curves which correspond to (from top to bottom) radio (1.6 GHz),  optical (1 eV) and X-ray (1 keV) bands, respectively, and the right-hand panels show the SSC light curves which correspond to (from top to bottom)  gamma-ray fluxes  at 100 keV, 10 GeV and 100 GeV, respectively.  The coalescence of the two NSs  launches  significant non-relativistic masses with different velocities which will contribute at distinct timescales, frequencies and intensities. These ejecta masses interact with the  circumstellar medium generating non-thermal emission.   The predicted synchrotron fluxes generated from: i) the shock breakout material peak at timescales from hours to days, ii) the cocoon material  peak at timescales from weeks to months, iii)  the dynamical ejecta peak at years and iv) the disk wind ejecta peak at centuries.  Similar timescales with different intensities are shown in the  SSC light curves.\\ 
This figure shows that the afterglow emission originated from the deceleration of the on-axis relativistic jet would have to decrease so that the electromagnetic emission from the non-relativistic masses could be observed.   \cite{2014ApJ...788L...8M, 2014ApJ...784L..28N} estimated the necessary conditions for sGRB production in the coalescence of the NS binary. In the case that such conditions are not satisfied, a relativistic jet is not expected and the electromagnetic emission from the non-relativistic masses could be observed since early times.\\
Because the shock breakout material is described with a velocity of $\beta=0.8$   \citep{2015MNRAS.446.1115M},  the trans-relativistic regime introduced in \cite{1998A&A...336L..69H} is used. In this regime,   the kinetic energy of the shock in the uniform-density medium is given by $E_{\rm K}=\frac{4\pi}{3}\sigma m_p c^2 \beta^2\Gamma^2 r^3 n$ \citep{1976PhFl...19.1130B} with $\sigma=0.73-0.38\beta$ and $r\approx \beta t$ \citep{1998A&A...336L..69H}.  Taking into account this regime,  the equivalent kinetic energy distribution will be $E=\tilde{E}\left(\beta\Gamma \right)^{-\alpha}$, and the deceleration time becomes 
\be\label{t_dec_new}
t_{\rm dec}=\left( \frac{3}{4\pi\sigma m_p c^5} \right)^\frac13\,(1+z)\,A^{-\frac13}_0\,\tilde{E}^{\frac13}\,\beta^{-\frac{\alpha+5}{3}}\Gamma^{-\frac{\alpha+2}{3}}.
\ee
In the trans-relativistic regime, the velocity obtained from eq. (\ref{t_dec_new}) is used to find the synchrotron and SSC quantities.  It is worth noting that in the limit $\Gamma \to 1$, the equations  (\ref{beta_dec}) and (\ref{t_dec}) are recovered.\\
In the ``deep Newtonian" regime the Lorentz factor of the lowest-energy electrons is $\gamma_{\rm m}\simeq 2$ \citep{2013ApJ...778..107S, 2016MNRAS.461.1568K, 2020arXiv200413028M}.   Using eq. (\ref{gamma_dec}), the deceleration time in this regime becomes 
\be
t_{\rm DN}\simeq 2\times 10^5\,{\rm days}\,  \left(\frac{1+z}{1.022}\right)\,\epsilon^{\frac{\alpha+5}{6}}_{\rm e, -1}\, A^{-\frac13}_{\rm 0}\,\tilde{E}^\frac13_{51}\,,
\ee
and the velocity of the ejecta is $\beta\simeq0.05$.  In this case the characteristic break frequency evolves as $\nu^{\rm syn}_{\rm m}\propto t^{-\frac{3}{\alpha+5}}$,  the spectral peak flux density as  $F^{\rm syn}_{\rm \nu, max}\propto t^{\frac{3(\alpha-1)}{\alpha+5}}$ and  the predicted  flux evolves as {\small $F_\nu= F^{\rm syn}_{\nu, max} \left(\frac{\nu}{\nu_{\rm m}} \right)^{-\frac{p-1}{2}} \propto t^{-\frac{3(1+p-\alpha)}{2(\alpha+5)}}\nu^{-\frac{p-1}{2}}$} for {\small $\nu^{\rm syn}_{\rm m} <\nu< \nu^{\rm syn}_{\rm c}$}.  Therefore, it is worth noting that due to the range of velocities considered ($0.07\leq \beta\leq 0.8$) and the deceleration time scales, the ``deep Newtonian" regime is only required for the wind ejecta.
The coalescence of the NS binaries launches  significant non-relativistic masses with different velocities which will contribute at distinct timescales, frequencies and intensities. These non-relativistic masses interact with the  circumstellar medium generating non-thermal electromagnetic emission.  We calculated the expected gamma-ray, X-ray, optical and radio fluxes via  SSC and synchrotron emission from electrons accelerated in the forward shocks. These electromagnetic signatures at different timescales and frequencies would be similar to those detected around  SNe for a uniform density medium and also  be present together with the GW detections.\\ 
\vspace{1cm}
\section{The non-relativistic masses ejected from the core-collapses of dying massive stars}
The origin of lGRBs is widely accepted to be connected to the death of massive stars leading to SNe, where the afterglow emission from bursts is detected together with  a type Ic SN with broad lines \citep[e.g., see][]{2005NatPh...1..147W,  2017AdAst2017E...5C}.  Sub-energetic GRBs are believed to be quasi-spherical explosions dominated by the non-relativistic ejecta mass. The non-relativistic material carries $\approx 99.9\%$ of the explosion energy,  and the mildly relativistic ejecta only $\approx0.1\%$.  The energy carried by the non-relativistic ejecta is similar to that exhibited by the sub-energetic bursts and comparable to the most energetic type Ib/c SNe.  \cite{2014ApJ...797..107M} showed the equivalent kinetic energy profile as a function of ejecta mass velocity in the non-relativistic and relativistic regime of ordinary type Ibc SNe, engine-driven SNe (E-SNe), sub-E GRBs and relativistic SNe.   The velocity of the ejecta mass was divided into the non-relativistic phase $0.02\lesssim\beta\lesssim 0.3$, mildly relativistic $\Gamma\beta\approx 0.6-0.8$ and relativistic $2\lesssim\beta\lesssim20$. It is worth noting that broadened lines in their spectra, indicate that a diverse range of materials with distinct non-relativistic expansion velocities are present \citep[e.g., $O_{\rm II}$, $N_{\rm II}$, $S_{\rm II}$, etc; ][]{2020ApJ...892..153M}.\\
Depending on the range of values in the observables such as luminosity, duration and bulk Lorentz factor,  lGRBs could be successful or choked \citep[e.g., see][]{2001ApJ...550..410M, 2001PhRvL..87q1102M, 2014MNRAS.437.2187F, 2017MNRAS.472..616S, 2011ApJ...739L..55B}. For instance, choked GRBs might be more frequent than successful ones, only limited by the ratio of SNe (types Ib/c and II) to lGRB rates \citep{2003ApJ...598.1151T, 2005PhRvL..95f1103A}. Some SN of type Ic-BL not connected with GRBs have been suggested to arise from events such as off-axis GRBs or failed jets \citep[e.g., see][]{2019Natur.565..324I, 2020arXiv200405941I,2020MNRAS.493.3521B}.  This is the case of the failed burst GRB 171205A, which besides being associated to SN 2017iuk,  exhibited material with high expansion velocities $\beta \sim 0.4$ interpreted as mildly relativistic cocoon material  \citep{2019Natur.565..324I}.   \cite{2020arXiv200405941I}  found that the X-ray observations from the nearby SN 2020bvc were consistent with the afterglow emission generated by an off-axis jet with viewing angle of $23^\circ$ when it decelerated in a circumburst medium  with a density profile with ${\rm k}=1.5$.\\
\\
Figure \ref{lc_cc} shows the synchrotron and SSC light curves when the non-relativistic ejecta mass decelerates in a stratified wind medium.   The light curves correspondent to the off-axis jet  are plotted in accordance with the afterglow model introduced in  \cite{2019ApJ...884...71F}.  The left-hand panels show the synchrotron light curves which correspond to (from top to bottom) radio (1.6 GHz),  optical (1 eV) and X-ray (1 keV) bands, respectively, and the right-hand panels show the SSC light curves which correspond to (from top to bottom)  gamma-ray fluxes  at 100 keV, 10 GeV and 100 GeV, respectively.   This figure shows that the afterglow emission originated from deceleration of the on-axis relativistic jet would have to decrease so that the 
afterglow emission from the non-relativistic ejecta could be observed.\\   
In our model, the afterglow emission from the non-relativistic ejecta in the radio band is observed as a flaring event while the afterglow from the optical and X-ray bands is seem as tails.  This figure displays that, depending on the parameters and the viewing angle,  the afterglow emission from the non-relativistic ejecta can be detected at early times. Similarly, when the jet is choked,  the afterglow emission from the non-relativistic ejecta can be observed at early times.\\
The way to distinguish the electromagnetic emission between  the non-relativistic ejecta and the relativistic jet, for example,  could be done through the evolution of the synchrotron flux derived in eqs. (\ref{fc_coast}) and (\ref{sc_coast}) for the coasting phase and in eqs. (\ref{fc_dec}) and (\ref{sc_dec}) for deceleration phase.\\
In order to analyze the mildly relativistic SNe with velocities in the range of $\Gamma\beta\approx 0.6-0.8$ as SN 2012ap, the synchrotron process in the trans-relativistic regime as discussed in Section \ref{analysis_sgrbs} for the shock breakout material through eq. (\ref{t_dec_new})  would be required.\\
\section{The non-relativistic ejecta mass and the kilonova afterglow in GW170817 and S190814\lowercase{bv}}
%
%
%
The coalescence of NS-NS and BH-NS binaries is the most natural candidate for the radiation of continuous GWs \citep[e.g., see][]{2017LRR....20....3M}. These coalescences are predicted to be accompanied by a kilonova \citep{1998ApJ...507L..59L, 2005ApJ...634.1202R, 2010MNRAS.406.2650M, 2013ApJ...774...25K, 2017LRR....20....3M}. This  transient is expected to be observed in near-IR/optical/UV bands throughout a timescale from days to weeks.\\ 
All coalescence of NS-NS, and only a fraction of BH-NS binaries can unbind at least some extremely neutron-rich material  which is able to form heavy r-process nuclei.   This Lanthanide-bearing matter with high opacity  is associated with the ``red" kilonova which is located inside the tidal tail in the equatorial plane.  The ``blue" kilonova is associated with the  low-opacity ejecta free of Lanthanide group elements which is located in the polar regions \citep{2014MNRAS.441.3444M, 2014MNRAS.443.3134P, 2014ApJ...789L..39W, 2019PhRvD.100b3008M}. While the  ``red" kilonova is associated with a slow ejecta material $\beta\simeq 0.1$, the ``blue" KN is characterized by being ejected with a faster velocity $\beta \simeq 0.3$.\\ 
In the following section, we give a brief introduction of multi-wavelength observations, and also constrain the parameter space of the faster ``blue" kilonova afterglow in GW170817 and  the KN afterglow possibly generated by the coalescence of a BH-NS system in S190814bv.\\
\subsection{Multi-wavelength observations}
\subsubsection{GW170817}
On 2017 August 17,  a GW signal \citep[GW170817;][]{PhysRevLett.119.161101,2041-8205-848-2-L12} together with a faint gamma-ray counterpart \citep[GRB 170817A;][]{2017ApJ...848L..14G, 2017ApJ...848L..15S} and kilonova emission were detected \citep{2017ApJ...848L..16S, 2017Sci...358.1556C, 2017ApJ...848L..17C, 2017ApJ...848L..27T, 2017Natur.551...75S, 2018MNRAS.473..576G}, for the first time, identifying the coalescence of a binary NS system. GRB 170817A exhibited observational evidence for extended emission in X-ray \citep{2017Natur.551...71T, 2018ApJ...856L..18M, 2018A&A...613L...1D, 2018ApJ...863L..18A}, optical  \citep{2018NatAs...2..751L, 2018ApJ...856L..18M} and radio  \citep{2041-8205-848-2-L12, 2018Natur.554..207M, 2018ApJ...858L..15D, 2017Natur.547..425T} bands.     About 11  hours  post-merger,  the optical transient coincident  with  the  quiescent  galaxy  NGC 4993  at  a  distance  of 40 Mpc ($z\approx0.01$) was associated with the kilonova \citep[AT2017gfo;][]{2017ApJ...848L..16S, 2017Sci...358.1556C, 2017ApJ...848L..17C, 2017ApJ...848L..27T, 2017Natur.551...75S, 2018MNRAS.473..576G}.\\
The temporarily extended X-ray and radio emission was fitted with a simple PL function which increased steadily as $\sim t^{0.8}$ \citep{2018ApJ...856L..18M, 2018Natur.554..207M} to reach its maximum value around $\sim$140 days post-merger, then decreased subsequently as $t^{-p}$ with $p\approx 2.2$.   A variety set of off-axis jet models \citep[with an opening angle $\lesssim 5^\circ$ and a viewing angle $\lesssim 28^\circ$;][]{PhysRevLett.119.161101, 2017MNRAS.472..616S, 2017Sci...358.1565E, 2017ApJ...848L..25H, 2018ApJ...853L..12M} such as top-hat jets \citep{2017Natur.551...71T, 2018ApJ...856L..18M,  2018ApJ...863L..18A, 2017ApJ...848L..21A, 2019ApJ...871..123F,2019arXiv190600502F}, radially stratified outflows \citep{2018Natur.554..207M, 2019ApJ...871..200F, 2018ApJ...867...95H} and structured outflows \citep{2017Sci...358.1559K, 2017MNRAS.472.4953L, 2018PhRvL.120x1103L, 2018Natur.561..355M, 2019ApJ...884...71F, 2018MNRAS.473..576G,2018MNRAS.473L.121K} were proposed to describe the temporarily extended electromagnetic emissions.\\ 
In order to constrain the KN afterglow, we use the data points displayed in  \cite{2019ApJ...884...71F} and complemented with those presented by \cite{2019ApJ...883L...1F} in the optical F606W filter, and by  \cite{2019ApJ...886L..17H} and \cite{2020arXiv200601150T} in the energy range of 0.3 - 10 keV.   Table \ref{table2} shows the latest Chandra afterglow observations in units of ${\rm erg\,cm^{-2}\,s^{-1}}$ (0.3 - 10 keV)  and ${\rm mJy}$ normalized at $1\,{\rm keV}$.  The data points in  ${\rm erg\,cm^{-2}\,s^{-1}}$ are taken from \cite{2019ApJ...886L..17H} and \cite{2020arXiv200601150T}.
\subsubsection{S190814bv}
%
On 2019 August 14 the LIGO and Virgo interferometer detected a GW signal \citep[S190814bv;][]{2019GCN.25324....1L, 2019GCN.25333....1L} associated with a NS-BH merger located at a distance of $267\pm52\,{\rm Mpc}$ \citep{2019ApJ...884L..55G}. Immediately, S190814bv was followed up by a large observational campaign that covered a large fraction of the electromagnetic spectrum \citep[e.g., see][]{2020ApJ...890..131A, 2019ApJ...887L..13D, 2020MNRAS.492.5916W, 2020arXiv200201950A, 2020arXiv200309437V}.  No counts were registered in any wavelengths and upper limits  were  reported.\\
Using the MegaCam instrument on the Canada-France-Hawaii Telescope (CFHT), \cite{2020arXiv200201950A} placed optical upper limits on the presence of any counterpart and discussed the implications for the KN possibly being generated by the coalescence of the BH-NS system. They excluded a KN with large ejecta mass $\gtrsim0.1 M_{\odot}$,  and, considering off-axis jet models, the authors ruled out  circumstellar densities $\gtrsim 1\,{\rm cm^{-3}}$ for typical values of microphysical parameters.   \cite{2020arXiv200309437V} reported observational constraints on the near-infrared (IR) and optical with the ElectromagNetic  counterparts of GRAvitational wave sources at the VERy Large Telescope (ENGRAVE). They constrained the KN ejecta mass to be $\lesssim 1.5\times 10^{-2}\,M_{\odot}$ for a ``blue" KN, and   $\lesssim 4\times 10^{-2}\,M_{\odot}$ for a ``red" KN.   \cite{2019ApJ...884L..55G} reported a Galaxy-targeted search for the optical emission with the Magellan Baade 6.5 m telescope and ruled out the on-axis typical jet of sGRBs.   \cite{2019ApJ...887L..13D} presented  upper limits in the radio band with the Australian Square Kilometre Array Pathfinder (ASKAP) and constrained the circumstellar density and inclination angle of the system. \\
In order to constrain the KN afterglow,  the upper limits on radio at 935 MHz  \citep{2019ApJ...887L..13D}, near-IR and optical at the ${\rm K_s}$ and ${\rm R}$ filters \citep{2020arXiv200201950A} and  X-rays at 1 keV \citep{2019GCN.25400....1E} are considered.\\
\subsection{Constraining the KN afterglow}
Figure \ref{grb170817A-kn} shows the multi-wavelength observations of GW170817 and S190814bv at the radio, optical and X-ray bands with the synchrotron light curves shown in Figure \ref{k_0} which are generated by the deceleration of the non-relativistic ejecta in the homogeneous-density medium.  Each colour of the synchrotron light curves corresponds to the set of parameter values reported in Section 3.   In addition, we include the synchrotron light curves at 3 GHz. The data points with the upper limits in black correspond to the observations performed for GW170817 and the upper limits in gray correspond to S190814bv.  Concerning GW170817,  the upper panels show the radio observations at 3 GHz (left) and 6 GHz  (right), and the lower panels show the optical observations at the F606W filter and the X-ray observations at 1 keV (right).   We include the upper limits collected from S190814bv on radio at 935 MHz (top left), near-IR and optical at ${\rm K_s}$ and ${\rm R}$ filters (bottom left), respectively, and  X-rays at 1 keV (bottom right).\\
As indicated in Section 3,  the synchrotron light curves are shown for different values of $p$ and $\alpha$, and the same values of $A_{0}$, $\epsilon_{\rm e}$ and $\epsilon_{\rm B}$. Therefore, any variation of these parameters will increase or decrease the intensity of the  observed flux in radio, optical and X-ray bands. Given the multiwavelength observations of GW170817 and S190814bv,  we constrain the parameter space of $A_{0}$, $\alpha$, $\epsilon_{\rm e}$ and $\epsilon_{\rm B}$ as a function of the velocity $\beta$.    In order to compare the density parameter of the uniform density medium $A_0$ found in this model with others, hereafter we use the usual notation $n$ for $A_0$.\\
\subsubsection{GW170817}
Figure \ref{grb170817a_lc} shows the multi-wavelength observations of GW170817 and the  parameter space  allowed with the synchrotron model presented in this work. The upper left-hand  panel   shows the multiwavelength data points of GW170817 with the best-fit curves obtained with the structured jet model presented in \cite{2019ApJ...884...71F} and a possible synchrotron contribution emitted by the deceleration of the non-relativistic ejecta.  The light curves are exhibited in radio (3 and 6 GHz), optical (at the F606W filter) and X-ray (1 keV) bands.  The upper right-hand panel and the lower panels show the allowed  parameter space of the uniform density of the circumstellar medium (${\rm n}$), the velocity of the non-relativistic ejecta ($\beta$),  the index of the PL distribution ($\alpha$) and the microphysical parameters ($\epsilon_{\rm e}$ and  $\epsilon_{\rm B}$) for  the fiducial energy $\tilde{E}_{\rm K}=10^{49}\,{\rm erg}$ \citep{2015MNRAS.450.1430H,2019ApJ...884...71F},   and  the spectral index $p=2.15$ \citep{2019ApJ...886L..17H,2018PhRvL.120x1103L}. We use the value of the microphysical parameter $\epsilon_{\rm e}=10^{-1}$ and $\alpha=3.0$  in the upper left-hand panel,  $\epsilon_{\rm B}=10^{-3}$ and $\alpha=3.0$ in the lower right-hand panel  and  $\epsilon_{\rm e}=10^{-1}$ and  $\epsilon_{\rm B}=10^{-3}$ in the lower left-hand panel.  The  allowed parameter spaces are below the relevant colored contours and  obtained using the data points in radio,  optical and X-ray as upper limits.    In order to illustrate the synchrotron light curves generated by the deceleration of the non-relativistic ejecta as shown in the upper left-hand panel, we consider a set of values from the parameter space.\\
Taking into account  the velocity of $\beta=0.3$ and $\alpha=3$ which are the reported values for the ``blue" KN \citep{2019LRR....23....1M},  the ranges of parameters allowed in our model ($5\times 10^{-2}\lesssim \epsilon_{\rm e}\lesssim 0.2$,  $10^{-3}\lesssim\epsilon_{\rm B}\lesssim10^{-2}$,  $10^{-3}\lesssim n\lesssim10^{-2}\,{\rm cm^{-3}}$) are similar to those reported from the description of the multi-wavelegth observations by deceleration of relativistic structured/top-hat outflows.   For instance, the values  $\epsilon_{\rm e}=0.15$, $\epsilon_{\rm B}=5\times 10^{-3}$, and $n=4\times 10^{-3}\,{\rm cm^{-3}}$  for $\beta\approx0.3$ strongly agree with the values reported in \cite{2019ApJ...886L..17H}. 
\subsubsection{S190814bv}
Figure \ref{lc_gw190814} shows the multi-wavelength upper limits of S190814bv and the parameter space ruled out with the synchrotron model presented in this work.  The upper left-hand panel  shows the multi-wavelength upper limits and the light curves in X-ray, optical, near-IR  and radio bands at 1 keV, ${\rm R}$-band,  ${\rm K_ s}$-band and 943 MHz, respectively.     In order to rule out  the parameters  ${\rm n}$, $\alpha$, $\beta$, $\epsilon_{\rm e}$ and  $\epsilon_{\rm B}$,  these upper limits are considered.  The ruled out regions are above the relevant colored contours.  The upper right-hand and lower panels show the parameter spaces which are ruled out in our model for  the fiducial energy $\tilde{E}=10^{50}\,{\rm erg}$ and  the spectral index $p=2.6$.    We use the value of the microphysical parameters $\epsilon_{\rm e}=10^{-1}$ and $\alpha=3$ in the upper right-hand panel, $\epsilon_{\rm B}=10^{-2}$ and $\alpha=3$ in the lower left-hand panel and $\epsilon_{\rm e}=10^{-1}$ and $\epsilon_{\rm B}=10^{-2}$ in the lower right-hand panel.  These panels display that a uniform-density $n\gtrsim 0.6\,{\rm cm^{-3}}$ is ruled out in our model for the parameters in the range of  $3\lesssim \alpha\lesssim 5.2$,  $0.1\lesssim \epsilon_{e}\lesssim0.9$ and $10^{-2}\lesssim\epsilon_{B}\lesssim 0.1$ for $\beta>0.33$.    In order to illustrate the synchrotron light curves in the X-ray, optical, near-IR and radio bands  generated by the deceleration of the non-relativistic ejecta as shown in the upper panel, we consider a set of parameters selected from the lower panels ($\beta=0.39$,  $\alpha=3$, $\epsilon_{\rm B}=0.1$, $\epsilon_{\rm e}=0.3$ and $n=1\,{\rm cm^{-3}}$).  In this panel it can be observed that the predicted radio flux is above the upper limit at $37\,{\rm days}$.\\
The value of the uniform-density medium ruled out in our model is consistent with the value of densities derived by \cite{2019ApJ...887L..13D,2020arXiv200201950A, 2019ApJ...884L..55G} using distinct off-axis jet models. Further observations on timescales of years post-merger are needed to derive tighter constraints and therefore, to increase (decrease) the parameter space ruled out (allowed) in our model. The value allowed of the power index $\alpha=3$ in our theoretical model agrees with the values found in numerical simulations \citep[e.g., see][]{2013ApJ...773...78B} and used for describing the KN emission \citep{2017LRR....20....3M, 2019LRR....23....1M}.
%
%
\subsection{A diversity range of KN features}
Disentangling the properties of KNe is an important point especially given the association of sGRBs with the GWs. The detection of GRB 170817A,  AT 2017gfo and GW170817 has paved the way on the nature of sGRBs as the coalescence of NS mergers. The evident KN signature in GW170817 provided the chance to estimate their detectability in sGRBs and the variability in their features.   Since sGRBs are usually discovered via detection of the $\gamma$-ray prompt emission from the relativistic jet, they are typically observed where the afterglow is brighter and thus most probably obscure the KN. This emission with its more isotropic component is easier to be seen at angles far away from the sGRB jet \citep{2012ApJ...746...48M}. Despite this, it has been possible to determine only four claimed KNe  with different features to AT 2017gfo.  SGRBs associated to the claimed KNe  are GRB 050709 \citep{2016NatCo...712898J}, GRB 060614 \citep{2015NatCo...6.7323Y}, GRB 130603B \citep{2013Natur.500..547T, 2013ApJ...774L..23B} and GRB 160821B  \citep{2017ApJ...843L..34K, 2019MNRAS.489.2104T}.  For instance, while KN associated to GRB 060614 is much brighter (2 or 3 times) than the interpolated KN model fit at the time of the observations, KN associated to GRB 160821B is less bright than AT 2017gfo.\\
%
\cite{2018ApJ...860...62G} analyzed a sample of 23 short nearby ($z \leq 0.5$) GRBs to compare the optical and near-IR light curves with  AT 2017gfo.    They considered short bursts, following the historical classification, the ones with $T_{90} \leq 2$ s and also the class of the sGRBs with extended emission \citep{2006ApJ...643..266N, 2010ApJ...722L.215D, 2016ApJ...825L..20D, 2017ApJ...848...88D, 2017A&A...600A..98D}. This comparison enables to characterize their diversity in terms of their brightness distribution.   \cite{2018ApJ...860...62G}  found that for four sGRBs: 050509B, 051210, 061201, and 080905A, a KN of the same brightness of AT 2017gfo could have been observed. For these bursts, deep 3 $\sigma$ upper limits, two times or more dimmer than the detections of AT 2017gfo at comparable rest-frame times, seem to exclude the presence of a KN like AT 2017gfo. In each case, a KN like AT 2017gfo could have been detected if it had been present. The authors also found that the afterglows in GRBs 150424A, 140903A and 150101B  were too bright for an AT 2017gfo-like KN to be detected. Finally, they reported that the host galaxies of sGRBs 061006, 071227 and 170428A were to bright, and in six bursts there was no sufficient constraining observations regarding the presence of KN.\\
Covering 14 years of operations with Swift, \cite{2020MNRAS.492.5011D} presented a systematic search for sGRBs in the local Universe. The authors found no events at a distance $\lesssim 100\,{\rm Mpc}$ and four candidates located at  $\lesssim 200\,{\rm Mpc}$.  They derived, in each case, constraining optical upper limits on the onset of a ``blue"  KN,  implying low mass ejecta ($\lesssim 10^{-3}\,M_\odot$).\\
The bursts that exclude the evidence of a KN similar to AT 2017gfo by several magnitudes together with the properties of previously claimed KNe in sGRBs support the hypothesis that a significant diversity exist in the properties of KN drawn from the coalescence of compact object mergers.  Therefore, a diversity range of KN features leads a wide parameter space of velocities, PL indices, masses, microphysical parameters and circumburst densities as discussed in this section.\\
Continuous energy injection by the central engine on the afterglow can produce a refreshed shock, and modifies the dynamics leading to rich radiation signatures. The problem of additional energy injection from the central engine has been studied by some authors \cite[e.g. see][]{2013ApJ...771...86G, 2020arXiv200601150T}.  Although this scenario is beyond the scope of the current paper,  these signatures,  in our model,  would usually appear in timescales from weeks to hundred of days.   In a forthcoming paper, we will present a detailed analysis of a refreshed shock in the energy injection scenario.

\vspace{1cm}

\section{Discussion and Summary}

We derived, based on analytic arguments, the dynamics of deceleration of a non-relativistic ejecta in a circumstellar medium with a density profile $A_{\rm k} r^{-k}$ with ${\rm k}=0$, $1$, $1.5$, $2$ and $2.5$ that covers short and long GRB progenitors. While the uniform-density  medium ($k=0$) is expected in the coalescence of binary compact objects and in CC-SNe, the stratified medium ($1\leq {\rm k} \leq 2.5$) is only connected with the death of massive stars with different mass-loss evolution at the end of their lives. Taking into account that electrons are accelerated during the forward shocks with a spectral index in the range $2.2\leq p \leq 3.2$,  we calculated the synchrotron and SSC light curves in the fast- and slow-cooling regime during the coasting and the deceleration phase.   During the coasting phase we considered velocities  in the range of $0.07 \leq \beta \leq  0.8$ and during the deceleration phase we assumed a PL velocity distribution $\propto \beta^{-\alpha}$ with $ 3\leq \alpha\leq 5.2$ for a generic source located at $100\,{\rm Mpc}$.\\
%
\\
We showed the predicted synchrotron light curves  in radio  at 6 GHz, optical at 1 eV and X-rays at 1 keV for typical values of GRB afterglows. All the light curves peak on timescales from several months to a few years, similar to those observed  in  some SNe such as SN2014C and SN2016aps. However, if the ejecta mass is extremely energetic or decelerates in a very dense medium, a peak in the light curve could be expected in weeks.\\ 
\\
We showed that when the non-relativistic ejecta decelerates in a uniform density medium a flattening or rebrightening in the light curve is expected, and when this ejecta decelerates in a stratified medium the rebrightening in the light curves is not so evident.  Therefore, a flattening or rebrightening at  timescales from months to years  in the light curve together with GW detection would be associated with the deceleration of a non-relativistic ejecta launched during the coalescence of a binary compact object. Otherwise, we showed that an observed flux that gradually decreases on timescales from months to years could be associated with the deceleration of a non-relativistic ejecta launched during the death of a massive star with different mass-loss evolution at the end of its life.\\ 
\\
The coalescence of the NS binaries launches  significant non-relativistic masses with different velocities which will contribute at distinct timescales, frequencies and intensities. These ejecta masses (the dynamical ejecta,  the cocoon material, the shock breakout material and the wind ejecta) interact with the  circumstellar medium generating non-thermal emission.  The shock breakout material peaks at timescales from hours to days, the cocoon  material  peaks at timescales from weeks to months,  the dynamical ejecta peaks at years and the disk wind ejecta peaks at centuries.   We calculated the expected gamma-ray, X-ray, optical and radio fluxes via  SSC and synchrotron emission from electrons accelerated in the forward shocks. These electromagnetic signatures at different timescales and frequencies would be similar to those detected around  SNe for a uniform density medium and also  be present together with the GW detections.\\ 
\\
We showed that variations in the density parameter could be observed more easily  i) in the radio  than in the X-ray light curve, ii) in a stratified than in a uniform density medium and iii) for larger values of $\alpha$. Therefore, a transition phase from stellar-wind to uniform-density medium is more noticeable in radio than X-ray bands.\\
\\
We showed that, in the case of a failed or an off-axis GRB, the non-thermal emission generated by the deceleration of non-relativistic ejecta could be detected at early times.   In the case of an on-axis GRB, the afterglow emission originated from deceleration of the relativistic jet would have to decrease substantially so that  the afterglow emission from the non-relativistic ejecta could be observed.    In addition, we gave an important tool  to distinguish the afterglow emission among the non-relativistic ejecta from  the relativistic jet through the evolution of the synchrotron flux derived in eqs. (\ref{fc_coast}) and (\ref{sc_coast}) for the coasting phase and in eqs. (\ref{fc_dec}) and (\ref{sc_dec}) for the deceleration phase.\\
\\
We computed the predicted SSC light curves from the deceleration of the non-relativistic ejecta mass for a density profile with $k=0$, $1$, $1.5$, $2$ and $2.5$.   The effect of the extragalactic background light (EBL) absorption modelled in \cite{2017A&A...603A..34F} was assumed. We showed that, when the non-relativistic ejecta decelerates in a uniform density medium, a flattening or rebrightening in the light curve is expected, and when, this ejecta decelerates in a stratified medium, the rebrightening in the light curves is not so evident.  Similarly, we showed that  the SSC flux is less sensitive to changes in the density parameter for higher frequencies than for lower ones and is more sensitive to the density parameter for larger values of $k$ and $\alpha$.\\
\\
In particular, using the multi-wavelength observations and upper limits of GW170817 and S190814bv, we constrained the parameter space of the uniform density of the circumstellar medium, the velocity of the non-relativistic ejecta, the index of the PL distribution and the microphysical parameters. In the case of GW170817,  we found similar values to those reported from the description of the multi-wavelength observations by the deceleration of relativistic structured/top-hat outflows for typical values of  KN ejecta mass  $\beta\approx0.3$ and $\alpha=3$ \citep{2017LRR....20....3M, 2019LRR....23....1M}. Therefore, we conclude that the KN afterglow scenario can be used to constrain the afterglow parameters of relativistic structured/top-hat outflows. In particular, the values of $\epsilon_{\rm e}=0.15$, $\epsilon_{\rm B}=5\times 10^{-3}$, and $n=4\times 10^{-3}\,{\rm cm^{-3}}$  strongly agree with the values reported in \cite{2019ApJ...886L..17H}.   For the case of S190814bv, we found that  the value of the uniform-density medium ruled out in our model is consistent with the value of density derived by \cite{2019ApJ...887L..13D,2020arXiv200201950A} and \cite{2019ApJ...884L..55G} using distinct off-axis jet models.  Further observations on timescales of years post-merger are needed to derive tighter constraints and therefore, to increase (decrease) the parameter space ruled out (allowed) in our model. The value allowed of the power index $\alpha=3$ in our theoretical model agrees with the values found in numerical simulations \citep[e.g., see][]{2013ApJ...773...78B} and used for describing the KN emission \citep{2017LRR....20....3M, 2019LRR....23....1M}.\\
\\
\acknowledgements
NF  acknowledges  financial  support  from UNAM-DGAPA-PAPIIT  through  grant  IA102019.  RBD  acknowledges support  from the National Science Foundation under grant 1816694. M.G.D. acknowledges funding from the AAS Chretienne Fellowship and the MINIATURA2 grant.
%
%
%
\newpage


\begin{thebibliography}{}

\bibitem[\protect\astroncite{Abbott et~al.}{2017a}]{PhysRevLett.119.161101}
Abbott, B.~P., Abbott, R., Abbott, T.~D., and et~al. (2017a).
\newblock Gw170817: Observation of gravitational waves from a binary neutron
  star inspiral.
\newblock {\em Phys. Rev. Lett.}, 119:161101.

\bibitem[\protect\astroncite{Abbott et~al.}{2017b}]{2041-8205-848-2-L12}
Abbott, B.~P., Abbott, R., Abbott, T.~D., and et~al. (2017b).
\newblock Multi-messenger observations of a binary neutron star merger.
\newblock {\em The Astrophysical Journal Letters}, 848(2):L12.

\bibitem[\protect\astroncite{{Ackley} et~al.}{2020}]{2020arXiv200201950A}
{Ackley}, K., {Amati}, L., {Barbieri}, C., {Bauer}, F.~E., {Benetti}, S.,
  {Bernardini}, M.~G., and {et al.} (2020).
\newblock {Observational constraints on the optical and near-infrared emission
  from the neutron star-black hole binary merger S190814bv}.
\newblock {\em arXiv e-prints}, page arXiv:2002.01950.

\bibitem[\protect\astroncite{{Alexander} et~al.}{2017}]{2017ApJ...848L..21A}
{Alexander}, K.~D., {Berger}, E., {Fong}, W., {Williams}, P.~K.~G., {Guidorzi},
  C., {Margutti}, R., {Metzger}, B.~D., {Annis}, J., {Blanchard}, P.~K.,
  {Brout}, D., {Brown}, D.~A., {Chen}, H.-Y., {Chornock}, R., {Cowperthwaite},
  P.~S., {Drout}, M., {Eftekhari}, T., {Frieman}, J., {Holz}, D.~E., {Nicholl},
  M., {Rest}, A., {Sako}, M., {Soares-Santos}, M., and {Villar}, V.~A. (2017).
\newblock {The Electromagnetic Counterpart of the Binary Neutron Star Merger
  LIGO/Virgo GW170817. VI. Radio Constraints on a Relativistic Jet and
  Predictions for Late-time Emission from the Kilonova Ejecta}.
\newblock {\em \apjl}, 848:L21.

\bibitem[\protect\astroncite{{Alexander} et~al.}{2018}]{2018ApJ...863L..18A}
{Alexander}, K.~D., {Margutti}, R., {Blanchard}, P.~K., {Fong}, W., {Berger},
  E., {Hajela}, A., and {et.} (2018).
\newblock {A Decline in the X-Ray through Radio Emission from GW170817
  Continues to Support an Off-axis Structured Jet}.
\newblock {\em \apjl}, 863(2):L18.

\bibitem[\protect\astroncite{{Ando} and {Beacom}}{2005}]{2005PhRvL..95f1103A}
{Ando}, S. and {Beacom}, J.~F. (2005).
\newblock {Revealing the Supernova Gamma-Ray Burst Connection with TeV
  Neutrinos}.
\newblock {\em \prl}, 95(6):061103.

\bibitem[\protect\astroncite{{Andreoni} et~al.}{2020}]{2020ApJ...890..131A}
{Andreoni}, I., {Goldstein}, D.~A., {Kasliwal}, M.~M., {Nugent}, P.~E., {Zhou},
  R., {Newman}, J.~A., and {et al.} (2020).
\newblock {GROWTH on S190814bv: Deep Synoptic Limits on the
  Optical/Near-infrared Counterpart to a Neutron Star?Black Hole Merger}.
\newblock {\em \apj}, 890(2):131.

\bibitem[\protect\astroncite{{Arcavi} et~al.}{2017}]{2017Natur.551...64A}
{Arcavi}, I., {Hosseinzadeh}, G., {Howell}, D.~A., {McCully}, C., {Poznanski},
  D., {Kasen}, D., {Barnes}, J., {Zaltzman}, M., {Vasylyev}, S., {Maoz}, D.,
  and {Valenti}, S. (2017).
\newblock {Optical emission from a kilonova following a
  gravitational-wave-detected neutron-star merger}.
\newblock {\em \nat}, 551(7678):64--66.

\bibitem[\protect\astroncite{{Barniol Duran} and
  {Giannios}}{2015}]{2015MNRAS.454.1711B}
{Barniol Duran}, R. and {Giannios}, D. (2015).
\newblock {Radio rebrightening of the GRB afterglow by the accompanying
  supernova}.
\newblock {\em \mnras}, 454(2):1711--1718.

\bibitem[\protect\astroncite{{Barniol Duran}
  et~al.}{2015}]{2015MNRAS.448..417B}
{Barniol Duran}, R., {Nakar}, E., {Piran}, T., and {Sari}, R. (2015).
\newblock {The afterglow of a relativistic shock breakout and low-luminosity
  GRBs}.
\newblock {\em \mnras}, 448(1):417--428.

\bibitem[\protect\astroncite{{Barthelmy} et~al.}{2005}]{2005ApJ...635L.133B}
{Barthelmy}, S.~D., {Cannizzo}, J.~K., {Gehrels}, N., {Cusumano}, G.,
  {Mangano}, V., {O'Brien}, P.~T., and {et al.} (2005).
\newblock {Discovery of an Afterglow Extension of the Prompt Phase of Two
  Gamma-Ray Bursts Observed by Swift}.
\newblock {\em \apjl}, 635:L133--L136.

\bibitem[\protect\astroncite{{Bauswein} et~al.}{2013}]{2013ApJ...773...78B}
{Bauswein}, A., {Goriely}, S., and {Janka}, H.~T. (2013).
\newblock {Systematics of Dynamical Mass Ejection, Nucleosynthesis, and
  Radioactively Powered Electromagnetic Signals from Neutron-star Mergers}.
\newblock {\em \apj}, 773(1):78.

\bibitem[\protect\astroncite{{Beniamini} et~al.}{2020}]{2020MNRAS.493.3521B}
{Beniamini}, P., {Granot}, J., and {Gill}, R. (2020).
\newblock {Afterglow light curves from misaligned structured jets}.
\newblock {\em \mnras}, 493(3):3521--3534.

\bibitem[\protect\astroncite{{Berger} et~al.}{2013}]{2013ApJ...774L..23B}
{Berger}, E., {Fong}, W., and {Chornock}, R. (2013).
\newblock {An r-process Kilonova Associated with the Short-hard GRB 130603B}.
\newblock {\em \apjl}, 774(2):L23.

\bibitem[\protect\astroncite{{Bj{\"o}rnsson} and
  {Fransson}}{2004}]{2004ApJ...605..823B}
{Bj{\"o}rnsson}, C.-I. and {Fransson}, C. (2004).
\newblock {The X-Ray and Radio Emission from SN 2002ap: The Importance of
  Compton Scattering}.
\newblock {\em \apj}, 605(2):823--829.

\bibitem[\protect\astroncite{{Blandford} and
  {McKee}}{1976}]{1976PhFl...19.1130B}
{Blandford}, R.~D. and {McKee}, C.~F. (1976).
\newblock {Fluid dynamics of relativistic blast waves}.
\newblock {\em Physics of Fluids}, 19:1130--1138.

\bibitem[\protect\astroncite{{Blondin} et~al.}{1996}]{1996ApJ...472..257B}
{Blondin}, J.~M., {Lundqvist}, P., and {Chevalier}, R.~A. (1996).
\newblock {Axisymmetric circumstellar interaction in supernovae}.
\newblock {\em \apj}, 472:257--266.

\bibitem[\protect\astroncite{{Bloom} et~al.}{1999}]{1999Natur.401..453B}
{Bloom}, J.~S., {Kulkarni}, S.~R., {Djorgovski}, S.~G., {Eichelberger}, A.~C.,
  {C{\^o}t{\'e}}, P., and {et al.} (1999).
\newblock {The unusual afterglow of the {\ensuremath{\gamma}}-ray burst of 26
  March 1998 as evidence for a supernova connection}.
\newblock {\em \nat}, 401(6752):453--456.

\bibitem[\protect\astroncite{{Bromberg} et~al.}{2011}]{2011ApJ...739L..55B}
{Bromberg}, O., {Nakar}, E., and {Piran}, T. (2011).
\newblock {Are Low-luminosity Gamma-Ray Bursts Generated by Relativistic Jets?}
\newblock {\em \apjl}, 739(2):L55.

\bibitem[\protect\astroncite{{Cano} et~al.}{2017}]{2017AdAst2017E...5C}
{Cano}, Z., {Wang}, S.-Q., {Dai}, Z.-G., and {Wu}, X.-F. (2017).
\newblock {The Observer's Guide to the Gamma-Ray Burst Supernova Connection}.
\newblock {\em Advances in Astronomy}, 2017:8929054.

\bibitem[\protect\astroncite{{Chevalier}}{1982}]{1982ApJ...258..790C}
{Chevalier}, R.~A. (1982).
\newblock {Self-similar solutions for the interaction of stellar ejecta with an
  external medium.}
\newblock {\em \apj}, 258:790--797.

\bibitem[\protect\astroncite{{Chevalier}}{1984}]{1984ApJ...285L..63C}
{Chevalier}, R.~A. (1984).
\newblock {The circumstellar interaction model for the radio emission from a
  Type I supernova}.
\newblock {\em \apj}, 285:L63--L66.

\bibitem[\protect\astroncite{{Chevalier} and
  {Fransson}}{2006}]{2006ApJ...651..381C}
{Chevalier}, R.~A. and {Fransson}, C. (2006).
\newblock {Circumstellar Emission from Type Ib and Ic Supernovae}.
\newblock {\em \apj}, 651(1):381--391.

\bibitem[\protect\astroncite{{Chevalier} et~al.}{2006}]{2006ApJ...641.1029C}
{Chevalier}, R.~A., {Fransson}, C., and {Nymark}, T.~K. (2006).
\newblock {Radio and X-Ray Emission as Probes of Type IIP Supernovae and Red
  Supergiant Mass Loss}.
\newblock {\em \apj}, 641(2):1029--1038.

\bibitem[\protect\astroncite{{Chevalier} and
  {Irwin}}{2011}]{2011ApJ...729L...6C}
{Chevalier}, R.~A. and {Irwin}, C.~M. (2011).
\newblock {Shock Breakout in Dense Mass Loss: Luminous Supernovae}.
\newblock {\em \apjl}, 729(1):L6.

\bibitem[\protect\astroncite{{Coulter} et~al.}{2017}]{2017Sci...358.1556C}
{Coulter}, D.~A., {Foley}, R.~J., {Kilpatrick}, C.~D., {Drout}, M.~R., {Piro},
  A.~L., {Shappee}, B.~J., {Siebert}, M.~R., {Simon}, J.~D., {Ulloa}, N.,
  {Kasen}, D., {Madore}, B.~F., {Murguia-Berthier}, A., {Pan}, Y.~C.,
  {Prochaska}, J.~X., {Ramirez-Ruiz}, E., {Rest}, A., and {Rojas-Bravo}, C.
  (2017).
\newblock {Swope Supernova Survey 2017a (SSS17a), the optical counterpart to a
  gravitational wave source}.
\newblock {\em Science}, 358(6370):1556--1558.

\bibitem[\protect\astroncite{{Cowperthwaite}
  et~al.}{2017}]{2017ApJ...848L..17C}
{Cowperthwaite}, P.~S., {Berger}, E., {Villar}, V.~A., {Metzger}, B.~D.,
  {Nicholl}, M., {Chornock}, R., and {et al.} (2017).
\newblock {The Electromagnetic Counterpart of the Binary Neutron Star Merger
  LIGO/Virgo GW170817. II. UV, Optical, and Near-infrared Light Curves and
  Comparison to Kilonova Models}.
\newblock {\em \apjl}, 848(2):L17.

\bibitem[\protect\astroncite{{Dai} and {Lu}}{1999}]{1999ApJ...519L.155D}
{Dai}, Z.~G. and {Lu}, T. (1999).
\newblock {The Afterglow of GRB 990123 and a Dense Medium}.
\newblock {\em \apjl}, 519:L155--L158.

\bibitem[\protect\astroncite{{Dainotti} et~al.}{2017a}]{2017ApJ...848...88D}
{Dainotti}, M.~G., {Hernandez}, X., {Postnikov}, S., {Nagataki}, S., {O'brien},
  P., {Willingale}, R., and {Striegel}, S. (2017a).
\newblock {A Study of the Gamma-Ray Burst Fundamental Plane}.
\newblock {\em \apj}, 848(2):88.

\bibitem[\protect\astroncite{{Dainotti} et~al.}{2017b}]{2017A&A...600A..98D}
{Dainotti}, M.~G., {Nagataki}, S., {Maeda}, K., {Postnikov}, S., and {Pian}, E.
  (2017b).
\newblock {A study of gamma ray bursts with afterglow plateau phases associated
  with supernovae}.
\newblock {\em \aap}, 600:A98.

\bibitem[\protect\astroncite{{Dainotti} et~al.}{2016}]{2016ApJ...825L..20D}
{Dainotti}, M.~G., {Postnikov}, S., {Hernandez}, X., and {Ostrowski}, M.
  (2016).
\newblock {A Fundamental Plane for Long Gamma-Ray Bursts with X-Ray Plateaus}.
\newblock {\em \apjl}, 825(2):L20.

\bibitem[\protect\astroncite{{Dainotti} et~al.}{2010}]{2010ApJ...722L.215D}
{Dainotti}, M.~G., {Willingale}, R., {Capozziello}, S., {Fabrizio Cardone}, V.,
  and {Ostrowski}, M. (2010).
\newblock {Discovery of a Tight Correlation for Gamma-ray Burst Afterglows with
  ``Canonical'' Light Curves}.
\newblock {\em \apjl}, 722(2):L215--L219.

\bibitem[\protect\astroncite{{D'Avanzo} et~al.}{2018}]{2018A&A...613L...1D}
{D'Avanzo}, P., {Campana}, S., {Salafia}, O.~S., {Ghirland a}, G.,
  {Ghisellini}, G., {Melandri}, A., {Bernardini}, M.~G., {Branchesi}, M.,
  {Chassande-Mottin}, E., {Covino}, S., {D'Elia}, V., {Nava}, L., {Salvaterra},
  R., {Tagliaferri}, G., and {Vergani}, S.~D. (2018).
\newblock {The evolution of the X-ray afterglow emission of GW 170817/ GRB
  170817A in XMM-Newton observations}.
\newblock {\em \aap}, 613:L1.

\bibitem[\protect\astroncite{{Davies} et~al.}{1994}]{1994ApJ...431..742D}
{Davies}, M.~B., {Benz}, W., {Piran}, T., and {Thielemann}, F.~K. (1994).
\newblock {Merging Neutron Stars. I. Initial Results for Coalescence of
  Noncorotating Systems}.
\newblock {\em \apj}, 431:742.

\bibitem[\protect\astroncite{{Dessart} et~al.}{2009}]{2009ApJ...690.1681D}
{Dessart}, L., {Ott}, C.~D., {Burrows}, A., {Rosswog}, S., and {Livne}, E.
  (2009).
\newblock {Neutrino Signatures and the Neutrino-Driven Wind in Binary Neutron
  Star Mergers}.
\newblock {\em \apj}, 690(2):1681--1705.

\bibitem[\protect\astroncite{{Dichiara} et~al.}{2020}]{2020MNRAS.492.5011D}
{Dichiara}, S., {Troja}, E., {O'Connor}, B., {Marshall}, F.~E., {Beniamini},
  P., {Cannizzo}, J.~K., {Lien}, A.~Y., and {Sakamoto}, T. (2020).
\newblock {Short gamma-ray bursts within 200 Mpc}.
\newblock {\em \mnras}, 492(4):5011--5022.

\bibitem[\protect\astroncite{{Dobie} et~al.}{2018}]{2018ApJ...858L..15D}
{Dobie}, D., {Kaplan}, D.~L., {Murphy}, T., {Lenc}, E., {Mooley}, K.~P.,
  {Lynch}, C., {Corsi}, A., {Frail}, D., {Kasliwal}, M., and {Hallinan}, G.
  (2018).
\newblock {A Turnover in the Radio Light Curve of GW170817}.
\newblock {\em \apjl}, 858(2):L15.

\bibitem[\protect\astroncite{{Dobie} et~al.}{2019}]{2019ApJ...887L..13D}
{Dobie}, D., {Stewart}, A., {Murphy}, T., {Lenc}, E., {Wang}, Z., {Kaplan},
  D.~L., and {et al.} (2019).
\newblock {An ASKAP Search for a Radio Counterpart to the First
  High-significance Neutron Star?Black Hole Merger LIGO/Virgo S190814bv}.
\newblock {\em \apjl}, 887(1):L13.

\bibitem[\protect\astroncite{{Evans} et~al.}{2017}]{2017Sci...358.1565E}
{Evans}, P.~A., {Cenko}, S.~B., {Kennea}, J.~A., {Emery}, S.~W.~K., {Kuin},
  N.~P.~M., {Korobkin}, O., and {et al.} (2017).
\newblock {Swift and NuSTAR observations of GW170817: Detection of a blue
  kilonova}.
\newblock {\em Science}, 358(6370):1565--1570.

\bibitem[\protect\astroncite{{Evans} et~al.}{2019}]{2019GCN.25400....1E}
{Evans}, P.~A., {Kennea}, J.~A., {Tohuvavohu}, A., {Barthelmy}, S.~D.,
  {Beardmore}, A.~P., {Bernardini}, M.~G., and {Swift Team} (2019).
\newblock {LIGO/Virgo S190814bv: No strong counterpart candidates in Swift/XRT
  observations}.
\newblock {\em GRB Coordinates Network}, 25400:1.

\bibitem[\protect\astroncite{{Fern{\'a}ndez}
  et~al.}{2015}]{2015MNRAS.446..750F}
{Fern{\'a}ndez}, R., {Kasen}, D., {Metzger}, B.~D., and {Quataert}, E. (2015).
\newblock {Outflows from accretion discs formed in neutron star mergers: effect
  of black hole spin}.
\newblock {\em \mnras}, 446(1):750--758.

\bibitem[\protect\astroncite{{Fong} et~al.}{2019}]{2019ApJ...883L...1F}
{Fong}, W., {Blanchard}, P.~K., {Alexander}, K.~D., {Strader}, J., {Margutti},
  R., {Hajela}, A., {Villar}, V.~A., {Wu}, Y., {Ye}, C.~S., {Berger}, E.,
  {Chornock}, R., {Coppejans}, D., {Cowperthwaite}, P.~S., {Eftekhari}, T.,
  {Giannios}, D., {Guidorzi}, C., {Kathirgamaraju}, A., {Laskar}, T.,
  {Macfadyen}, A., {Metzger}, B.~D., {Nicholl}, M., {Paterson}, K., {Terreran},
  G., {Sand}, D.~J., {Sironi}, L., {Williams}, P.~K.~G., {Xie}, X., and
  {Zrake}, J. (2019).
\newblock {The Optical Afterglow of GW170817: An Off-axis Structured Jet and
  Deep Constraints on a Globular Cluster Origin}.
\newblock {\em \apjl}, 883(1):L1.

\bibitem[\protect\astroncite{{Fong} et~al.}{2016}]{2016ApJ...831..141F}
{Fong}, W., {Metzger}, B.~D., {Berger}, E., and {{\"O}zel}, F. (2016).
\newblock {Radio Constraints on Long-lived Magnetar Remnants in Short Gamma-Ray
  Bursts}.
\newblock {\em \apj}, 831(2):141.

\bibitem[\protect\astroncite{{Fraija}}{2014}]{2014MNRAS.437.2187F}
{Fraija}, N. (2014).
\newblock {GeV-PeV neutrino production and oscillation in hidden jets from
  gamma-ray bursts}.
\newblock {\em \mnras}, 437:2187--2200.

\bibitem[\protect\astroncite{{Fraija} et~al.}{2019a}]{2019ApJ...871..123F}
{Fraija}, N., {De Colle}, F., {Veres}, P., {Dichiara}, S., {Barniol Duran}, R.,
  {Galvan-Gamez}, A., and {Pedreira}, A.~C.~C.~d.~E.~S. (2019a).
\newblock {The Short GRB 170817A: Modeling the Off-axis Emission and
  Implications on the Ejecta Magnetization}.
\newblock {\em \apj}, 871:123.

\bibitem[\protect\astroncite{{Fraija} et~al.}{2019b}]{2019arXiv190600502F}
{Fraija}, N., {De Colle}, F., {Veres}, P., {Dichiara}, S., {Barniol Duran}, R.,
  {Pedreira}, A.~C. C. d. E.~S., {Galvan-Gamez}, A., and {Betancourt
  Kamenetskaia}, B. (2019b).
\newblock {Description of atypical bursts seen slightly off-axis}.
\newblock {\em arXiv e-prints}, page arXiv:1906.00502.

\bibitem[\protect\astroncite{{Fraija} et~al.}{2019c}]{2019ApJ...879L..26F}
{Fraija}, N., {Dichiara}, S., {Pedreira}, A.~C. C. d. E.~S., {Galvan-Gamez},
  A., {Becerra}, R.~L., {Barniol Duran}, R., and {Zhang}, B.~B. (2019c).
\newblock {Analysis and Modeling of the Multi-wavelength Observations of the
  Luminous GRB 190114C}.
\newblock {\em \apjl}, 879(2):L26.

\bibitem[\protect\astroncite{{Fraija} et~al.}{2019d}]{2019ApJ...884...71F}
{Fraija}, N., {Lopez-Camara}, D., {Pedreira}, A.~C. C. d. E.~S., {Betancourt
  Kamenetskaia}, B., {Veres}, P., and {Dichiara}, S. (2019d).
\newblock {Signatures from a Quasi-spherical Outflow and an Off-axis Top-hat
  Jet Launched in a Merger of Compact Objects: An Analytical Approach}.
\newblock {\em \apj}, 884(1):71.

\bibitem[\protect\astroncite{{Fraija} et~al.}{2019e}]{2019ApJ...871..200F}
{Fraija}, N., {Pedreira}, A.~C.~C.~d.~E.~S., and {Veres}, P. (2019e).
\newblock {Light Curves of a Shock-breakout Material and a Relativistic
  Off-axis Jet from a Binary Neutron Star System}.
\newblock {\em \apj}, 871:200.

\bibitem[\protect\astroncite{{Fraija} et~al.}{2017}]{2017ApJ...848...15F}
{Fraija}, N., {Veres}, P., {Zhang}, B.~B., {Barniol Duran}, R., {Becerra},
  R.~L., {Zhang}, B., {Lee}, W.~H., {Watson}, A.~M., {Ordaz-Salazar}, C., and
  {Galvan-Gamez}, A. (2017).
\newblock {Theoretical Description of GRB 160625B with Wind-to-ISM Transition
  and Implications for a Magnetized Outflow}.
\newblock {\em \apj}, 848:15.

\bibitem[\protect\astroncite{{Franceschini} and
  {Rodighiero}}{2017}]{2017A&A...603A..34F}
{Franceschini}, A. and {Rodighiero}, G. (2017).
\newblock {The extragalactic background light revisited and the cosmic photon-
  photon opacity}.
\newblock {\em \aap}, 603:A34.

\bibitem[\protect\astroncite{{Gal-Yam}}{2017}]{2017hsn..book..195G}
{Gal-Yam}, A. (2017).
\newblock {\em {Observational and Physical Classification of Supernovae}}, page
  195.

\bibitem[\protect\astroncite{{Galama} et~al.}{1998}]{1998Natur.395..670G}
{Galama}, T.~J., {Vreeswijk}, P.~M., {van Paradijs}, J., {Kouveliotou}, C.,
  {Augusteijn}, T., {B{\"o}hnhardt}, H., and {et al.} (1998).
\newblock {An unusual supernova in the error box of the {$\gamma$}-ray burst of
  25 April 1998}.
\newblock {\em \nat}, 395:670--672.

\bibitem[\protect\astroncite{{Gao} et~al.}{2013a}]{2013ApJ...771...86G}
{Gao}, H., {Ding}, X., {Wu}, X.-F., {Zhang}, B., and {Dai}, Z.-G. (2013a).
\newblock {Bright Broadband Afterglows of Gravitational Wave Bursts from
  Mergers of Binary Neutron Stars}.
\newblock {\em \apj}, 771(2):86.

\bibitem[\protect\astroncite{{Gao} et~al.}{2013b}]{2013MNRAS.435.2520G}
{Gao}, H., {Lei}, W.-H., {Wu}, X.-F., and {Zhang}, B. (2013b).
\newblock {Compton scattering of self-absorbed synchrotron emission}.
\newblock {\em \mnras}, 435:2520--2531.

\bibitem[\protect\astroncite{{Gao} et~al.}{2015}]{2015ApJ...810..160G}
{Gao}, H., {Wang}, X.-G., {M{\'e}sz{\'a}ros}, P., and {Zhang}, B. (2015).
\newblock {A Morphological Analysis of Gamma-Ray Burst Early-optical
  Afterglows}.
\newblock {\em \apj}, 810:160.

\bibitem[\protect\astroncite{{Geng} et~al.}{2019}]{2019ApJ...877L..40G}
{Geng}, J.-J., {Zhang}, B., {K{\"o}lligan}, A., {Kuiper}, R., and {Huang},
  Y.-F. (2019).
\newblock {Propagation of a Short GRB Jet in the Ejecta: Jet Launching Delay
  Time, Jet Structure, and GW170817/GRB 170817A}.
\newblock {\em \apjl}, 877(2):L40.

\bibitem[\protect\astroncite{{Giblin} et~al.}{1999}]{1999ApJ...524L..47G}
{Giblin}, T.~W., {van Paradijs}, J., {Kouveliotou}, C., {Connaughton}, V.,
  {Wijers}, R.~A.~M.~J., {Briggs}, M.~S., {Preece}, R.~D., and {Fishman}, G.~J.
  (1999).
\newblock {Evidence for an Early High-Energy Afterglow Observed with BATSE from
  GRB 980923}.
\newblock {\em \apjl}, 524:L47--L50.

\bibitem[\protect\astroncite{{Goldstein} et~al.}{2017}]{2017ApJ...848L..14G}
{Goldstein}, A., {Veres}, P., {Burns}, E., {Briggs}, M.~S., {Hamburg}, R.,
  {Kocevski}, D., and {et al.} (2017).
\newblock {An Ordinary Short Gamma-Ray Burst with Extraordinary Implications:
  Fermi-GBM Detection of GRB 170817A}.
\newblock {\em \apjl}, 848:L14.

\bibitem[\protect\astroncite{{Gomez} et~al.}{2019}]{2019ApJ...884L..55G}
{Gomez}, S., {Hosseinzadeh}, G., {Cowperthwaite}, P.~S., {Villar}, V.~A.,
  {Berger}, E., {Gardner}, T., and {et al.} (2019).
\newblock {A Galaxy-targeted Search for the Optical Counterpart of the
  Candidate NS-BH Merger S190814bv with Magellan}.
\newblock {\em \apjl}, 884(2):L55.

\bibitem[\protect\astroncite{{Gompertz} et~al.}{2018}]{2018ApJ...860...62G}
{Gompertz}, B.~P., {Levan}, A.~J., {Tanvir}, N.~R., {Hjorth}, J., {Covino}, S.,
  {Evans}, P.~A., {Fruchter}, A.~S., {Gonz{\'a}lez-Fern{\'a}ndez}, C., {Jin},
  Z.~P., {Lyman}, J.~D., {Oates}, S.~R., {O'Brien}, P.~T., and {Wiersema}, K.
  (2018).
\newblock {The Diversity of Kilonova Emission in Short Gamma-Ray Bursts}.
\newblock {\em \apj}, 860(1):62.

\bibitem[\protect\astroncite{{Goriely} et~al.}{2011}]{2011ApJ...738L..32G}
{Goriely}, S., {Bauswein}, A., and {Janka}, H.-T. (2011).
\newblock {r-process Nucleosynthesis in Dynamically Ejected Matter of Neutron
  Star Mergers}.
\newblock {\em \apjl}, 738(2):L32.

\bibitem[\protect\astroncite{{Gottlieb} et~al.}{2018}]{2018MNRAS.473..576G}
{Gottlieb}, O., {Nakar}, E., and {Piran}, T. (2018).
\newblock {The cocoon emission - an electromagnetic counterpart to
  gravitational waves from neutron star mergers}.
\newblock {\em \mnras}, 473(1):576--584.

\bibitem[\protect\astroncite{{Grossman} et~al.}{2014}]{2014MNRAS.439..757G}
{Grossman}, D., {Korobkin}, O., {Rosswog}, S., and {Piran}, T. (2014).
\newblock {The long-term evolution of neutron star merger remnants - II.
  Radioactively powered transients}.
\newblock {\em \mnras}, 439(1):757--770.

\bibitem[\protect\astroncite{{Haggard} et~al.}{2017}]{2017ApJ...848L..25H}
{Haggard}, D., {Nynka}, M., {Ruan}, J.~J., {Kalogera}, V., {Cenko}, S.~B.,
  {Evans}, P., and {Kennea}, J.~A. (2017).
\newblock {A Deep Chandra X-Ray Study of Neutron Star Coalescence GW170817}.
\newblock {\em \apjl}, 848(2):L25.

\bibitem[\protect\astroncite{{Hajela} et~al.}{2019}]{2019ApJ...886L..17H}
{Hajela}, A., {Margutti}, R., {Alexander}, K.~D., {Kathirgamaraju}, A.,
  {Baldeschi}, A., {Guidorzi}, C., {Giannios}, D., {Fong}, W., {Wu}, Y.,
  {MacFadyen}, A., {Paggi}, A., {Berger}, E., {Blanchard}, P.~K., {Chornock},
  R., {Coppejans}, D.~L., {Cowperthwaite}, P.~S., {Eftekhari}, T., {Gomez}, S.,
  {Hosseinzadeh}, G., {Laskar}, T., {Metzger}, B.~D., {Nicholl}, M.,
  {Paterson}, K., {Radice}, D., {Sironi}, L., {Terreran}, G., {Villar}, V.~A.,
  {Williams}, P.~K.~G., {Xie}, X., and {Zrake}, J. (2019).
\newblock {Two Years of Nonthermal Emission from the Binary Neutron Star Merger
  GW170817: Rapid Fading of the Jet Afterglow and First Constraints on the
  Kilonova Fastest Ejecta}.
\newblock {\em \apjl}, 886(1):L17.

\bibitem[\protect\astroncite{{Horesh} et~al.}{2013}]{2013ApJ...778...63H}
{Horesh}, A., {Kulkarni}, S.~R., {Corsi}, A., {Frail}, D.~A., {Cenko}, S.~B.,
  {Ben-Ami}, S., {Gal-Yam}, A., {Yaron}, O., {Arcavi}, I., {Kasliwal}, M.~M.,
  and {Ofek}, E.~O. (2013).
\newblock {PTF 12gzk{\textemdash}A Rapidly Declining, High-velocity Type Ic
  Radio Supernova}.
\newblock {\em \apj}, 778(1):63.

\bibitem[\protect\astroncite{{Hotokezaka} et~al.}{2018}]{2018ApJ...867...95H}
{Hotokezaka}, K., {Kiuchi}, K., {Shibata}, M., {Nakar}, E., and {Piran}, T.
  (2018).
\newblock {Synchrotron Radiation from the Fast Tail of Dynamical Ejecta of
  Neutron Star Mergers}.
\newblock {\em \apj}, 867(2):95.

\bibitem[\protect\astroncite{{Hotokezaka} et~al.}{2013}]{2013ApJ...778L..16H}
{Hotokezaka}, K., {Kyutoku}, K., {Tanaka}, M., {Kiuchi}, K., {Sekiguchi}, Y.,
  {Shibata}, M., and {Wanajo}, S. (2013).
\newblock {Progenitor Models of the Electromagnetic Transient Associated with
  the Short Gamma Ray Burst 130603B}.
\newblock {\em \apjl}, 778:L16.

\bibitem[\protect\astroncite{{Hotokezaka} and
  {Piran}}{2015}]{2015MNRAS.450.1430H}
{Hotokezaka}, K. and {Piran}, T. (2015).
\newblock {Mass ejection from neutron star mergers: different components and
  expected radio signals}.
\newblock {\em \mnras}, 450:1430--1440.

\bibitem[\protect\astroncite{{Huang} and {Cheng}}{2003}]{2003MNRAS.341..263H}
{Huang}, Y.~F. and {Cheng}, K.~S. (2003).
\newblock {Gamma-ray bursts: optical afterglows in the deep Newtonian phase}.
\newblock {\em \mnras}, 341:263--269.

\bibitem[\protect\astroncite{{Huang} et~al.}{1998}]{1998A&A...336L..69H}
{Huang}, Y.~F., {Dai}, Z.~G., and {Lu}, T. (1998).
\newblock {GRB afterglows: from ultra-relativistic to non-relativistic phase}.
\newblock {\em \aap}, 336:L69--L72.

\bibitem[\protect\astroncite{{Huang} et~al.}{1999}]{1999MNRAS.309..513H}
{Huang}, Y.~F., {Dai}, Z.~G., and {Lu}, T. (1999).
\newblock {A generic dynamical model of gamma-ray burst remnants}.
\newblock {\em \mnras}, 309:513--516.

\bibitem[\protect\astroncite{{Izzo} et~al.}{2020}]{2020arXiv200405941I}
{Izzo}, L., {Auchettl}, K., {Hjorth}, J., {De Colle}, F., {Gall}, C., {Angus},
  C.~R., {Raimundo}, S.~I., and {Ramirez-Ruiz}, E. (2020).
\newblock {The broad-line type Ic SN 2020bvc: signatures of an off-axis
  gamma-ray burst afterglow}.
\newblock {\em arXiv e-prints}, page arXiv:2004.05941.

\bibitem[\protect\astroncite{{Izzo} et~al.}{2019}]{2019Natur.565..324I}
{Izzo}, L., {de Ugarte Postigo}, A., {Maeda}, K., {Th{\"o}ne}, C.~C., {Kann},
  D.~A., {Della Valle}, M., {Sagues Carracedo}, A., {Micha{\l}owski}, M.~J.,
  {Schady}, P., {Schmidl}, S., {Selsing}, J., {Starling}, R.~L.~C., {Suzuki},
  A., {Bensch}, K., {Bolmer}, J., {Campana}, S., {Cano}, Z., {Covino}, S.,
  {Fynbo}, J.~P.~U., {Hartmann}, D.~H., {Heintz}, K.~E., {Hjorth}, J.,
  {Japelj}, J., {Kami{\'n}ski}, K., {Kaper}, L., {Kouveliotou}, C.,
  {Kru{\.Z}y{\'n}ski}, M., {Kwiatkowski}, T., {Leloudas}, G., {Levan}, A.~J.,
  {Malesani}, D.~B., {Micha{\l}owski}, T., {Piranomonte}, S., {Pugliese}, G.,
  {Rossi}, A., {S{\'a}nchez-Ram{\'\i}rez}, R., {Schulze}, S., {Steeghs}, D.,
  {Tanvir}, N.~R., {Ulaczyk}, K., {Vergani}, S.~D., and {Wiersema}, K. (2019).
\newblock {Signatures of a jet cocoon in early spectra of a supernova
  associated with a {\ensuremath{\gamma}}-ray burst}.
\newblock {\em \nat}, 565(7739):324--327.

\bibitem[\protect\astroncite{{Jin} et~al.}{2016}]{2016NatCo...712898J}
{Jin}, Z.-P., {Hotokezaka}, K., {Li}, X., {Tanaka}, M., {D'Avanzo}, P., {Fan},
  Y.-Z., {Covino}, S., {Wei}, D.-M., and {Piran}, T. (2016).
\newblock {The Macronova in GRB 050709 and the GRB-macronova connection}.
\newblock {\em Nature Communications}, 7:12898.

\bibitem[\protect\astroncite{{Jin} et~al.}{2009}]{2009MNRAS.400.1829J}
{Jin}, Z.~P., {Xu}, D., {Covino}, S., {D'Avanzo}, P., {Antonelli}, A., {Fan},
  Y.~Z., and {Wei}, D.~M. (2009).
\newblock {The X-ray afterglow of GRB 081109A: clue to the wind bubble
  structure}.
\newblock {\em \mnras}, 400:1829--1834.

\bibitem[\protect\astroncite{{Kamble} et~al.}{2007}]{2007ApJ...664L...5K}
{Kamble}, A., {Resmi}, L., and {Misra}, K. (2007).
\newblock {Observations of the Optical Afterglow of GRB 050319: The Wind-to-ISM
  Transition in View}.
\newblock {\em \apjl}, 664:L5--L8.

\bibitem[\protect\astroncite{{Kasen} et~al.}{2013}]{2013ApJ...774...25K}
{Kasen}, D., {Badnell}, N.~R., and {Barnes}, J. (2013).
\newblock {Opacities and Spectra of the r-process Ejecta from Neutron Star
  Mergers}.
\newblock {\em \apj}, 774:25.

\bibitem[\protect\astroncite{{Kasliwal} et~al.}{2017a}]{2017ApJ...843L..34K}
{Kasliwal}, M.~M., {Korobkin}, O., {Lau}, R.~M., {Wollaeger}, R., and {Fryer},
  C.~L. (2017a).
\newblock {Infrared Emission from Kilonovae: The Case of the Nearby Short Hard
  Burst GRB 160821B}.
\newblock {\em \apjl}, 843(2):L34.

\bibitem[\protect\astroncite{{Kasliwal} et~al.}{2017b}]{2017Sci...358.1559K}
{Kasliwal}, M.~M., {Nakar}, E., {Singer}, L.~P., {Kaplan}, D.~L., {Cook},
  D.~O., {Van Sistine}, A., and {et al.} (2017b).
\newblock {Illuminating gravitational waves: A concordant picture of photons
  from a neutron star merger}.
\newblock {\em Science}, 358:1559--1565.

\bibitem[\protect\astroncite{{Kathirgamaraju}
  et~al.}{2016}]{2016MNRAS.461.1568K}
{Kathirgamaraju}, A., {Barniol Duran}, R., and {Giannios}, D. (2016).
\newblock {GRB off-axis afterglows and the emission from the accompanying
  supernovae}.
\newblock {\em \mnras}, 461(2):1568--1575.

\bibitem[\protect\astroncite{{Kathirgamaraju}
  et~al.}{2018}]{2018MNRAS.473L.121K}
{Kathirgamaraju}, A., {Barniol Duran}, R., and {Giannios}, D. (2018).
\newblock {Off-axis short GRBs from structured jets as counterparts to GW
  events}.
\newblock {\em \mnras}, 473(1):L121--L125.

\bibitem[\protect\astroncite{{Kathirgamaraju}
  et~al.}{2019}]{2019MNRAS.487.3914K}
{Kathirgamaraju}, A., {Giannios}, D., and {Beniamini}, P. (2019).
\newblock {Observable features of GW170817 kilonova afterglow}.
\newblock {\em \mnras}, 487(3):3914--3921.

\bibitem[\protect\astroncite{{Kotak} et~al.}{2004}]{2004MNRAS.354L..13K}
{Kotak}, R., {Meikle}, W.~P.~S., {Adamson}, A., and {Leggett}, S.~K. (2004).
\newblock {On the nature of the circumstellar medium of the remarkable Type
  Ia/IIn supernova SN 2002ic}.
\newblock {\em \mnras}, 354:L13--L17.

\bibitem[\protect\astroncite{{Kouveliotou} et~al.}{1993}]{1993ApJ...413L.101K}
{Kouveliotou}, C., {Meegan}, C.~A., {Fishman}, G.~J., {Bhat}, N.~P., {Briggs},
  M.~S., {Koshut}, T.~M., {Paciesas}, W.~S., and {Pendleton}, G.~N. (1993).
\newblock {Identification of Two Classes of Gamma-Ray Bursts}.
\newblock {\em \apjl}, 413:L101.

\bibitem[\protect\astroncite{{Kulkarni} et~al.}{1998}]{1998Natur.395..663K}
{Kulkarni}, S.~R., {Frail}, D.~A., {Wieringa}, M.~H., {Ekers}, R.~D., {Sadler},
  E.~M., {Wark}, R.~M., {Higdon}, J.~L., {Phinney}, E.~S., and {Bloom}, J.~S.
  (1998).
\newblock {Radio emission from the unusual supernova 1998bw and its association
  with the {\ensuremath{\gamma}}-ray burst of 25 April 1998}.
\newblock {\em \nat}, 395(6703):663--669.

\bibitem[\protect\astroncite{{Kyutoku} et~al.}{2014}]{2014MNRAS.437L...6K}
{Kyutoku}, K., {Ioka}, K., and {Shibata}, M. (2014).
\newblock {Ultrarelativistic electromagnetic counterpart to binary neutron star
  mergers}.
\newblock {\em \mnras}, 437:L6--L10.

\bibitem[\protect\astroncite{{Lamb} and
  {Kobayashi}}{2017}]{2017MNRAS.472.4953L}
{Lamb}, G.~P. and {Kobayashi}, S. (2017).
\newblock {Electromagnetic counterparts to structured jets from gravitational
  wave detected mergers}.
\newblock {\em \mnras}, 472:4953--4964.

\bibitem[\protect\astroncite{{Lazzati} et~al.}{2017}]{2017ApJ...848L...6L}
{Lazzati}, D., {L{\'o}pez-C{\'a}mara}, D., {Cantiello}, M., {Morsony}, B.~J.,
  {Perna}, R., and {Workman}, J.~C. (2017).
\newblock {Off-axis Prompt X-Ray Transients from the Cocoon of Short Gamma-Ray
  Bursts}.
\newblock {\em \apjl}, 848:L6.

\bibitem[\protect\astroncite{{Lazzati} et~al.}{2012}]{2012ApJ...750...68L}
{Lazzati}, D., {Morsony}, B.~J., {Blackwell}, C.~H., and {Begelman}, M.~C.
  (2012).
\newblock {Unifying the Zoo of Jet-driven Stellar Explosions}.
\newblock {\em \apj}, 750(1):68.

\bibitem[\protect\astroncite{{Lazzati} et~al.}{2018}]{2018PhRvL.120x1103L}
{Lazzati}, D., {Perna}, R., {Morsony}, B.~J., {Lopez-Camara}, D., {Cantiello},
  M., {Ciolfi}, R., {Giacomazzo}, B., and {Workman}, J.~C. (2018).
\newblock {Late Time Afterglow Observations Reveal a Collimated Relativistic
  Jet in the Ejecta of the Binary Neutron Star Merger GW170817}.
\newblock {\em \prl}, 120(24):241103.

\bibitem[\protect\astroncite{{Li} and
  {Paczy{\'n}ski}}{1998}]{1998ApJ...507L..59L}
{Li}, L.-X. and {Paczy{\'n}ski}, B. (1998).
\newblock {Transient Events from Neutron Star Mergers}.
\newblock {\em \apjl}, 507:L59--L62.

\bibitem[\protect\astroncite{{Liang} et~al.}{2013}]{2013ApJ...774...13L}
{Liang}, E.-W., {Li}, L., {Gao}, H., {Zhang}, B., {Liang}, Y.-F., {Wu}, X.-F.,
  {Yi}, S.-X., {Dai}, Z.-G., {Tang}, Q.-W., {Chen}, J.-M., {L{\"u}}, H.-J.,
  {Zhang}, J., {Lu}, R.-J., {L{\"u}}, L.-Z., and {Wei}, J.-Y. (2013).
\newblock {A Comprehensive Study of Gamma-Ray Burst Optical Emission. II.
  Afterglow Onset and Late Re-brightening Components}.
\newblock {\em \apj}, 774(1):13.

\bibitem[\protect\astroncite{{LIGO Scientific Collaboration} and {Virgo
  Collaboration}}{2019a}]{2019GCN.25324....1L}
{LIGO Scientific Collaboration} and {Virgo Collaboration} (2019a).
\newblock {LIGO/Virgo S190814bv: Identification of a GW compact binary merger
  candidate}.
\newblock {\em GRB Coordinates Network}, 25324:1.

\bibitem[\protect\astroncite{{LIGO Scientific Collaboration} and {Virgo
  Collaboration}}{2019b}]{2019GCN.25333....1L}
{LIGO Scientific Collaboration} and {Virgo Collaboration} (2019b).
\newblock {LIGO/Virgo S190814bv: Update on Sky-Localization and
  Source-Classification}.
\newblock {\em GRB Coordinates Network}, 25333:1.

\bibitem[\protect\astroncite{{Liu} et~al.}{2020}]{2020ApJ...890..102L}
{Liu}, L.-D., {Gao}, H., and {Zhang}, B. (2020).
\newblock {Constraining the Long-lived Magnetar Remnants in Short Gamma-Ray
  Bursts from Late-time Radio Observations}.
\newblock {\em \apj}, 890(2):102.

\bibitem[\protect\astroncite{{Livio} and {Waxman}}{2000}]{2000ApJ...538..187L}
{Livio}, M. and {Waxman}, E. (2000).
\newblock {Toward a Model for the Progenitors of Gamma-Ray Bursts}.
\newblock {\em \apj}, 538:187--191.

\bibitem[\protect\astroncite{{Lyman} et~al.}{2018}]{2018NatAs...2..751L}
{Lyman}, J.~D., {Lamb}, G.~P., {Levan}, A.~J., {Mandel}, I., {Tanvir}, N.~R.,
  {Kobayashi}, S., and {et al.} (2018).
\newblock {The optical afterglow of the short gamma-ray burst associated with
  GW170817}.
\newblock {\em Nature Astronomy}, 2:751--754.

\bibitem[\protect\astroncite{{MacFadyen} et~al.}{2001}]{2001ApJ...550..410M}
{MacFadyen}, A.~I., {Woosley}, S.~E., and {Heger}, A. (2001).
\newblock {Supernovae, Jets, and Collapsars}.
\newblock {\em \apj}, 550(1):410--425.

\bibitem[\protect\astroncite{{Mandel}}{2018}]{2018ApJ...853L..12M}
{Mandel}, I. (2018).
\newblock {The Orbit of GW170817 Was Inclined by Less Than 28$^\circ$ to the
  Line of Sight}.
\newblock {\em \apjl}, 853(1):L12.

\bibitem[\protect\astroncite{{Margalit} and
  {Piran}}{2020}]{2020arXiv200413028M}
{Margalit}, B. and {Piran}, T. (2020).
\newblock {Shock within a shock: revisiting the radio flares of NS merger
  ejecta and GRB-supernovae}.
\newblock {\em arXiv e-prints}, page arXiv:2004.13028.

\bibitem[\protect\astroncite{{Margutti} et~al.}{2018}]{2018ApJ...856L..18M}
{Margutti}, R., {Alexander}, K.~D., {Xie}, X., {Sironi}, L., {Metzger}, B.~D.,
  {Kathirgamaraju}, A., {Fong}, W., {Blanchard}, P.~K., {Berger}, E.,
  {MacFadyen}, A., {Giannios}, D., {Guidorzi}, C., {Hajela}, A., {Chornock},
  R., {Cowperthwaite}, P.~S., {Eftekhari}, T., {Nicholl}, M., {Villar}, V.~A.,
  {Williams}, P.~K.~G., and {Zrake}, J. (2018).
\newblock {The Binary Neutron Star Event LIGO/Virgo GW170817 160 Days after
  Merger: Synchrotron Emission across the Electromagnetic Spectrum}.
\newblock {\em \apjl}, 856(1):L18.

\bibitem[\protect\astroncite{{Margutti} et~al.}{2017}]{2017ApJ...835..140M}
{Margutti}, R., {Kamble}, A., {Milisavljevic}, D., {Zapartas}, E., {de Mink},
  S.~E., {Drout}, M., {Chornock}, R., {Risaliti}, G., {Zauderer}, B.~A.,
  {Bietenholz}, M., {Cantiello}, M., {Chakraborti}, S., {Chomiuk}, L., {Fong},
  W., {Grefenstette}, B., {Guidorzi}, C., {Kirshner}, R., {Parrent}, J.~T.,
  {Patnaude}, D., {Soderberg}, A.~M., {Gehrels}, N.~C., and {Harrison}, F.
  (2017).
\newblock {Ejection of the Massive Hydrogen-rich Envelope Timed with the
  Collapse of the Stripped SN 2014C}.
\newblock {\em \apj}, 835(2):140.

\bibitem[\protect\astroncite{{Margutti} et~al.}{2014}]{2014ApJ...797..107M}
{Margutti}, R., {Milisavljevic}, D., {Soderberg}, A.~M., {Guidorzi}, C.,
  {Morsony}, B.~J., {Sanders}, N., {Chakraborti}, S., {Ray}, A., {Kamble}, A.,
  {Drout}, M., {Parrent}, J., {Zauderer}, A., and {Chomiuk}, L. (2014).
\newblock {Relativistic Supernovae have Shorter-lived Central Engines or More
  Extended Progenitors: The Case of SN 2012ap}.
\newblock {\em \apj}, 797(2):107.

\bibitem[\protect\astroncite{{Margutti} et~al.}{2012}]{2012ApJ...751..134M}
{Margutti}, R., {Soderberg}, A.~M., {Chomiuk}, L., {Chevalier}, R., {Hurley},
  K., and {et al.} (2012).
\newblock {Inverse Compton X-Ray Emission from Supernovae with Compact
  Progenitors: Application to SN2011fe}.
\newblock {\em \apj}, 751(2):134.

\bibitem[\protect\astroncite{{Margutti} et~al.}{2013}]{2013ApJ...778...18M}
{Margutti}, R., {Soderberg}, A.~M., {Wieringa}, M.~H., {Edwards}, P.~G.,
  {Chevalier}, R.~A., and {et al.} (2013).
\newblock {The Signature of the Central Engine in the Weakest Relativistic
  Explosions: GRB 100316D}.
\newblock {\em \apj}, 778(1):18.

\bibitem[\protect\astroncite{{M{\'e}sz{\'a}ros} and
  {Waxman}}{2001}]{2001PhRvL..87q1102M}
{M{\'e}sz{\'a}ros}, P. and {Waxman}, E. (2001).
\newblock {TeV Neutrinos from Successful and Choked Gamma-Ray Bursts}.
\newblock {\em \prl}, 87(17):171102.

\bibitem[\protect\astroncite{{Metzger}}{2017}]{2017LRR....20....3M}
{Metzger}, B.~D. (2017).
\newblock {Kilonovae}.
\newblock {\em Living Reviews in Relativity}, 20:3.

\bibitem[\protect\astroncite{{Metzger}}{2019}]{2019LRR....23....1M}
{Metzger}, B.~D. (2019).
\newblock {Kilonovae}.
\newblock {\em Living Reviews in Relativity}, 23(1):1.

\bibitem[\protect\astroncite{{Metzger} et~al.}{2015}]{2015MNRAS.446.1115M}
{Metzger}, B.~D., {Bauswein}, A., {Goriely}, S., and {Kasen}, D. (2015).
\newblock {Neutron-powered precursors of kilonovae}.
\newblock {\em \mnras}, 446:1115--1120.

\bibitem[\protect\astroncite{{Metzger} and
  {Berger}}{2012}]{2012ApJ...746...48M}
{Metzger}, B.~D. and {Berger}, E. (2012).
\newblock {What is the Most Promising Electromagnetic Counterpart of a Neutron
  Star Binary Merger?}
\newblock {\em \apj}, 746(1):48.

\bibitem[\protect\astroncite{{Metzger} and {Bower}}{2014}]{2014MNRAS.437.1821M}
{Metzger}, B.~D. and {Bower}, G.~C. (2014).
\newblock {Constraints on long-lived remnants of neutron star binary mergers
  from late-time radio observations of short duration gamma-ray bursts}.
\newblock {\em \mnras}, 437(2):1821--1827.

\bibitem[\protect\astroncite{{Metzger} and
  {Fern{\'a}ndez}}{2014}]{2014MNRAS.441.3444M}
{Metzger}, B.~D. and {Fern{\'a}ndez}, R. (2014).
\newblock {Red or blue? A potential kilonova imprint of the delay until black
  hole formation following a neutron star merger}.
\newblock {\em \mnras}, 441(4):3444--3453.

\bibitem[\protect\astroncite{{Metzger} et~al.}{2010}]{2010MNRAS.406.2650M}
{Metzger}, B.~D., {Mart{\'{\i}}nez-Pinedo}, G., {Darbha}, S., {Quataert}, E.,
  {Arcones}, A., {Kasen}, D., {Thomas}, R., {Nugent}, P., {Panov}, I.~V., and
  {Zinner}, N.~T. (2010).
\newblock {Electromagnetic counterparts of compact object mergers powered by
  the radioactive decay of r-process nuclei}.
\newblock {\em \mnras}, 406:2650--2662.

\bibitem[\protect\astroncite{{Metzger} et~al.}{2008}]{2008MNRAS.385.1455M}
{Metzger}, B.~D., {Quataert}, E., and {Thompson}, T.~A. (2008).
\newblock {Short-duration gamma-ray bursts with extended emission from
  protomagnetar spin-down}.
\newblock {\em \mnras}, 385(3):1455--1460.

\bibitem[\protect\astroncite{{Miller} et~al.}{2019}]{2019PhRvD.100b3008M}
{Miller}, J.~M., {Ryan}, B.~R., {Dolence}, J.~C., {Burrows}, A., {Fontes},
  C.~J., {Fryer}, C.~L., {Korobkin}, O., {Lippuner}, J., {Mumpower}, M.~R., and
  {Wollaeger}, R.~T. (2019).
\newblock {Full transport model of GW170817-like disk produces a blue
  kilonova}.
\newblock {\em \prd}, 100(2):023008.

\bibitem[\protect\astroncite{{Modjaz} et~al.}{2020}]{2020ApJ...892..153M}
{Modjaz}, M., {Bianco}, F.~B., {Siwek}, M., {Huang}, S., {Perley}, D.~A.,
  {Fierroz}, D., {Liu}, Y.-Q., {Arcavi}, I., {Gal-Yam}, A., {Filippenko},
  A.~V., {Blagorodnova}, N., {Cenko}, B.~S., {Kasliwal}, M., {Kulkarni}, S.,
  {Schulze}, S., {Taggart}, K., and {Zheng}, W. (2020).
\newblock {Host Galaxies of Type Ic and Broad-lined Type Ic Supernovae from the
  Palomar Transient Factory: Implications for Jet Production}.
\newblock {\em \apj}, 892(2):153.

\bibitem[\protect\astroncite{{Mooley} et~al.}{2018a}]{2018Natur.561..355M}
{Mooley}, K.~P., {Deller}, A.~T., {Gottlieb}, O., {Nakar}, E., {Hallinan}, G.,
  {Bourke}, S., {Frail}, D.~A., {Horesh}, A., {Corsi}, A., and {Hotokezaka}, K.
  (2018a).
\newblock {Superluminal motion of a relativistic jet in the neutron-star merger
  GW170817}.
\newblock {\em \nat}, 561:355--359.

\bibitem[\protect\astroncite{{Mooley} et~al.}{2018b}]{2018Natur.554..207M}
{Mooley}, K.~P., {Nakar}, E., {Hotokezaka}, K., {Hallinan}, G., {Corsi}, A.,
  {Frail}, D.~A., and {et al.} (2018b).
\newblock {A mildly relativistic wide-angle outflow in the neutron-star merger
  event GW170817}.
\newblock {\em \nat}, 554(7691):207--210.

\bibitem[\protect\astroncite{{Moriya} and
  {Tominaga}}{2012}]{2012ApJ...747..118M}
{Moriya}, T.~J. and {Tominaga}, N. (2012).
\newblock {Diversity of luminous supernovae from non-steady mass loss}.
\newblock {\em \apj}, 747:118--124.

\bibitem[\protect\astroncite{{Murguia-Berthier}
  et~al.}{2014}]{2014ApJ...788L...8M}
{Murguia-Berthier}, A., {Montes}, G., {Ramirez-Ruiz}, E., {De Colle}, F., and
  {Lee}, W.~H. (2014).
\newblock {Necessary Conditions for Short Gamma-Ray Burst Production in Binary
  Neutron Star Mergers}.
\newblock {\em \apjl}, 788:L8.

\bibitem[\protect\astroncite{{Nagakura} et~al.}{2014}]{2014ApJ...784L..28N}
{Nagakura}, H., {Hotokezaka}, K., {Sekiguchi}, Y., {Shibata}, M., and {Ioka},
  K. (2014).
\newblock {Jet Collimation in the Ejecta of Double Neutron Star Mergers: A New
  Canonical Picture of Short Gamma-Ray Bursts}.
\newblock {\em \apjl}, 784:L28.

\bibitem[\protect\astroncite{{Nakar} and {Piran}}{2017}]{2017ApJ...834...28N}
{Nakar}, E. and {Piran}, T. (2017).
\newblock {The Observable Signatures of GRB Cocoons}.
\newblock {\em \apj}, 834(1):28.

\bibitem[\protect\astroncite{{Nicholl} et~al.}{2017}]{2017ApJ...848L..18N}
{Nicholl}, M., {Berger}, E., {Kasen}, D., {Metzger}, B.~D., {Elias}, J.,
  {Brice{\~n}o}, C., and {et al.} (2017).
\newblock {The Electromagnetic Counterpart of the Binary Neutron Star Merger
  LIGO/Virgo GW170817. III. Optical and UV Spectra of a Blue Kilonova from Fast
  Polar Ejecta}.
\newblock {\em \apjl}, 848(2):L18.

\bibitem[\protect\astroncite{{Nicholl} et~al.}{2020}]{2020NatAs.tmp...78N}
{Nicholl}, M., {Blanchard}, P.~K., {Berger}, E., {Chornock}, R., {Margutti},
  R., {Gomez}, S., and {et al.} (2020).
\newblock {An extremely energetic supernova from a very massive star in a dense
  medium}.
\newblock {\em Nature Astronomy}.

\bibitem[\protect\astroncite{{Norris} and
  {Bonnell}}{2006}]{2006ApJ...643..266N}
{Norris}, J.~P. and {Bonnell}, J.~T. (2006).
\newblock {Short Gamma-Ray Bursts with Extended Emission}.
\newblock {\em \apj}, 643:266--275.

\bibitem[\protect\astroncite{{Panaitescu} and
  {Kumar}}{2000}]{2000ApJ...543...66P}
{Panaitescu}, A. and {Kumar}, P. (2000).
\newblock {Analytic Light Curves of Gamma-Ray Burst Afterglows: Homogeneous
  versus Wind External Media}.
\newblock {\em \apj}, 543:66--76.

\bibitem[\protect\astroncite{{Perego} et~al.}{2014}]{2014MNRAS.443.3134P}
{Perego}, A., {Rosswog}, S., {Cabez{\'o}n}, R.~M., {Korobkin}, O.,
  {K{\"a}ppeli}, R., {Arcones}, A., and {Liebend{\"o}rfer}, M. (2014).
\newblock {Neutrino-driven winds from neutron star merger remnants}.
\newblock {\em \mnras}, 443(4):3134--3156.

\bibitem[\protect\astroncite{{Piran} et~al.}{2013}]{2013MNRAS.430.2121P}
{Piran}, T., {Nakar}, E., and {Rosswog}, S. (2013).
\newblock {The electromagnetic signals of compact binary mergers}.
\newblock {\em \mnras}, 430:2121--2136.

\bibitem[\protect\astroncite{{Planck Collaboration}
  et~al.}{2016}]{2016A&A...594A..13P}
{Planck Collaboration}, {Ade}, P.~A.~R., {Aghanim}, N., {Arnaud}, M.,
  {Ashdown}, M., {Aumont}, J., and {et al.} (2016).
\newblock {Planck 2015 results. XIII. Cosmological parameters}.
\newblock {\em \aap}, 594:A13.

\bibitem[\protect\astroncite{{Popham} et~al.}{1999}]{1999ApJ...518..356P}
{Popham}, R., {Woosley}, S.~E., and {Fryer}, C. (1999).
\newblock {Hyperaccreting Black Holes and Gamma-Ray Bursts}.
\newblock {\em \apj}, 518(1):356--374.

\bibitem[\protect\astroncite{{Ramirez-Ruiz} et~al.}{2005}]{2005ApJ...631..435R}
{Ramirez-Ruiz}, E., {Garc{\'\i}a-Segura}, G., {Salmonson}, J.~D., and
  {P{\'e}rez-Rend{\'o}n}, B. (2005).
\newblock {The State of the Circumstellar Medium Surrounding Gamma-Ray Burst
  Sources and Its Effect on the Afterglow Appearance}.
\newblock {\em \apj}, 631(1):435--445.

\bibitem[\protect\astroncite{{Rosswog}}{2005}]{2005ApJ...634.1202R}
{Rosswog}, S. (2005).
\newblock {Mergers of Neutron Star-Black Hole Binaries with Small Mass Ratios:
  Nucleosynthesis, Gamma-Ray Bursts, and Electromagnetic Transients}.
\newblock {\em \apj}, 634:1202--1213.

\bibitem[\protect\astroncite{{Rosswog} et~al.}{2014}]{2014MNRAS.439..744R}
{Rosswog}, S., {Korobkin}, O., {Arcones}, A., {Thielemann}, F.~K., and {Piran},
  T. (2014).
\newblock {The long-term evolution of neutron star merger remnants - I. The
  impact of r-process nucleosynthesis}.
\newblock {\em \mnras}, 439(1):744--756.

\bibitem[\protect\astroncite{{Rosswog} et~al.}{1999}]{1999A&A...341..499R}
{Rosswog}, S., {Liebend{\"o}rfer}, M., {Thielemann}, F.~K., {Davies}, M.~B.,
  {Benz}, W., and {Piran}, T. (1999).
\newblock {Mass ejection in neutron star mergers}.
\newblock {\em \aap}, 341:499--526.

\bibitem[\protect\astroncite{{Ruffert} et~al.}{1997}]{1997A&A...319..122R}
{Ruffert}, M., {Janka}, H.~T., {Takahashi}, K., and {Schaefer}, G. (1997).
\newblock {Coalescing neutron stars - a step towards physical models. II.
  Neutrino emission, neutron tori, and gamma-ray bursts.}
\newblock {\em \aap}, 319:122--153.

\bibitem[\protect\astroncite{{Sari} and {Esin}}{2001}]{2001ApJ...548..787S}
{Sari}, R. and {Esin}, A.~A. (2001).
\newblock {On the Synchrotron Self-Compton Emission from Relativistic Shocks
  and Its Implications for Gamma-Ray Burst Afterglows}.
\newblock {\em \apj}, 548:787--799.

\bibitem[\protect\astroncite{{Sari} and
  {M{\'e}sz{\'a}ros}}{2000}]{2000ApJ...535L..33S}
{Sari}, R. and {M{\'e}sz{\'a}ros}, P. (2000).
\newblock {Impulsive and Varying Injection in Gamma-Ray Burst Afterglows}.
\newblock {\em \apjl}, 535(1):L33--L37.

\bibitem[\protect\astroncite{{Savchenko} et~al.}{2017}]{2017ApJ...848L..15S}
{Savchenko}, V., {Ferrigno}, C., {Kuulkers}, E., {Bazzano}, A., {Bozzo}, E.,
  {Brandt}, S., and {et al.} (2017).
\newblock {INTEGRAL Detection of the First Prompt Gamma-Ray Signal Coincident
  with the Gravitational-wave Event GW170817}.
\newblock {\em \apjl}, 848:L15.

\bibitem[\protect\astroncite{{Schroeder} et~al.}{2020}]{2020arXiv200607434S}
{Schroeder}, G., {Margalit}, B., {Fong}, W.-f., {Metzger}, B.~D., {Williams},
  P. K.~G., {Paterson}, K., {Alexander}, K.~D., {Laskar}, T., {Goyal}, A.~V.,
  and {Berger}, E. (2020).
\newblock {A Late-time Radio Survey of Short GRBs at $z<0.5$: New Constraints
  on the Remnants of Neutron Star Mergers}.
\newblock {\em arXiv e-prints}, page arXiv:2006.07434.

\bibitem[\protect\astroncite{{Shibata} and
  {Taniguchi}}{2006}]{2006PhRvD..73f4027S}
{Shibata}, M. and {Taniguchi}, K. (2006).
\newblock {Merger of binary neutron stars to a black hole: Disk mass, short
  gamma-ray bursts, and quasinormal mode ringing}.
\newblock {\em \prd}, 73(6):064027.

\bibitem[\protect\astroncite{{Siegel} and
  {Metzger}}{2017}]{2017PhRvL.119w1102S}
{Siegel}, D.~M. and {Metzger}, B.~D. (2017).
\newblock {Three-Dimensional General-Relativistic Magnetohydrodynamic
  Simulations of Remnant Accretion Disks from Neutron Star Mergers: Outflows
  and r -Process Nucleosynthesis}.
\newblock {\em \prl}, 119(23):231102.

\bibitem[\protect\astroncite{{Sironi} and
  {Giannios}}{2013}]{2013ApJ...778..107S}
{Sironi}, L. and {Giannios}, D. (2013).
\newblock {A Late-time Flattening of Light Curves in Gamma-Ray Burst
  Afterglows}.
\newblock {\em \apj}, 778(2):107.

\bibitem[\protect\astroncite{{Smartt} et~al.}{2017}]{2017Natur.551...75S}
{Smartt}, S.~J., {Chen}, T.~W., {Jerkstrand}, A., {Coughlin}, M., {Kankare},
  E., {Sim}, S.~A., and {et al.} (2017).
\newblock {A kilonova as the electromagnetic counterpart to a
  gravitational-wave source}.
\newblock {\em \nat}, 551(7678):75--79.

\bibitem[\protect\astroncite{{Soares-Santos}
  et~al.}{2017}]{2017ApJ...848L..16S}
{Soares-Santos}, M., {Holz}, D.~E., {Annis}, J., {Chornock}, R., {Herner}, K.,
  {Berger}, E., and {Dark Energy Camera GW-EM Collaboration} (2017).
\newblock {The Electromagnetic Counterpart of the Binary Neutron Star Merger
  LIGO/Virgo GW170817. I. Discovery of the Optical Counterpart Using the Dark
  Energy Camera}.
\newblock {\em \apjl}, 848(2):L16.

\bibitem[\protect\astroncite{{Sobacchi} et~al.}{2017}]{2017MNRAS.472..616S}
{Sobacchi}, E., {Granot}, J., {Bromberg}, O., and {Sormani}, M.~C. (2017).
\newblock {A common central engine for long gamma-ray bursts and Type Ib/c
  supernovae}.
\newblock {\em \mnras}, 472(1):616--627.

\bibitem[\protect\astroncite{{Soderberg} et~al.}{2006}]{2006ApJ...651.1005S}
{Soderberg}, A.~M., {Chevalier}, R.~A., {Kulkarni}, S.~R., and {Frail}, D.~A.
  (2006).
\newblock {The radio and X-ray luminous SN 2003bg and the circumstellar density
  variation around radio supernovae}.
\newblock {\em \apj}, 651:1005--1018.

\bibitem[\protect\astroncite{{Surman} et~al.}{2008}]{2008ApJ...679L.117S}
{Surman}, R., {McLaughlin}, G.~C., {Ruffert}, M., {Janka}, H.~T., and {Hix},
  W.~R. (2008).
\newblock {r-Process Nucleosynthesis in Hot Accretion Disk Flows from Black
  Hole-Neutron Star Mergers}.
\newblock {\em \apjl}, 679(2):L117.

\bibitem[\protect\astroncite{{Tan} et~al.}{2001}]{2001ApJ...551..946T}
{Tan}, J.~C., {Matzner}, C.~D., and {McKee}, C.~F. (2001).
\newblock {Trans-Relativistic Blast Waves in Supernovae as Gamma-Ray Burst
  Progenitors}.
\newblock {\em \apj}, 551:946--972.

\bibitem[\protect\astroncite{{Tanvir} et~al.}{2013}]{2013Natur.500..547T}
{Tanvir}, N.~R., {Levan}, A.~J., {Fruchter}, A.~S., {Hjorth}, J., {Hounsell},
  R.~A., {Wiersema}, K., and {Tunnicliffe}, R.~L. (2013).
\newblock {A `kilonova' associated with the short-duration {$\gamma$}-ray burst
  GRB 130603B}.
\newblock {\em \nat}, 500:547--549.

\bibitem[\protect\astroncite{{Tanvir} et~al.}{2017}]{2017ApJ...848L..27T}
{Tanvir}, N.~R., {Levan}, A.~J., {Gonz{\'a}lez-Fern{\'a}ndez}, C., {Korobkin},
  O., {Mandel}, I., {Rosswog}, S., and {et al.} (2017).
\newblock {The Emergence of a Lanthanide-rich Kilonova Following the Merger of
  Two Neutron Stars}.
\newblock {\em \apjl}, 848(2):L27.

\bibitem[\protect\astroncite{{Totani}}{2003}]{2003ApJ...598.1151T}
{Totani}, T. (2003).
\newblock {A Failed Gamma-Ray Burst with Dirty Energetic Jets Spirited Away?
  New Implications for the Gamma-Ray Burst-Supernova Connection from SN
  2002ap}.
\newblock {\em \apj}, 598(2):1151--1162.

\bibitem[\protect\astroncite{{Troja} et~al.}{2019}]{2019MNRAS.489.2104T}
{Troja}, E., {Castro-Tirado}, A.~J., {Becerra Gonz{\'a}lez}, J., {Hu}, Y.,
  {Ryan}, G.~S., {Cenko}, S.~B., and {et al.} (2019).
\newblock {The afterglow and kilonova of the short GRB 160821B}.
\newblock {\em \mnras}, 489(2):2104--2116.

\bibitem[\protect\astroncite{{Troja} et~al.}{2017a}]{2017Natur.547..425T}
{Troja}, E., {Lipunov}, V.~M., {Mundell}, C.~G., and et~al. (2017a).
\newblock {Significant and variable linear polarization during the prompt
  optical flash of GRB 160625B.}
\newblock {\em \nat}, 547:425--427.

\bibitem[\protect\astroncite{{Troja} et~al.}{2020a}]{2020GCN.27411....1T}
{Troja}, E., {Piro}, L., {Ryan}, G., {van Eerten}, H., and {Zhang}, B. (2020a).
\newblock {GW170817: Continued X-ray emission detected with Chandra at 940 days
  post-merger}.
\newblock {\em GRB Coordinates Network}, 27411:1.

\bibitem[\protect\astroncite{{Troja} et~al.}{2017b}]{2017Natur.551...71T}
{Troja}, E., {Piro}, L., {van Eerten}, H., {Wollaeger}, R.~T., {Im}, M., {Fox},
  O.~D., and {et al.} (2017b).
\newblock {The X-ray counterpart to the gravitational-wave event GW170817}.
\newblock {\em \nat}, 551(7678):71--74.

\bibitem[\protect\astroncite{{Troja} et~al.}{2020b}]{2020arXiv200601150T}
{Troja}, E., {van Eerten}, H., {Zhang}, B., {Ryan}, G., {Piro}, L., {Ricci},
  R., {O'Connor}, B., {Wieringa}, M.~H., {Cenko}, S.~B., and {Sakamoto}, T.
  (2020b).
\newblock {A thousand days after the merger: continued X-ray emission from
  GW170817}.
\newblock {\em arXiv e-prints}, page arXiv:2006.01150.

\bibitem[\protect\astroncite{{Valenti} et~al.}{2008}]{2008MNRAS.383.1485V}
{Valenti}, S., {Benetti}, S., {Cappellaro}, E., {Patat}, F., {Mazzali}, P.,
  {Turatto}, M., {Hurley}, K., {Maeda}, K., {Gal-Yam}, A., {Foley}, R.~J.,
  {Filippenko}, A.~V., {Pastorello}, A., {Challis}, P., {Frontera}, F.,
  {Harutyunyan}, A., {Iye}, M., {Kawabata}, K., {Kirshner}, R.~P., {Li}, W.,
  {Lipkin}, Y.~M., {Matheson}, T., {Nomoto}, K., {Ofek}, E.~O., {Ohyama}, Y.,
  {Pian}, E., {Poznanski}, D., {Salvo}, M., {Sauer}, D.~N., {Schmidt}, B.~P.,
  {Soderberg}, A., and {Zampieri}, L. (2008).
\newblock {The broad-lined Type Ic supernova 2003jd}.
\newblock {\em \mnras}, 383(4):1485--1500.

\bibitem[\protect\astroncite{{van Marle} et~al.}{2006}]{2006A&A...460..105V}
{van Marle}, A.~J., {Langer}, N., {Achterberg}, A., and {Garc{\'\i}a-Segura},
  G. (2006).
\newblock {Forming a constant density medium close to long gamma-ray bursts}.
\newblock {\em \aap}, 460(1):105--116.

\bibitem[\protect\astroncite{{Vieira} et~al.}{2020}]{2020arXiv200309437V}
{Vieira}, N., {Ruan}, J.~J., {Haggard}, D., {Drout}, M.~R., {Nynka}, M.~C.,
  {Boyce}, H., {Spekkens}, K., {Safi-Harb}, S., {Carlberg}, R.~G.,
  {Fern{\'a}ndez}, R., {Piro}, A.~L., {Afsariardchi}, N., and {Moon}, D.-S.
  (2020).
\newblock {A Deep CFHT Optical Search for a Counterpart to the Possible Neutron
  Star -- Black Hole Merger GW190814}.
\newblock {\em arXiv e-prints}, page arXiv:2003.09437.

\bibitem[\protect\astroncite{{Wanajo} et~al.}{2014}]{2014ApJ...789L..39W}
{Wanajo}, S., {Sekiguchi}, Y., {Nishimura}, N., {Kiuchi}, K., {Kyutoku}, K.,
  and {Shibata}, M. (2014).
\newblock {Production of All the r-process Nuclides in the Dynamical Ejecta of
  Neutron Star Mergers}.
\newblock {\em \apjl}, 789(2):L39.

\bibitem[\protect\astroncite{{Watson} et~al.}{2020}]{2020MNRAS.492.5916W}
{Watson}, A.~M., {Butler}, N.~R., {Lee}, W.~H., {Becerra}, R.~L., {Pereyra},
  M., {Angeles}, F., {Farah}, A., {Figueroa}, L., {G{\'o}nzalez-Buitrago}, D.,
  {Quir{\'o}s}, F., {Ru{\'\i}z-D{\'\i}az-Soto}, J., {Tejada de Vargas}, C.,
  {Tinoco}, S.~J., and {Wolfram}, T. (2020).
\newblock {Limits on the electromagnetic counterpart to S190814bv}.
\newblock {\em \mnras}, 492(4):5916--5921.

\bibitem[\protect\astroncite{{Weinberg}}{1972}]{1972gcpa.book.....W}
{Weinberg}, S. (1972).
\newblock {\em {Gravitation and Cosmology}}.

\bibitem[\protect\astroncite{{Wijers} et~al.}{1997}]{1997MNRAS.288L..51W}
{Wijers}, R.~A.~M.~J., {Rees}, M.~J., and {Meszaros}, P. (1997).
\newblock {Shocked by GRB 970228: the afterglow of a cosmological fireball}.
\newblock {\em \mnras}, 288:L51--L56.

\bibitem[\protect\astroncite{{Woosley} and {Janka}}{2005}]{2005NatPh...1..147W}
{Woosley}, S. and {Janka}, T. (2005).
\newblock {The physics of core-collapse supernovae}.
\newblock {\em Nature Physics}, 1(3):147--154.

\bibitem[\protect\astroncite{{Woosley}}{1993}]{1993ApJ...405..273W}
{Woosley}, S.~E. (1993).
\newblock {Gamma-Ray Bursts from Stellar Mass Accretion Disks around Black
  Holes}.
\newblock {\em \apj}, 405:273.

\bibitem[\protect\astroncite{{Woosley} and {Bloom}}{2006}]{2006ARA&A..44..507W}
{Woosley}, S.~E. and {Bloom}, J.~S. (2006).
\newblock {The Supernova Gamma-Ray Burst Connection}.
\newblock {\em \araa}, 44:507--556.

\bibitem[\protect\astroncite{{Yamazaki} et~al.}{2006}]{2006MNRAS.369..311Y}
{Yamazaki}, R., {Toma}, K., {Ioka}, K., and {Nakamura}, T. (2006).
\newblock {Tail emission of prompt gamma-ray burst jets}.
\newblock {\em \mnras}, 369:311--316.

\bibitem[\protect\astroncite{{Yang} et~al.}{2015}]{2015NatCo...6.7323Y}
{Yang}, B., {Jin}, Z.-P., {Li}, X., {Covino}, S., {Zheng}, X.-Z., {Hotokezaka},
  and {et al.} (2015).
\newblock {A possible macronova in the late afterglow of the long-short burst
  GRB 060614}.
\newblock {\em Nature Communications}, 6:7323.

\bibitem[\protect\astroncite{{Yi} et~al.}{2013}]{2013ApJ...776..120Y}
{Yi}, S.-X., {Wu}, X.-F., and {Dai}, Z.-G. (2013).
\newblock {Early Afterglows of Gamma-Ray Bursts in a Stratified Medium with a
  Power-law Density Distribution}.
\newblock {\em \apj}, 776(2):120.

\bibitem[\protect\astroncite{{Zhang} and
  {M{\'e}sz{\'a}ros}}{2001}]{2001ApJ...559..110Z}
{Zhang}, B. and {M{\'e}sz{\'a}ros}, P. (2001).
\newblock {High-Energy Spectral Components in Gamma-Ray Burst Afterglows}.
\newblock {\em \apj}, 559:110--122.

\end{thebibliography}
%

%
\clearpage
\begin{table}
\centering \renewcommand{\arraystretch}{1.85}\addtolength{\tabcolsep}{1pt}
\caption{The relevant terms of the free-coasting  and deceleration phases.}
\label{table1}
\begin{tabular}{l   c  c  c c c}
 \hline \hline
\scriptsize{} &  \scriptsize{${\bf k=0}$  }  &\hspace{0.5cm}   \scriptsize{${\bf k=1.0}$ } &\hspace{0.5cm}   \scriptsize{${\bf k=1.5}$ } &\hspace{0.5cm}   \scriptsize{${\bf k=2.0}$ }  &\hspace{0.5cm}   \scriptsize{${\bf k=2.5}$ }  \\ 
\hline
$A_{\rm k}$ & $1\,{\rm cm^{-3}}$ & $1.5\times 10^{19}\,{\rm cm^{-2}}$ & $2.7\times 10^{28}\,{\rm cm^{-\frac32}}$ & $3\times 10^{36}\,{\rm cm^{-1}}$ & $1.3\times 10^{45}\,{\rm cm^{-\frac12}}$ \\
\hline\hline
Coasting Phase & & & & &\\
\hline
$r^0\, ( \times 10^{17}\,   \rm cm)$& $2.2$ & $1.8$ & $1.5$ & $1.3$ & $1.1$\\
$\gamma^0_{\rm m}$& $12.4$ & $12.4$ & $12.4$ & $12.4$ & $12.4$\\
$\gamma^0_{\rm c}\,  (\times 10^3) $& $3.1\times 10^2$ & $1.8$ & $0.3$ & 0.8& $0.5$\\
$\nu^{\rm syn,0}_{\rm a,1}\,(\rm Hz)$& $2.9\times10^9$ & $1.3\times10^{11}$ & $5.5\times10^{11}$ & $2.9\times10^{11}$ & $6.4\times10^{11}$\\
$\nu^{\rm syn,0}_{\rm a,2}\,(\rm Hz)$& $1.0\times10^8$ & $2.5\times10^9$ & $8.1\times10^{9}$ & $4.7\times10^9$ & $7.7\times10^9$\\
$\nu^{\rm syn,0}_{\rm a,3}\,(\rm Hz)$& $1.5\times10^6$ & $6.5\times10^9$ & $9.3\times10^{10}$ & $2.6\times10^{10}$ & $7.5\times10^{10}$\\
$\nu^{\rm syn,0}_{\rm m}\,(\rm Hz)$& $3.5\times10^6$ & $4.6\times10^7$ & $1.1\times10^8$ & $6.8\times10^7$ & $8.3\times10^7$\\
$\nu^{\rm syn,0}_{\rm c}\,(\rm Hz)$&$2.9\times10^{16}$ & $7.3\times10^{12}$ & $2.8\times10^{11}$ & $1.7\times10^{12}$    & $8.2\times10^{11}$\\
$F^{\rm syn,0}_{\rm \nu,max}\,({\rm mJy})    $ & $1.1\times10^2$  &  $6.5\times10^4$ & $5.5\times10^5$ & $8.1\times10^4$       &  $1.1\times10^5$   \\
$\nu^{\rm ssc,0}_{\rm mm}\,{\rm (eV)}$& $7.0\times 10^{-7}$  & $9.2\times 10^{-6}$ & $2.3\times 10^{-5}$ & $1.4\times 10^{-5}$  & $1.7\times 10^{-5}$\\
$\nu^{\rm ssc,0}_{\rm cc}\,({\rm MeV})$ &$3.8\times 10^6$ & $3.2\times10^{-2}$ & $3.4\times10^{-5}$ & $1.5\times 10^{-3}$ & $3.2\times 10^{-4}$\\
$\nu^{\rm ssc,0}_{\rm ma,1}\,({\rm eV})$ &$5.8\times 10^{-4}$ & $2.6\times10^{-2}$ & $0.1$ & $6.0\times 10^{-2}$ & $0.1$\\
$\nu^{\rm ssc,0}_{\rm mc,1}\,({\rm eV})$ &$5.9\times 10^3$ & $1.5$ & $5.7\times10^{-2}$ & $0.3$ & $0.2$\\
$\nu^{\rm ssc,0}_{\rm ma,2}\,({\rm eV})$ &$2.1\times10^{-5}$ & $5.1\times10^{-4}$ & $1.6\times10^{-3}$ & $9.5\times 10^{-4}$ & $1.6\times 10^{-3}$\\
$\nu^{\rm ssc,0}_{\rm mc,2}\,({\rm eV})$ &$5.9\times 10^{3}$ & $1.5$ & $5.7\times10^{-2}$ & $0.3$ & $0.2$\\
$\nu^{\rm ssc,0}_{\rm ca,2}\,({\rm eV})$ &$1.3\times 10^{4}$ & $11.1$ & $1.0$ & $4.1$ & $3.0$\\
$\nu^{\rm ssc,0}_{\rm ca,3}\,({\rm eV})$ &$1.9\times 10^{2}$ & $28.4$ & $11.4$ & $22.9$ & $29.3$\\
$\nu^{\rm ssc,0}_{\rm cm,3}\,({\rm eV})$ &$4.5\times 10^{2}$ & $0.2$ & $1.4\times10^{-2}$ & $5.9\times 10^{-2}$ & $3.2\times 10^{-2}$\\
$F^{\rm ssc,0}_{\rm \nu,max}\,({\rm mJy})$ & $5.8\times 10^{-5}$&  $2.3$ & $89.0$ & $4.4$ &  $9.6$ \\
\hline 
Deceleration & & & & &\\
\hline

$t^0 (\times 10^{3}\,   \rm day)$& $1.2$ & $1.1$ & $1.4$ & $28.4$ & $7.0\times 10^3$\\
$r^0 (\times 10^{17}\,   \rm cm)$& $7.2$ & $4.1$ & $3.2$ & $3.2$ & $2.6$ \\
$\gamma^0_{\rm m}\, (\times 10^{1})$& $13.2$ & $6.8$ & $5.2$ & $7.3$ & $6.7$\\
$\gamma^0_{\rm c}\,(\times 10^4)$& $3.0$ &  $0.2$ & $0.05$ & $0.2$ & $0.1$\\
$\nu^{\rm syn,0}_{\rm a,1}\,(\rm Hz)$& $8.8\times10^8$ & $2.5\times10^{10}$ & $9.5\times10^{10}$ & $2.5\times10^{10}$ & $4.6\times10^{10}$\\
$\nu^{\rm syn,0}_{\rm a,2}\,(\rm Hz)$& $1.1\times10^9$ & $6.1\times10^9$ & $1.2\times10^{10}$ & $6.4\times10^9$ & $8.4\times10^9$\\
$\nu^{\rm syn,0}_{\rm a,3}\,(\rm Hz)$& $3.2\times10^7$ & $4.3\times10^9$ & $2.9\times10^{10}$ & $4.2\times10^9$ & $8.5\times10^9$\\
$\nu^{\rm syn,0}_{\rm m}(\rm Hz)$& $1.3\times10^9$ & $1.5\times10^9$  & $1.5\times10^9$ & $1.6\times10^9$ & $1.5\times10^9$ \\
$\nu^{\rm syn,0}_{\rm c}(\rm Hz)$& $5.2\times10^{14}$ & $3.2\times10^{12}$ &  $4.0\times10^{11}$    & $4.2\times10^{12}$ & $2.8\times10^{12}$ \\
$F^{\rm syn,0}_{\rm \nu,max}\,({\rm mJy})$ & $1.3\times10^4$  &  $3.8\times10^5$ & $1.3\times10^6$ & $1.3\times10^{5}$ & $1.0\times10^5$ \\
$\nu^{\rm ssc,0}_{\rm mm}{\rm (eV)}$& $2.9\times10^{-2}$  & $8.8\times10^{-3}$ & $5.5\times10^{-3}$ & $1.1\times10^{-2}$ & $8.7\times10^{-3}$\\
$\nu^{\rm ssc,0}_{\rm cc}{\rm (MeV)}$ & $6.0\times 10^2$ & $1.0\times10^{-2}$ & $1.3\times10^{-4}$ & $1.8\times 10^{-2}$ & $7.7\times 10^{-3}$\\
$\nu^{\rm ssc,0}_{\rm ma,1}\,({\rm MeV})$ &$2.0\times 10^{-2}$ & $0.1$ & $0.3$ & $0.2$ & $0.3$\\
$\nu^{\rm ssc,0}_{\rm mc,1}\,({\rm MeV})$ &$1.2\times 10^{4}$ & $19.3$ & $1.4$ & $29.6$ & $16.3$\\
$\nu^{\rm ssc,0}_{\rm ma,2}\,({\rm MeV})$ &$2.4\times 10^{-2}$ & $3.7\times10^{-2}$ & $4.4\times10^{-2}$ & $4.5\times 10^{-2}$ & $4.9\times 10^{-2}$\\
$\nu^{\rm ssc,0}_{\rm mc,2}\,({\rm MeV})$ &$1.2\times 10^{4}$ & $19.3$ & $1.4$ & $29.6$ & $16.3$\\
$\nu^{\rm ssc,0}_{\rm ca,2}\,({\rm MeV})$ &$1.2\times 10^{3}$ & $19.6$ & $4.0$ & $28.2$ & $23.2$\\
$\nu^{\rm ssc,0}_{\rm ca,3}\,({\rm MeV})$ &$36.7$ & $14.0$ & $9.2$ & $18.5$ & $23.4$\\
$\nu^{\rm ssc,0}_{\rm cm,3}\,({\rm MeV})$ &$1.6\times 10^{3}$ & $10.2$ & $0.9$ & $7.0$ & $2.2$\\
$F^{\rm ssc,0}_{\rm \nu,max}\,({\rm mJy})$ & $2.1\times 10^{-2}$ &  $13.7$ & $1.4\times 10^2$ & $2.9$ & $2.5$ \\
\hline \hline

\end{tabular}
\end{table}
%

\begin{table}
\centering \renewcommand{\arraystretch}{2}\addtolength{\tabcolsep}{3pt}
\caption{The latest data points from Chandra Afterglow Observations of GRB 170817A.}
\label{table2}
\begin{tabular}{c   c  c  c  }
 \hline \hline
\scriptsize{$\delta t$} &  \scriptsize{$F_{\rm X}$ (0.3 - 10\,{\rm keV})  }  &\hspace{0.5cm}   \scriptsize{$F_\nu$ (1 keV) } &\hspace{0.5cm}   \scriptsize{$\Gamma_{\rm X}$ }  \\ 
\scriptsize{(days)} &  \scriptsize{$(\times 10^{15}\,{\rm erg\, cm^{-2}\,s^{-1}})$  }  &\hspace{0.5cm}   \scriptsize{$(\times 10^{-7}{\rm mJy})$ } & \hspace{0.5cm}     \\ 
\hline
\hline
\scriptsize{$358.6$} & \scriptsize{$7.75^{+2.70}_{-0.73}$} & \scriptsize{$6.77^{+2.36}_{-0.64}$} & \scriptsize{$1.69^{+0.49}_{-0.34}$} \\
\scriptsize{$582.2$} & \scriptsize{$3.25^{+0.85}_{-1.03}$} & \scriptsize{$2.76^{+0.72}_{-0.88}$} & \scriptsize{$1.57$} \\
\scriptsize{$741.9$} & \scriptsize{$2.21^{+0.85}_{-0.79}$} & \scriptsize{$1.88^{+0.72}_{-0.67}$} & \scriptsize{$1.57$} \\
\scriptsize{$940$} & \scriptsize{$1.10^{+0.60}_{-0.60}$} & \scriptsize{$0.95^{+0.52}_{-0.52}$} & \scriptsize{$1.585$} \\
\hline
\end{tabular}
\begin{flushleft}
\scriptsize{
\centering The data points in  ${\rm erg\,cm^{-2}\,s^{-1}}$ are taken from \cite{2019ApJ...886L..17H} and \cite{2020GCN.27411....1T}.\\}
\end{flushleft}
\end{table}

\clearpage

\appendix
\section{Light curves of SSC emission}
\subsection{The free-coasting phase}
%
%
The corresponding SSC break frequencies during the free-coasting phase are given by
%
%
{\small
\bary\label{nu_ssc_0}
h\nu^{\rm ssc}_{\rm mm}&=& \nu^{\rm ssc,0}_{\rm mm}\,g^4(p) \left(\frac{1+z}{1.022}\right)^{\frac{k-2}{2}}\epsilon_{\rm e,-1}^4\,\epsilon_{\rm B,-2}^\frac12 A^\frac12_{\rm k} \beta_{-0.5}^\frac{18-k}{2}\,t_6^{-\frac{k}{2}}\cr
h\nu^{\rm ssc}_{\rm cc}&=& \nu^{\rm ssc,0}_{\rm cc} \left(\frac{1+z}{1.022}\right)^{\frac{6-7k}{2}} (1+Y)^{-4}\,  \epsilon_{\rm B,-2}^{-\frac72}\, A^{-\frac72}_{\rm k} \beta_{-0.5}^\frac{7(k-2)}{2}\,t_6^{\frac{7k-8}{2}}.\,\,\,\,\,
\eary
}
For the case {\small $\nu^{\rm syn}_{\rm a,1}\leq \nu^{\rm syn}_{\rm m} \leq \nu^{\rm syn}_{\rm c}$}, the SSC break frequencies are
{\small
\bary\label{nu_ssc_1}
h\nu^{\rm ssc}_{\rm ma,1}&=& \nu^{\rm ssc,0}_{\rm ma,1}\,g(p) \left(\frac{1+z}{1.022}\right)^{\frac{4(k-2)}{5}}\epsilon_{\rm e,-1}\,\epsilon_{\rm B,-2}^\frac15 A^\frac45_{\rm k} \beta_{-0.5}^{-\frac{4k-15}{5}}\,t_6^{\frac{3-4k}{5}}\cr
h\nu^{\rm ssc}_{\rm mc,1}&=& \nu^{\rm ssc,0}_{\rm mc,1}\,g^2(p) \left(\frac{1+z}{1.022}\right)^{\frac{2-3k}{2}} (1+Y)^{-2}\, \epsilon_{\rm e,-1}^2\,\epsilon_{\rm B,-2}^{-\frac32} A^{-\frac32}_{\rm k} \beta_{-0.5}^{\frac{3k+2}{2}}\,t_6^{\frac{3k-4}{2}}\,.
\eary
}
For the case \small $\nu^{\rm syn}_{\rm m} \leq\nu^{\rm syn}_{\rm a,2}\leq \nu^{\rm syn}_{\rm c}$, the SSC break frequencies are
{\small
\bary\label{nu_ssc_2}
h\nu^{\rm ssc}_{\rm ma,2}&=& \nu^{\rm ssc,0}_{\rm ma,2}\,g(p)^{\frac{2(2p+3)}{p+4}}\,\left(\frac{1+z}{1.022}\right)^{\frac{(k-2)(p+6)}{2(p+4)}}\,\epsilon_{\rm e,-1}^{\frac{2(2p+3)}{p+4}} \,\epsilon_{\rm B,-2}^{\frac{p+2)}{2(p+4)}}A^{\frac{p+6}{2(p+4)}}_{\rm k} \beta_{-0.5}^{\frac{18p-kp-6k+32}{2(p+4)}}\,t_6^{\frac{4-kp-6k}{2(p+4)}}\cr
h\nu^{\rm ssc}_{\rm mc,2}&=& \nu^{\rm ssc,0}_{\rm mc,2}\,g^2(p) \left(\frac{1+z}{1.022}\right)^{\frac{2-3k}{2}} (1+Y)^{-2}\, \epsilon_{\rm e,-1}^2\,\epsilon_{\rm B,-2}^{-\frac32} A^{-\frac32}_{\rm k} \beta_{-0.5}^{\frac{3k+2}{2}}\,t_6^{\frac{3k-4}{2}}\cr
h\nu^{\rm ssc}_{\rm ca,2}&=& \nu^{\rm ssc,0}_{\rm ca,2}\,g(p)^{\frac{2(p-1)}{p+4}}\, \left(\frac{1+z}{1.022}\right)^{\frac{4-10k+2p-3kp}{2(p+4)}} (1+Y)^{-2}\, \epsilon_{\rm e,-1}^{\frac{2(p-1)}{p+4}}\epsilon_{\rm B,-2}^{-\frac{3p+14}{2(p+4)}} A^{-\frac{3p+10}{2(p+4)}}_{\rm k} \beta_{-0.5}^{\frac{3kp+10k+2p-32}{2(p+4)}}\,t_6^{\frac{10k-4p+3kp-12}{2(p+4)}},
\eary
}
and  for {\small $\nu^{\rm syn}_{\rm a,3}\leq \nu^{\rm syn}_{\rm c} \leq  \nu^{\rm syn}_{\rm m}$}, they are
{\small
\bary\label{nu_ssc_3}
h\nu^{\rm ssc}_{\rm ca,3}&=& \nu^{\rm ssc,0}_{\rm ca,3}\, \left(\frac{1+z}{1.022}\right)^{-\frac{k+3}{5}}\,(1+Y)^{-1}\,\epsilon_{\rm B,-2}^{-\frac45} A^{-\frac15}_{\rm k}\beta_{-0.5}^{\frac{k-5}{5}}\,t_6^{\frac{k-2}{5}}\cr
h\nu^{\rm ssc}_{\rm cm,3}&=& \nu^{\rm ssc,0}_{\rm cm,3}\,g^2(p) \left(\frac{1+z}{1.022}\right)^{\frac{2-3k}{2}} (1+Y)^{-2}\, \epsilon_{\rm e,-1}^2\,\epsilon_{\rm B,-2}^{-\frac32} A^{-\frac32}_{\rm k} \beta_{-0.5}^{\frac{3k+2}{2}}\,t_6^{\frac{3k-4}{2}}\,.
\eary
}

The spectral peak flux density of SSC emission $F^{\rm ssc}_{\rm max}\sim\, \frac43 g(p)^{-1} \sigma_T A_{\rm k} r^{1-k} \,F^{\rm syn}_{\rm max}$, with $\sigma_T$ the Thomson cross section is given by
{\small
\be\label{f_ssc}
F^{\rm ssc}_{\rm \nu,max}= F^{\rm ssc,0}_{\rm \nu,max}\, g(p)^{-1} \left(\frac{1+z}{1.022}\right)^{\frac{5k-6}{2}} \epsilon^{\frac12}_{\rm B,-2}\, d_{\rm z,26.5}^{-2}\, A^{\frac52}_{\rm k}\, \beta_{-0.5}^{\frac{5(2-k)}{2}}t_6^{\frac{8-5k}{2}}\,.
\ee
}
The terms of $\nu^{\rm ssc,0}_{\rm mm}$, $\nu^{\rm ssc,0}_{\rm cc}$,  $\nu^{\rm ssc,0}_{\rm ma,1}$, $\nu^{\rm ssc,0}_{\rm mc,1}$, $\nu^{\rm ssc,0}_{\rm ma,2}$, $\nu^{\rm ssc,0}_{\rm mc,2}$, $\nu^{\rm ssc,0}_{\rm ca,2}$, $\nu^{\rm ssc,0}_{\rm ma,3}$, $\nu^{\rm ssc,0}_{\rm mc,3}$  and $F^{\rm ssc,0}_{\rm \nu,max}$ are given in Table \ref{table1} for ${\rm k}=0$, $1$, $1.5$, $2$ and $2.5$.\\
Using the SSC break frequencies (eqs. \ref{nu_ssc_0}, \ref{nu_ssc_1}, \ref{nu_ssc_2} and \ref{nu_ssc_3}) and the spectral peak flux density (eq. \ref{f_ssc}),  the SSC light curves in the fast- and slow-cooling regime evolve as 
{\small
\begin{eqnarray}
\label{fcssc_t}
F^{\rm ssc}_{\rm \nu}\propto \cases{ 
t^{\frac{28-19k}{5}} \nu,\hspace{5.6cm} \nu<\nu^{\rm ssc}_{\rm ca,3}, \cr
t^{\frac{16-11k}{3}} \nu^{\frac13},\hspace{5.4cm} \nu^{\rm ssc}_{\rm ca,3}<\nu<\nu^{\rm ssc}_{\rm cc},\, \cr
t^{\frac{8-3k}{4}}\left(C_{\rm cf,11}+C_{\rm cf,12}\ln \left[t^{\frac{8-7k}{2}}\nu \right] \right)\, \nu^{-\frac{1}{2}},\hspace{1.3cm}\nu^{\rm ssc}_{\rm cc}<\nu<\nu^{\rm ssc}_{\rm cm,3},\,\,\,\,\, \cr
t^{\frac{8-3k}{4}}\left(C_{\rm cf,21}+C_{\rm cf,22}\ln \left[t^{-\frac{k}{2}}\nu^{-1} \right]\right)\nu^{-\frac{1}{2}},\,\,\,\,\hspace{1.1cm}   \nu^{\rm ssc}_{\rm cm,3}<\nu<\nu^{\rm ssc}_{\rm mm},\,\,\,\,\, \cr
t^{\frac{8-k(p+2)}{4}}\left(C_{\rm cf,31}+C_{\rm cf,32}\ln \left[t^{\frac{k}{2}}\nu \right]\right) \nu^{-\frac{p}{2}},\hspace{1.3cm} \nu^{\rm ssc}_{\rm mm}<\nu, \cr
}
\end{eqnarray}
}
and 
{\small
\begin{eqnarray}
\label{fcssc_t}
F^{\rm ssc}_{\rm \nu}\propto \cases{ 
	t^{-\frac{9(k-2)}{5}} \nu,\hspace{6.5cm} \nu<\nu^{\rm ssc}_{\rm ma,1}, \cr
	t^{\frac{12-7k}{3}} \nu^{\frac{1}{3}},\hspace{6.5cm} \nu^{\rm ssc}_{\rm ma,1}<\nu<\nu^{\rm ssc}_{\rm mm},\, \cr
	t^{\frac{16-9k-kp}{4}}\left(C_{\rm cs1,11}+C_{\rm cs1,12}\ln \left[t^{\frac{k}{2}}\nu \right] \right)\, \nu^{\frac{1-p}{2}},\hspace{1.6cm}\nu^{\rm ssc}_{\rm mm}<\nu<\nu^{\rm ssc}_{\rm mc,1},\,\,\,\,\, \cr
	t^{\frac{16-9k-kp}{4}}\left(C_{\rm cs1,21}+C_{\rm cs1,22}\ln \left[t^{\frac{7k-8}{2}}\nu^{-1} \right]\right)\nu^{\frac{1-p}{2}},\,\,\,\,\hspace{0.7cm}   \nu^{\rm ssc}_{\rm mc,1}<\nu<\nu^{\rm ssc}_{\rm cc},\,\,\,\,\, \cr
	t^{\frac{8-2k-kp}{4}}\left(C_{\rm cs1,31}+C_{\rm cs1,32}\ln \left[t^{\frac{8-7k}{2}}\nu \right]\right) \nu^{-\frac{p}{2}},\hspace{1.5cm} \nu^{\rm ssc}_{\rm cc}<\nu, \cr
}
\end{eqnarray}
}
{\small
\begin{eqnarray}
\label{fcssc_t}
F^{\rm ssc}_{\rm \nu}\propto \cases{ 
	t^{-\frac{3(k-2)(p+5)}{2(p+4)}} \nu,\hspace{7.1cm} \nu<\nu^{\rm ssc}_{\rm ma,2}, \cr
	t^{\frac{16-9k-kp}{4}}\left(C_{\rm cs2,11}+C_{\rm cs2,12}\ln \left[t^{\frac{k(p+6)-4}{2(p+4)}}\nu\right]\right)\,\nu^{\frac{1-p}{2}},\hspace{1.8cm} \nu^{\rm ssc}_{\rm ma,2}<\nu<\nu^{\rm ssc}_{\rm mc,2},\, \cr
	t^{\frac{16-9k-kp}{4}}\left(C_{\rm cs2,21}+C_{\rm cs2,22}\ln \left[t^{\frac{k(2p+9)-2(p+5)}{p+4}} \right]\right)\, \nu^{\frac{1-p}{2}},\hspace{1.2cm}\nu^{\rm ssc}_{\rm mc,2}<\nu<\nu^{\rm ssc}_{\rm ca,2},\,\,\,\,\, \cr
	t^{\frac{16-9k-kp}{4}}\left(C_{\rm cs2,31}+C_{\rm cs2,32}\ln \left[t^{\frac{7k-8}{2}}\nu^{-1} \right]\right)\nu^{\frac{1-p}{2}},\,\,\,\,\hspace{1.8cm}   \nu^{\rm ssc}_{\rm ca,2}<\nu<\nu^{\rm ssc}_{\rm cc},\,\,\,\,\, \cr
	t^{\frac{8-2k-kp}{4}}\left(C_{\rm cs2,41}+C_{\rm cs2,42}\ln \left[t^{\frac{8-7k}{2}}\nu\right] \right) \nu^{-\frac{p}{2}},\hspace{2.6cm} \nu^{\rm ssc}_{\rm cc}<\nu, \cr
}
\end{eqnarray}
}
respectively. The parameters $C_{\rm cf}$, $C_{\rm cs,1}$ and $C_{\rm cs,2}$  do not evolve with time.

\subsection{The deceleration phase}

The PL indices $m_{\rm ij}$ for $i,\,j=$1, 2 and 3 are  $m_{\rm 11}=-\frac{55+8\alpha-k(21+4\alpha)}{5(\alpha+5-k)}$, $m_{\rm 12}=\frac{-2p(\alpha-10)+kp(\alpha-6)+6k(\alpha+4)-12(\alpha+5)}{2(p+4)(\alpha+5-k)}$,\\
$m_{\rm 13}=\frac{k(16+9\alpha)-13\alpha-20}{5(\alpha+5-k)}$, $m_{\rm 21}=\frac{3\alpha-16k+30-4k\alpha}{ 5(\alpha+5-k)}$,  $m_{\rm 22}=\frac{4\alpha-6k\alpha-\alpha kp+8kp-30p-16k+20}{2(p+4)(\alpha+5-k)}$ and $m_{\rm 23}=\frac{8\alpha-9k\alpha-11k-5}{5(\alpha+5-k)}$.\\
The corresponding SSC break frequencies during the deceleration phase are given by

{\small
\bary\label{nu_ssc_de}
h\nu^{\rm ssc}_{\rm mm}&=&\nu^{\rm ssc,0}_{\rm mm}\,\left(\frac{1+z}{1.022}\right)^{\frac{44+k(\alpha-14)-2\alpha }{2(\alpha+5-k)}}\,g(p)^4 \epsilon^4_{\rm e,-1}\,\epsilon^\frac12_{\rm B,-2}\,  A^{\frac{\alpha-13}{2(\alpha+5-k)}}_{\rm k}\,\tilde{E}_{51}^{\frac{18-k}{2(\alpha+5-k)}}\,t_7^{-\frac{54 + k(\alpha-16)}{2(\alpha+5-k)}}\cr
h\nu^{\rm ssc}_{\rm cc}&=&\nu^{\rm ssc,0}_{\rm cc}\,\left(\frac{1+z}{1.022}\right)^{\frac{6\alpha-7k\alpha  - 6k - 12}{2(\alpha +5-k)}}\, \epsilon^{-\frac72}_{\rm B,-2}\, (1+Y)^{-4}\, A^{-\frac{7(\alpha+3)}{2(\alpha+5-k)}}_{\rm k}\,\tilde{E}_{51}^{\frac{7(k-2)}{2(\alpha+5-k)}}\,t_7^{\frac{2+8k-8\alpha+7 k\alpha}{2(\alpha+5-k)}}
\eary
}

For the case {\small $\nu^{\rm syn}_{\rm a,1}\leq \nu^{\rm syn}_{\rm m} \leq \nu^{\rm syn}_{\rm c}$}, the SSC break frequencies are

{\small
	\bary\label{nu1_ssc_de}
	h\nu^{\rm ssc}_{\rm ma,1}&=& \nu^{\rm ssc,0}_{\rm ma,1}\,g(p) \left(\frac{1+z}{1.022}\right)^{\frac{5+k-8\alpha+4k\alpha}{5(\alpha+5-k)}}\epsilon_{\rm e,-1}\,\epsilon_{\rm B,-2}^\frac15\, A^{\frac{4\alpha+5}{5(\alpha+5-k)}}_{\rm k} \tilde{E}_{51}^{\frac{15-4k}{5(\alpha+5-k)}}\,t_7^{\frac{3\alpha-30+4k(1-\alpha)}{5(\alpha+5-k)}}\cr
	h\nu^{\rm ssc}_{\rm mc,1}&=& \nu^{\rm ssc,0}_{\rm mc,1}\,g^2(p) \left(\frac{1+z}{1.022}\right)^{\frac{16-10k+2\alpha-3k\alpha}{2(\alpha+5-k)}} (1+Y)^{-2}\,\epsilon_{\rm e,-1}^2\,\epsilon_{\rm B,-2}^{-\frac32} A^{-\frac{3\alpha+17}{2(\alpha+5-k)}}_{\rm k} \tilde{E}_{51}^{\frac{3k+2}{2(\alpha+5-k)}}\,t_7^{-\frac{26+4\alpha-3k(\alpha+4)}{2(\alpha+5-k)}}
	\eary
}
For the case \small $\nu^{\rm syn}_{\rm m} \leq\nu^{\rm syn}_{\rm a,2}\leq \nu^{\rm syn}_{\rm c}$, the SSC break frequencies are

{\small
	\bary\label{nu2_ssc_de}
	h\nu^{\rm ssc}_{\rm ma,2}&=& \nu^{\rm ssc,0}_{\rm ma,2}\,g(p)^{\frac{2(2p+3)}{p+4}}\,\left(\frac{1+z}{1.022}\right)^{\frac{36+44p-12\alpha-2p\alpha+k(6\alpha-8+p(\alpha-14))}{2(p+4)(\alpha+5-k)}} \epsilon_{\rm e,-1}^{\frac{2(2p+3)}{p+4}} \,\epsilon_{\rm B,-2}^{\frac{p+2}{2(p+4)}}A^{\frac{6\alpha-2+p(\alpha-13)}{2(p+4)(\alpha+5-k)}}_{\rm k}\cr
	&&\hspace{6.9cm}	 \times\, \tilde{E}_{51}^{-\frac{k(p+6)-2(9p+16)}{2(p+4)(\alpha+5-k)}} \,t_7^{-\frac{76+54p-4\alpha+k(6\alpha-16+p(\alpha-16))}{2(p+4)(\alpha+5-k)}}\cr
	h\nu^{\rm ssc}_{\rm mc,2}&=& \nu^{\rm ssc,0}_{\rm mc,2}\,g^2(p) \left(\frac{1+z}{1.022}\right)^{\frac{16-10k+2\alpha-3k\alpha}{2(\alpha+5-k)}} (1+Y)^{-2}\, \epsilon_{\rm e,-1}^2\,\epsilon_{\rm B,-2}^{-\frac32} A^{-\frac{17+3\alpha}{2(\alpha+5-k)}}_{\rm k} \tilde{E}_{51}^{\frac{3k+2}{2(\alpha+5-k)}}\,t_7^{-\frac{26+4\alpha-3k(\alpha+4)}{2(\alpha+5-k)}}\cr
	h\nu^{\rm ssc}_{\rm ca,2}&=& \nu^{\rm ssc,0}_{\rm ca,2}\,g(p)^{\frac{2(p-1)}{p+4}}\, \left(\frac{1+z}{1.022}\right)^{\frac{2(2(\alpha-19)+p(\alpha+8))-k(10\alpha-8+p(3\alpha+10))}{2(p+4)(\alpha+5-k)}}\,(1+Y)^{-2}\, \epsilon_{\rm e,-1}^{\frac{2(p-1)}{p+4}}\epsilon_{\rm B,-2}^{-\frac{3p+14}{2(p+4)}} A^{-\frac{18+17p+10\alpha+3p\alpha}{2(p+4)(\alpha+5-k)}}_{\rm k}\,\cr
&&\hspace{6.9cm}	 \times \tilde{E}_{51}^{\frac{10k-32+2p+3kp}{2(p+4)(\alpha+5-k)}}\,t_7^{\frac{36+2(5k-6)\alpha+p(3k(\alpha+4)-26-4\alpha)}{2(p+4)(\alpha+5-k)}}
	\eary
}
and  for {\small $\nu^{\rm syn}_{\rm a,3}\leq \nu^{\rm syn}_{\rm c} \leq  \nu^{\rm syn}_{\rm m}$}, they are
{\small
	\bary\label{nu3_ssc_de}
	h\nu^{\rm ssc}_{\rm ca,3}&=& \nu^{\rm ssc,0}_{\rm ca,3}\, \left(\frac{1+z}{1.022}\right)^{-\frac{k(\alpha-6)+3(\alpha+10)}{5(\alpha+5-k)}}\,(1+Y)^{-1}\,\epsilon_{\rm B,-2}^\frac45 A^{-\frac{\alpha}{5(\alpha+5-k)}}_{\rm k} \tilde{E}_{51}^{\frac{k-5}{5(\alpha+5-k)}}\,t_7^{\frac{5-k-2\alpha+k\alpha}{5(\alpha+5-k)}}\cr
	h\nu^{\rm ssc}_{\rm cm,3}&=& \nu^{\rm ssc,0}_{\rm cm,3}\,g^2(p) \left(\frac{1+z}{1.022}\right)^{\frac{16-10k+2\alpha-3k\alpha}{2(\alpha+5-k)}} (1+Y)^{-2}\, \epsilon_{\rm e,-1}^2\epsilon_{\rm B,-2}^{-\frac32} A^{-\frac{3\alpha+17}{2(\alpha+5-k)}}_{\rm k} \tilde{E}_{51}^{\frac{3k+2}{2(\alpha+5-k)}}\,t_7^{\frac{3k(\alpha+4)-26-4\alpha}{2(\alpha+5-k)}}
	\eary
}
The terms of $\nu^{\rm ssc,0}_{\rm mm}$, $\nu^{\rm ssc,0}_{\rm cc}$,  $\nu^{\rm ssc,0}_{\rm ma,1}$, $\nu^{\rm ssc,0}_{\rm mc,1}$, $\nu^{\rm ssc,0}_{\rm ma,2}$, $\nu^{\rm ssc,0}_{\rm mc,2}$, $\nu^{\rm ssc,0}_{\rm ca,2}$, $\nu^{\rm ssc,0}_{\rm ma,3}$, $\nu^{\rm ssc,0}_{\rm mc,3}$  and $F^{\rm ssc,0}_{\rm \nu,max}$ are given in Table \ref{table1} for ${\rm k}=0$, $1$, $1.5$, $2$ and $2.5$.\\
The spectral peak flux density of SSC emission 
\be\label{SSC_fl}
F^{\rm ssc}_{\rm \nu,max}=F^{\rm ssc,0}_{\rm \nu,max}\,g(p)^{-1}\left(\frac{1+z}{1.022}\right)^{\frac{6k-6\alpha+5k\alpha}{2(\alpha+5-k)}}\, \epsilon^{\frac12}_{\rm B,-2}\, d_{\rm z,26.5}^{-2}\, A^{\frac{5(\alpha+3)}{2(\alpha+5-k)}}_{\rm k}\, \tilde{E}_{51}^{\frac{5(2-k)}{2(\alpha+5-k)}}\,t_7^{\frac{10-8k + 8\alpha-5k\alpha}{2(\alpha+5-k)}}\,.
\ee

Using the SSC break frequencies  (eqs. \ref{nu_ssc_de}, \ref{nu1_ssc_de}, \ref{nu2_ssc_de} and \ref{nu3_ssc_de}) and the spectral peak flux density (eq. \ref{SSC_fl}),  the SSC light curves in the fast- and slow-cooling regime are 
{\small
\begin{eqnarray}
\label{fc_ssc_t}
F^{\rm ssc}_{\rm \nu}\propto \cases{ 
	t^{\frac{20-26k+28\alpha-19k\alpha}{5(\alpha+5-k)}} \nu,\hspace{8.3cm} \nu<\nu^{\rm ssc}_{\rm ca,3}, \cr
	t^{\frac{14-16k+16\alpha-11k\alpha}{3(\alpha+5-k)}} \nu^{\frac13},\hspace{8.cm} \nu^{\rm ssc}_{\rm ca,3}<\nu<\nu^{\rm ssc}_{\rm cc},\, \cr
	t^{\frac{22-8k+8\alpha-3k\alpha}{4(\alpha+5-k)}}\left(C_{\rm df,11}+C_{\rm df,12}\ln \left[t^{\frac{8\alpha-2-k(8+7\alpha)}{2(\alpha+5-k)}}\nu \right]\right)\, \nu^{-\frac{1}{2}},\hspace{2.7cm}\nu^{\rm ssc}_{\rm cc}<\nu<\nu^{\rm ssc}_{\rm cm,3},\,\,\,\,\, \cr
	t^{\frac{22-8k+8\alpha-3k\alpha}{4(\alpha+5-k)}}\left(C_{\rm df,21}+C_{\rm df,22}\ln \left[t^{-\frac{54+k(\alpha-16)}{2(\alpha+5-k)}}\nu^{-1} \right]\right)\nu^{-\frac{1}{2}},\,\,\,\,\hspace{2.3cm}   \nu^{\rm ssc}_{\rm cm,3}<\nu<\nu^{\rm ssc}_{\rm mm},\,\,\,\,\, \cr
	t^{\frac{76-54p-kp(\alpha-16)+8\alpha-2k(\alpha+12)}{4(\alpha+5-k)}}\left(C_{\rm df,31}+C_{\rm df,32}\ln \left[t^{\frac{54+k(\alpha-16)}{2(\alpha+5-k)}}\nu \right]\right) \nu^{-\frac{p}{2}},\hspace{1.1cm} \nu^{\rm ssc}_{\rm mm}<\nu, \cr
}
\end{eqnarray}
}
{\small
\begin{eqnarray}
\label{sc_ssc_t}
F^{\rm ssc}_{\rm \nu}\propto \cases{ 
	t^{\frac{9(2(\alpha+5)-k(\alpha+4))}{5(\alpha+5-k)}} \nu,\hspace{9.2cm} \nu<\nu^{\rm ssc}_{\rm ma,1}, \cr
	t^{\frac{42-20k+12\alpha-7k\alpha}{3(\alpha+5-k)}} \nu^{\frac{1}{3}},\hspace{9.1cm} \nu^{\rm ssc}_{\rm ma,1}<\nu<\nu^{\rm ssc}_{\rm mm},\, \cr
	t^{\frac{74-54p+16\alpha-k(32+p(\alpha-16)+9\alpha)}{4(\alpha+5-k)}}\left(C_{\rm ds1,11}+C_{\rm ds1,12}\ln \left[t^{\frac{54+k(\alpha-16)}{2(\alpha+5-k)}}\nu \right]\right)\, \nu^{\frac{1-p}{2}},\hspace{1.6cm}\nu^{\rm ssc}_{\rm mm}<\nu<\nu^{\rm ssc}_{\rm mc,1},\,\,\,\,\, \cr
	t^{\frac{74-54p+16\alpha-k(32+p(\alpha-16)+9\alpha)}{4(\alpha+5-k)}}\left(C_{\rm ds1,21}+C_{\rm ds1,22}\ln \left[ t^{\frac{2+8k-8\alpha+7k\alpha}{2(\alpha+5-k)}}\nu^{-1} \right]\right)\nu^{\frac{1-p}{2}},\,\,\,\,\hspace{0.9cm}   \nu^{\rm ssc}_{\rm mc,1}<\nu<\nu^{\rm ssc}_{\rm cc},\,\,\,\,\, \cr
	t^{\frac{76-54p-kp(\alpha-16)+8\alpha-2k(\alpha+12)}{4(\alpha+5-k)}}\left(C_{\rm ds1,31}+C_{\rm ds1,32}\ln \left[t^{\frac{8\alpha-2-k(7\alpha+8)}{2(\alpha+5-k)}}\nu \right]\right) \nu^{-\frac{p}{2}},\hspace{1.3cm} \nu^{\rm ssc}_{\rm cc}<\nu, \cr
}
\end{eqnarray}
}
and 
{\small
	\begin{eqnarray}
	\label{fcssc_t}
	F^{\rm ssc}_{\rm \nu}\propto \cases{ 
		t^{\frac{6(31+p(\alpha-1)+5\alpha)-3k(24+\alpha(p+5))}{2(p+4)(\alpha+5-k)}} \nu,\hspace{9.1cm} \nu<\nu^{\rm ssc}_{\rm ma,2}, \cr
		t^{\frac{74-54p+16\alpha-k(32+p(\alpha-16)+9\alpha)}{4(\alpha+5-k)}}\left(C_{\rm ds2,11}+C_{\rm ds2,12}\ln\left[t^{\frac{76+54p-4\alpha+k(6\alpha-16+p(\alpha-16))}{2(p+4)(\alpha+5-k)}}\nu\right]\right)\,\nu^{\frac{1-p}{2}},\hspace{1.cm} \nu^{\rm ssc}_{\rm ma,2}<\nu<\nu^{\rm ssc}_{\rm mc,2},\, \cr
		t^{\frac{74-54p+16\alpha-k(32+p(\alpha-16)+9\alpha)}{4(\alpha+5-k)}}\left(C_{\rm ds2,21}+C_{\rm ds2,22}\ln \left[t^{\frac{k(16+2p(\alpha-1)+9\alpha)-2(7+p(\alpha-7)+5\alpha)}{(p+4)(\alpha+5-k)}} \right] \right)\, \nu^{\frac{1-p}{2}},\hspace{0.5cm}\nu^{\rm ssc}_{\rm mc,2}<\nu<\nu^{\rm ssc}_{\rm ca,2},\,\,\,\,\,\,\,\,\,\,\, \cr
		t^{\frac{74-54p+16\alpha-k(32+p(\alpha-16)+9\alpha)}{4(\alpha+5-k)}}\left(C_{\rm ds2,31}+C_{\rm ds2,32}\ln \left[t^{\frac{2+8k-8\alpha+7k\alpha}{2(\alpha+5-k)}}\nu^{-1} \right]\right)\nu^{\frac{1-p}{2}},\,\,\,\,\hspace{2.5cm}   \nu^{\rm ssc}_{\rm ca,2}<\nu<\nu^{\rm ssc}_{\rm cc},\,\,\,\,\,\,\, \cr
		t^{\frac{76-54p-kp(\alpha-16)+8\alpha-2k(\alpha+12)}{4(\alpha+5-k)}}\left(C_{\rm ds2,41}+C_{\rm ds2,42}\ln\left[t^{\frac{8\alpha-2-k(7\alpha+8)}{2(\alpha+5-k)}}\nu\right]\right) \nu^{-\frac{p}{2}},\hspace{2.9cm} \nu^{\rm ssc}_{\rm cc}<\nu, \cr
	}
	\end{eqnarray}
}

respectively.  The parameters $C_{\rm df}$, $C_{\rm ds,1}$ and $C_{\rm ds,2}$  do not evolve with time. 
\begin{figure}
{\centering
\resizebox*{0.9\textwidth}{0.65\textheight}
{\includegraphics{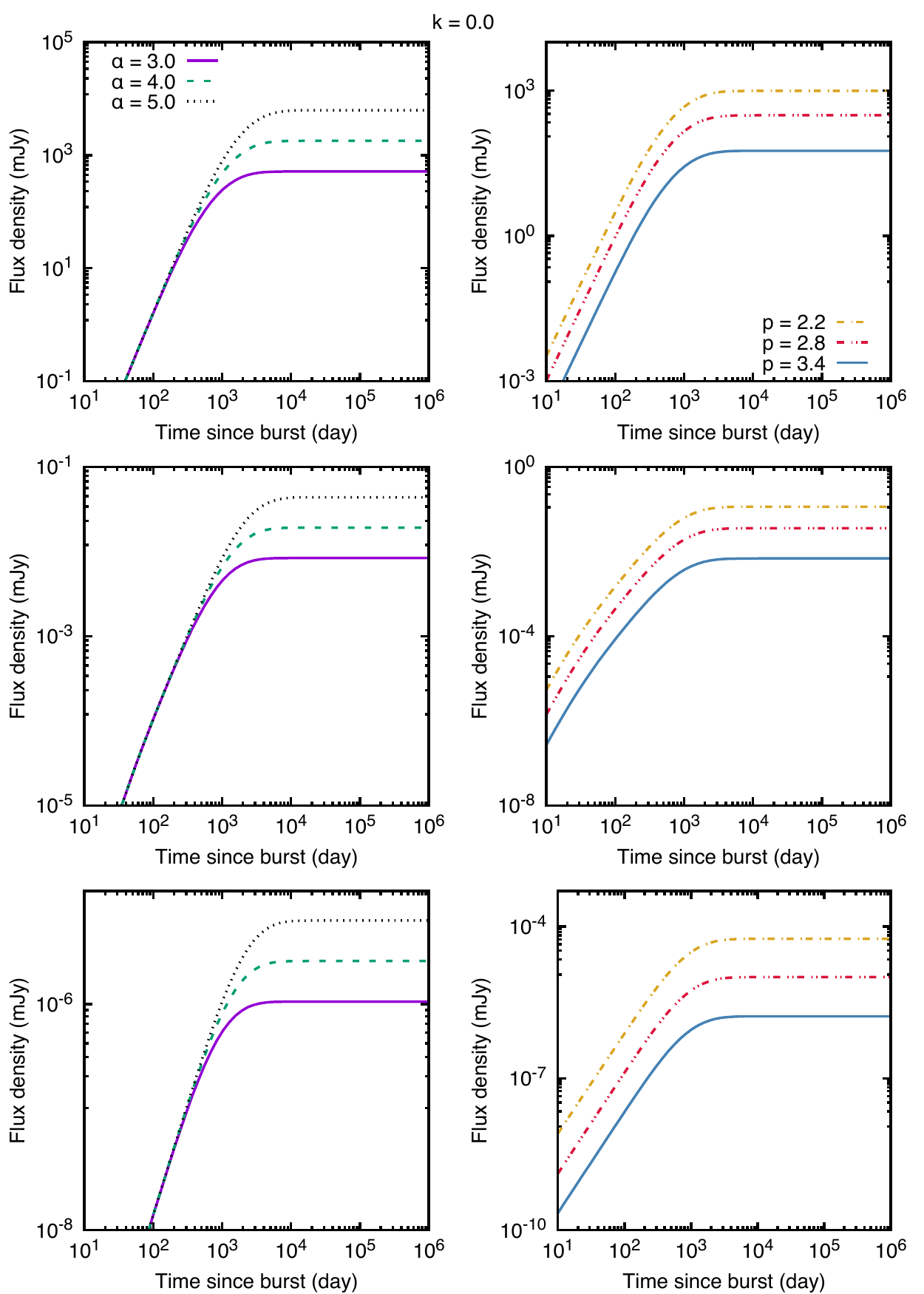}}
}   
\caption{Synchrotron light curves generated by the deceleration of the non-relativistic ejecta for $k=0$. Panels from top to bottom correspond to radio (1.6 GHz),  optical (1 eV) and X-ray (1 keV) bands, respectively.  The left-hand panels show the light curves for $p=2.6$ with $\alpha=3$, $4$ and $5$, and the right-hand panels show the light curves for $\alpha=3$ with $p=2.2$, $2.8$  and $3.4$.   The following parameters $\tilde{E}=10^{51}\,{\rm erg}$,  $A_0=1\,{\rm cm^{-3}}$, $\epsilon_{\rm B}=10^{-2}$, $\epsilon_{\rm B}=10^{-1}$ and $d=100\,{\rm Mpc}$ are used.  }
\label{k_0}
\end{figure} 
\begin{figure}
{\centering
\resizebox*{0.9\textwidth}{0.65\textheight}
{\includegraphics{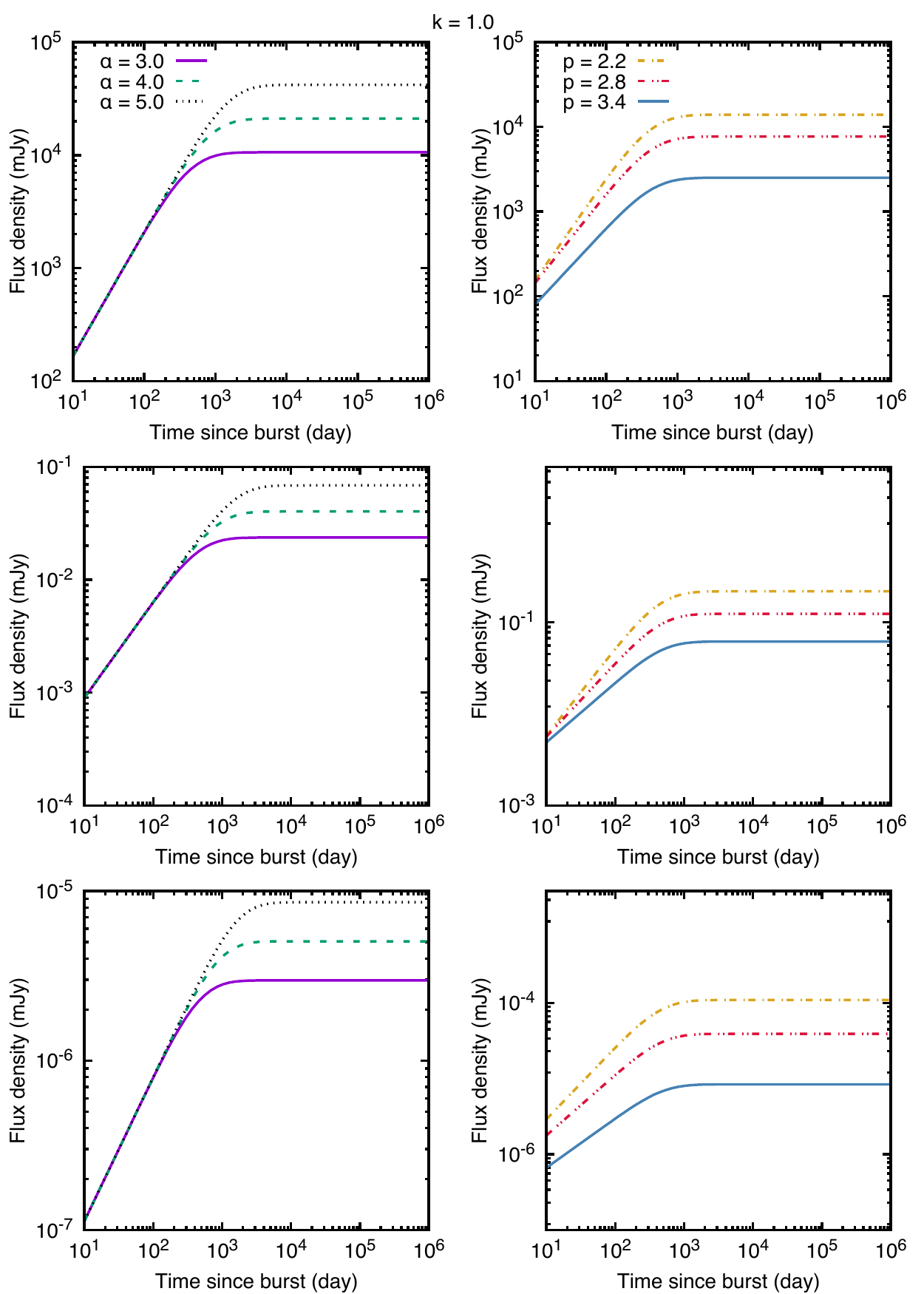}}
}   
\caption{The same as Figure \ref{k_0}, but for $k=1.0$ with $A_1=1.5\times 10^{19}\,{\rm cm^{-2}}$.}
\label{k_10}
\end{figure} 
\begin{figure}
{\centering
\resizebox*{0.9\textwidth}{0.65\textheight}
{\includegraphics{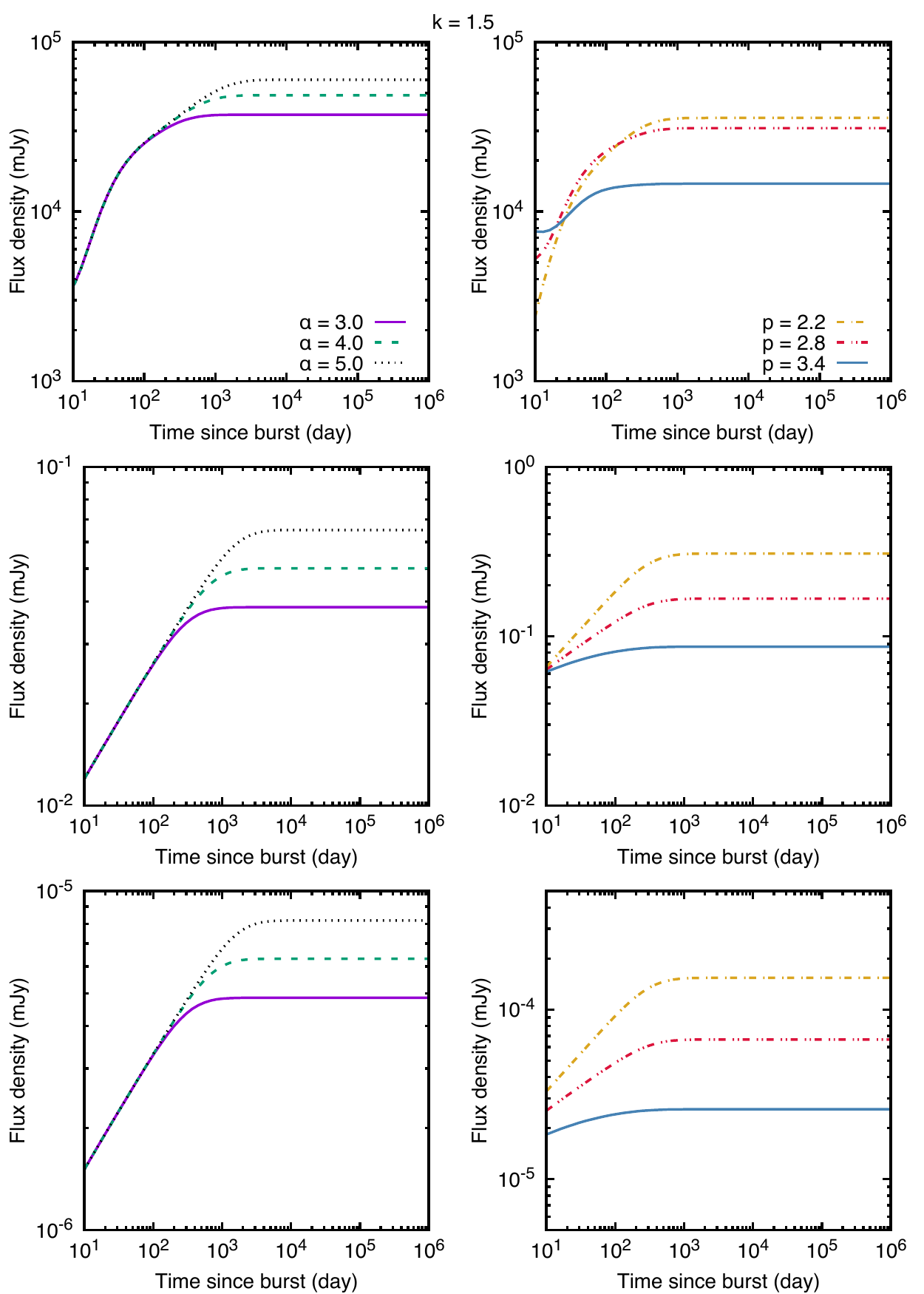}}
}   
\caption{The same as Figure \ref{k_0}, but for $k=1.5$ with $A_{1.5}=2.7\times 10^{28}\,{\rm cm^{-\frac32}}$.}
\label{k_15}
\end{figure} 
\begin{figure}
{\centering
\resizebox*{0.9\textwidth}{0.65\textheight}
{\includegraphics{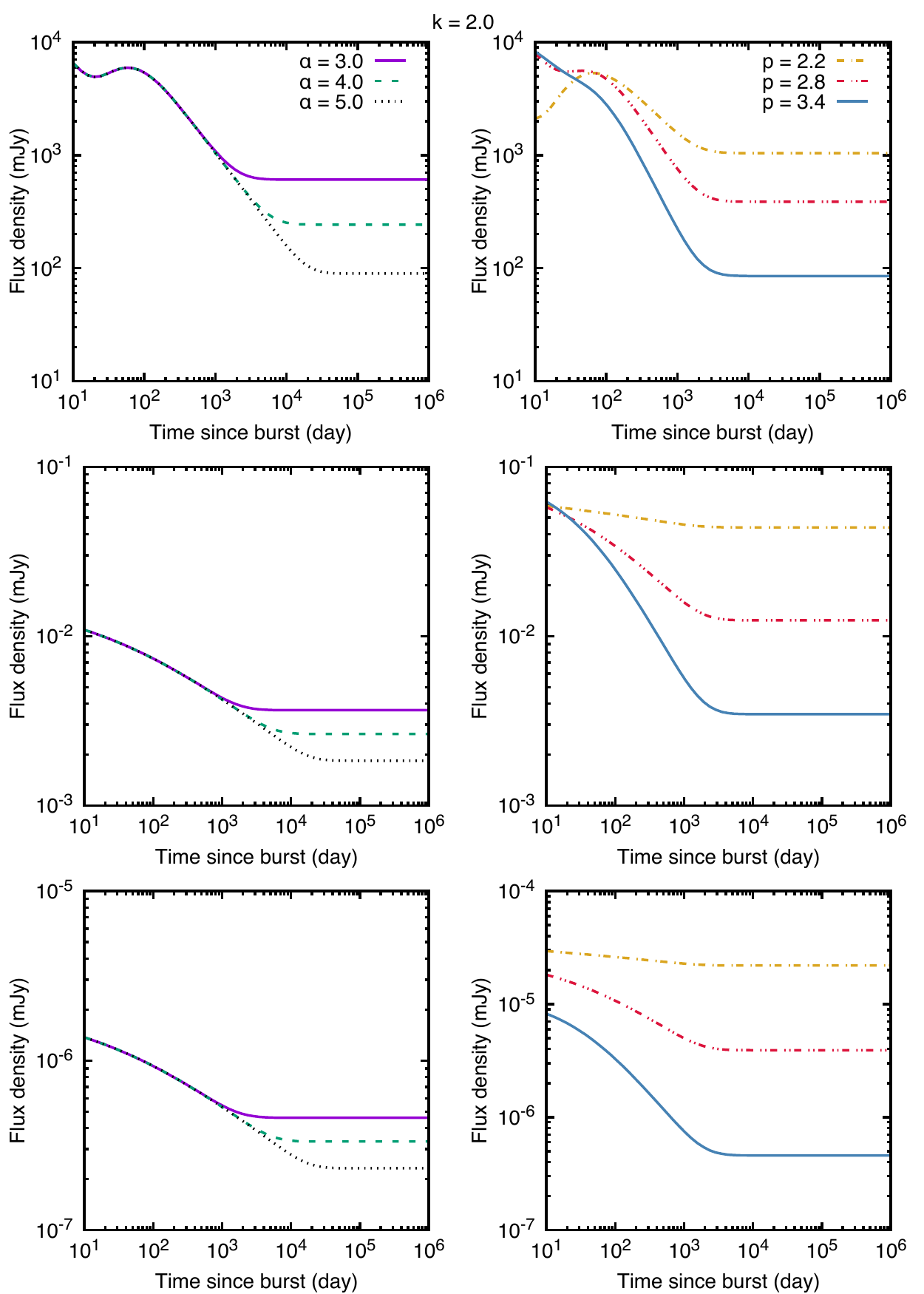}}
}   
\caption{The same as Figure \ref{k_0}, but for $k=2.0$ with $A_2=3\times 10^{36}\,{\rm cm^{-1}}$.}
\label{k_20}
\end{figure} 

\begin{figure}
{\centering
\resizebox*{0.9\textwidth}{0.65\textheight}
{\includegraphics{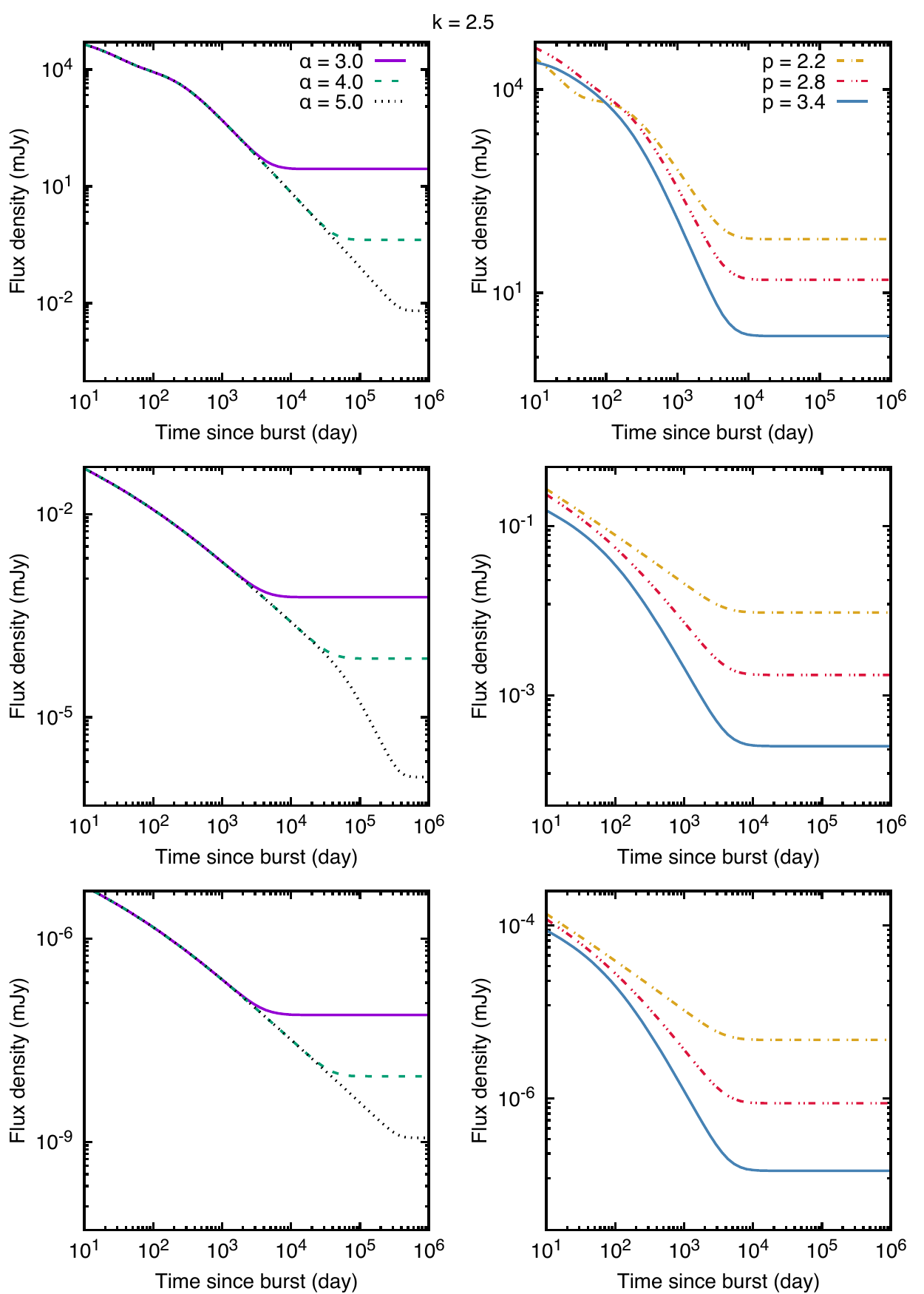}}
}   
\caption{The same as Figure \ref{k_0}, but for $k=2.5$ with $A_{2.5}=1.3\times 10^{45}\,{\rm cm^{-\frac12}}$.}
\label{k_25}
\end{figure} 
\begin{figure}
{\centering
\resizebox*{0.9\textwidth}{0.65\textheight}
{\includegraphics{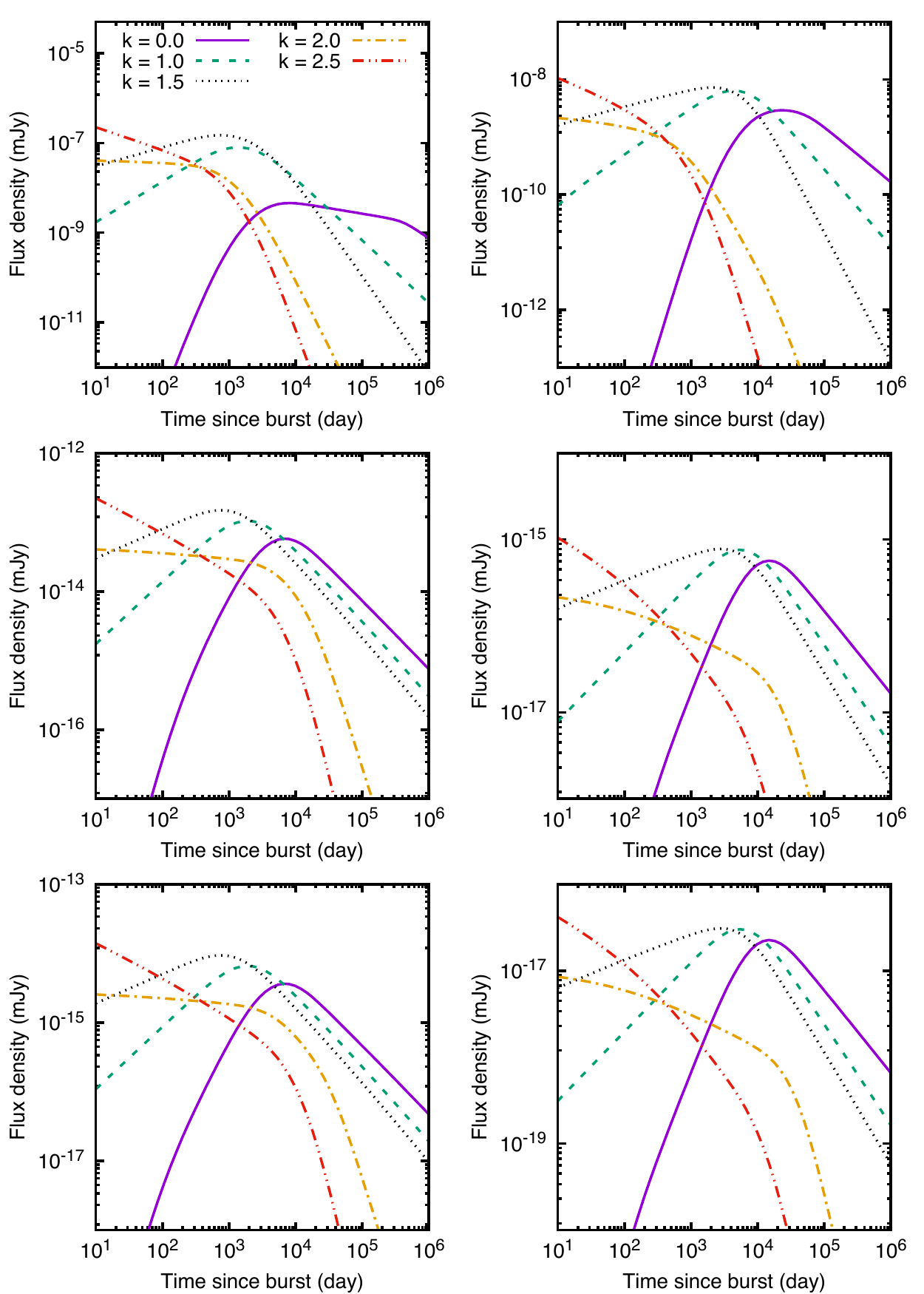}}
}   
\caption{SSC light curves generated by the deceleration of the non-relativistic ejecta for  $k=0$, $1$, $1.5$, $2$ and $2.5$. Panels from top to bottom correspond to gamma-ray fluxes  at 100 keV, 10 GeV and 100 GeV, respectively.  The left-hand panels show the light curves for $p=2.4$  and  $\alpha=3.0$, and the right-hand panels for $p=2.8$ and  $\alpha=5.0$.  We use the same typical values that were used for the synchrotron light curves.}
\label{ssc_all_k}
\end{figure} 

\begin{figure}
{\centering
\resizebox*{0.9\textwidth}{0.65\textheight}
{\includegraphics{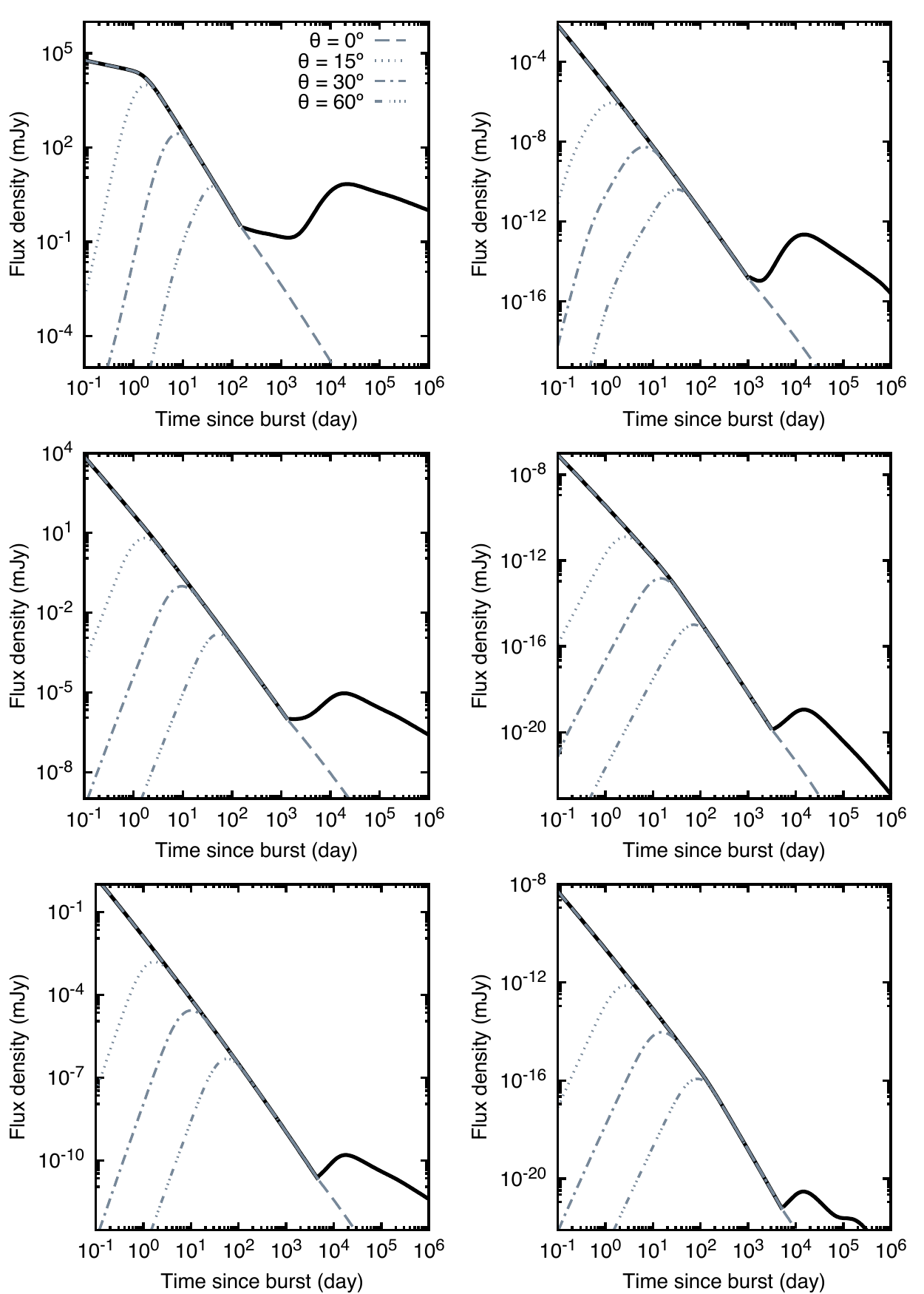}}
}   
\caption{SSC and synchrotron light curves generated by the non-relativistic masses  ejected from NS mergers such as  the dynamical ejecta,  the cocoon, the shock breakout and the wind. The left-hand panels show the synchrotron light curves which correspond to (from top to bottom) radio (1.6 GHz),  optical (1 eV) and X-ray (1 keV) bands, respectively, and the right-hand panels show the SSC light curves which correspond to (from top to bottom)  gamma-ray fluxes  at 100 keV, 10 GeV and 100 GeV, respectively.    The following parameters  $A_0=1\,{\rm cm^{-3}}$, $p=2.6$,   $\alpha=3.0$, $\epsilon_{\rm B}=10^{-1}$, $\epsilon_{\rm B}=10^{-1}$ and $d=100\,{\rm Mpc}$ are used.  The gray solid lines correspond to an on-axis and off-axis relativistic jet with viewing angles of $\theta=15^\circ$, $30^\circ$ and $60^\circ$. The black solid line represents the total contribution from the relativistic jet and the non-relativistic masses  ejected from NS mergers.  The expected fluxes from the relativistic jet are obtained in accordance with the afterglow model introduced in  \cite{2019ApJ...884...71F}.}
\label{lc_ns}
\end{figure} 
\begin{figure}
{\centering
\resizebox*{0.9\textwidth}{0.65\textheight}
{\includegraphics{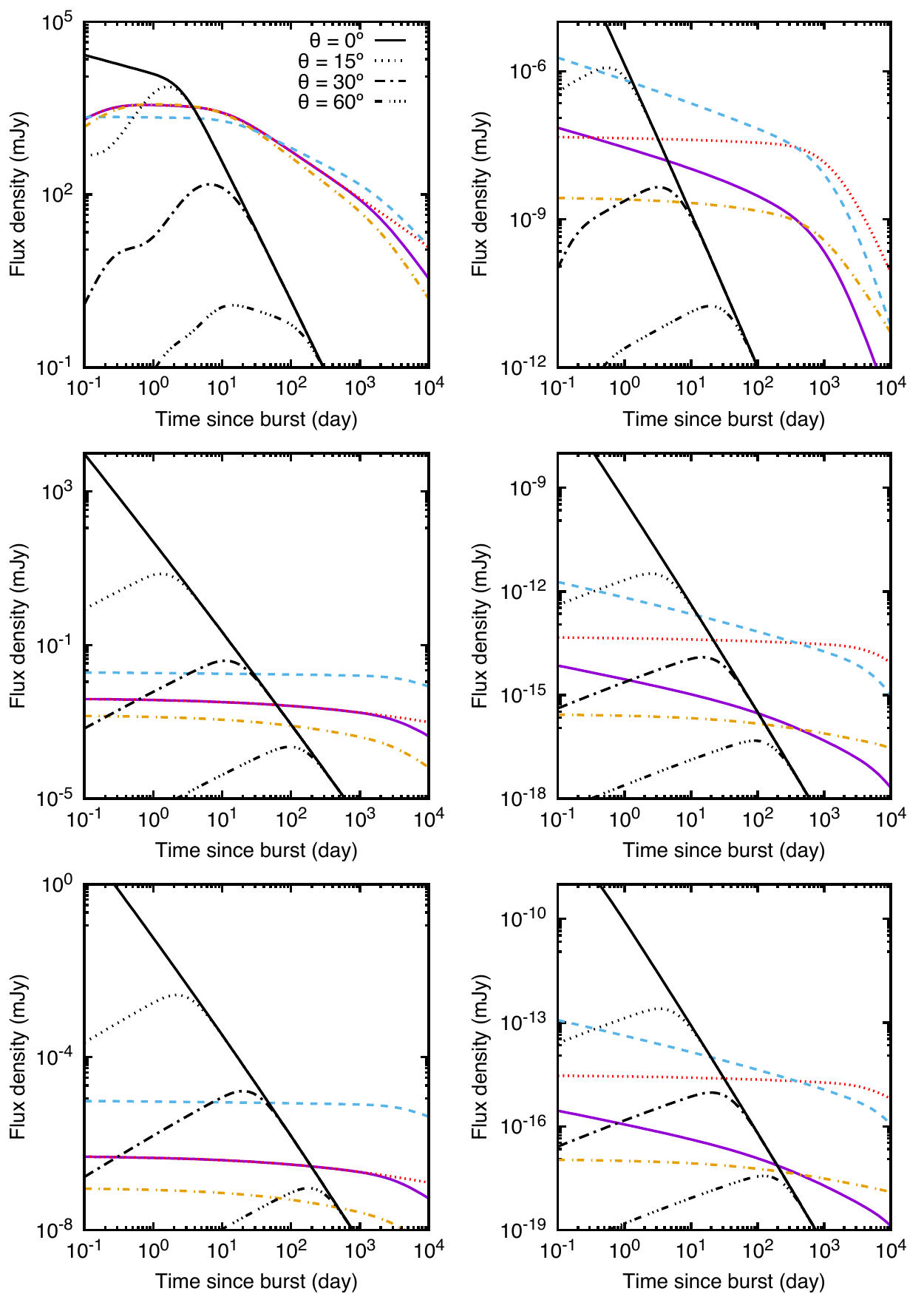}}
}   
\caption{SSC and synchrotron light curves generated by the deceleration of non-relativistic masses  ejected in a stellar wind medium. The left-hand panels show the synchrotron light curves which correspond to (from top to bottom) radio (1.6 GHz),  optical (1 eV) and X-ray (1 keV) bands, respectively, and the right-hand panels show the SSC light curves which correspond to (from top to bottom)  gamma-ray fluxes  at 100 keV, 10 GeV and 100 GeV, respectively.  The solid purple line is for $\alpha=3$, $p=2.6$, the dashed blue line  is for $\alpha=5$, $p=2.6$, the dotted-dashed yellow line is for  $\alpha=3.5$, $p=2.2$ and the dotted red line is for $\alpha=3.5$, $p=2.8$.  The following parameters $\tilde{E}=10^{51}\,{\rm erg}$,  $A_2=3\times 10^{36}\,{\rm cm^{-1}}$, $\epsilon_{\rm B}=10^{-2}$, $\epsilon_{\rm B}=10^{-1}$ and $d=100\,{\rm Mpc}$ are used. }
\label{lc_cc}
\end{figure} 
\begin{figure}
{\centering
\resizebox*{0.9\textwidth}{0.6\textheight}
{\includegraphics{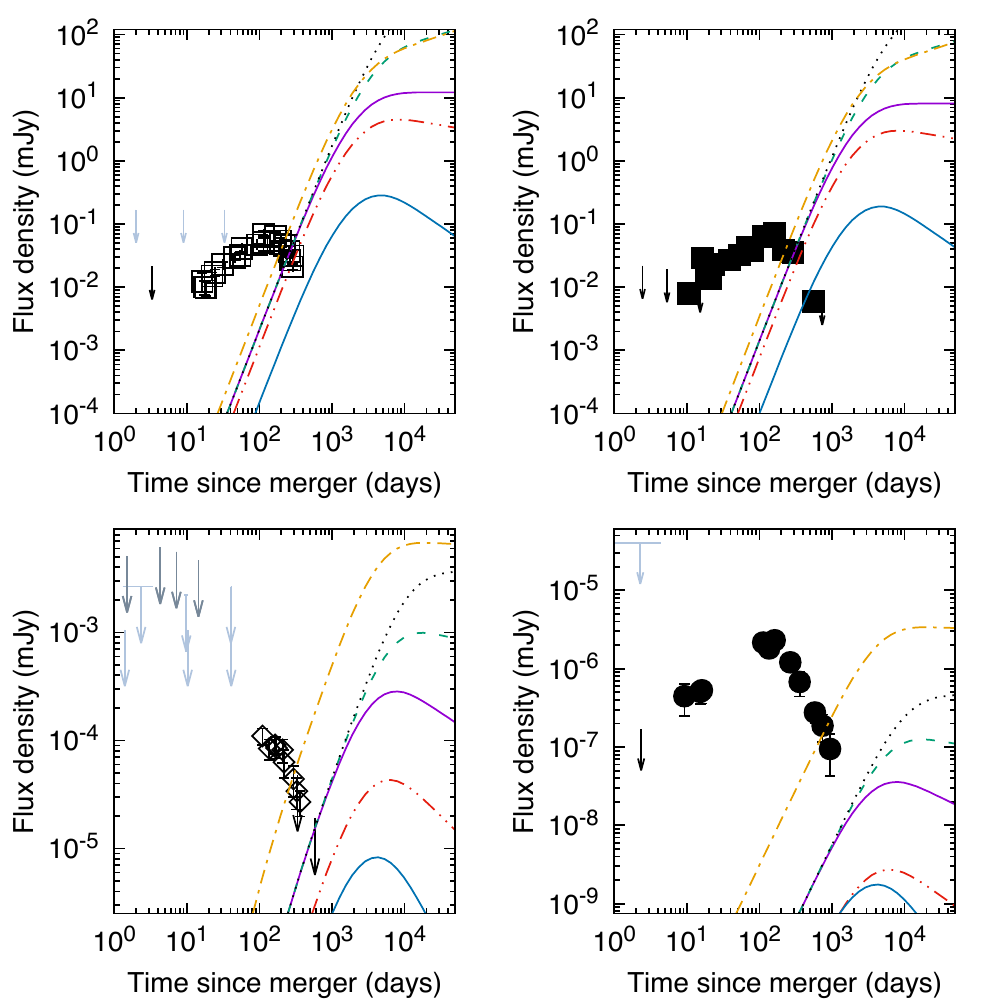}}
}   
\caption{The multi-wavelength observations and upper limits of GRB 170817A and S190814bv at the radio, optical and X-ray bands with the synchrotron light curves generated by the deceleration of the non-relativistic ejecta for $k=0$ (see Figure \ref{k_0}). The upper panels show the radio observations at 3 GHz (left) and 6 GHz (right), and the lower panels show the observations at  the F606W filter (left) and 1 keV (right). The solid curves in blue for optical and X-ray are multiplied by $60$ and $1.6\times 10^3$, respectively, to be illustrated in each panel.  The data points of GRB 170817A are taken from \cite{2019ApJ...883L...1F, 2019ApJ...884...71F, 2019ApJ...886L..17H, 2020GCN.27411....1T}, and upper limits of S190814bv are taken from  \cite{2019ApJ...887L..13D, 2020arXiv200201950A, 2019GCN.25400....1E}.  The data conversion of the latest Chandra afterglow observations between ${\rm erg\,cm^{-2}\,s^{-1}}$ (0.3 - 10 keV)  and ${\rm mJy}$ normalized at $1\,{\rm keV}$ is reported in Table \ref{table2}.}
\label{grb170817A-kn}
\end{figure} 
\begin{figure}
{\centering
\resizebox*{0.5\textwidth}{0.35\textheight}
{\includegraphics{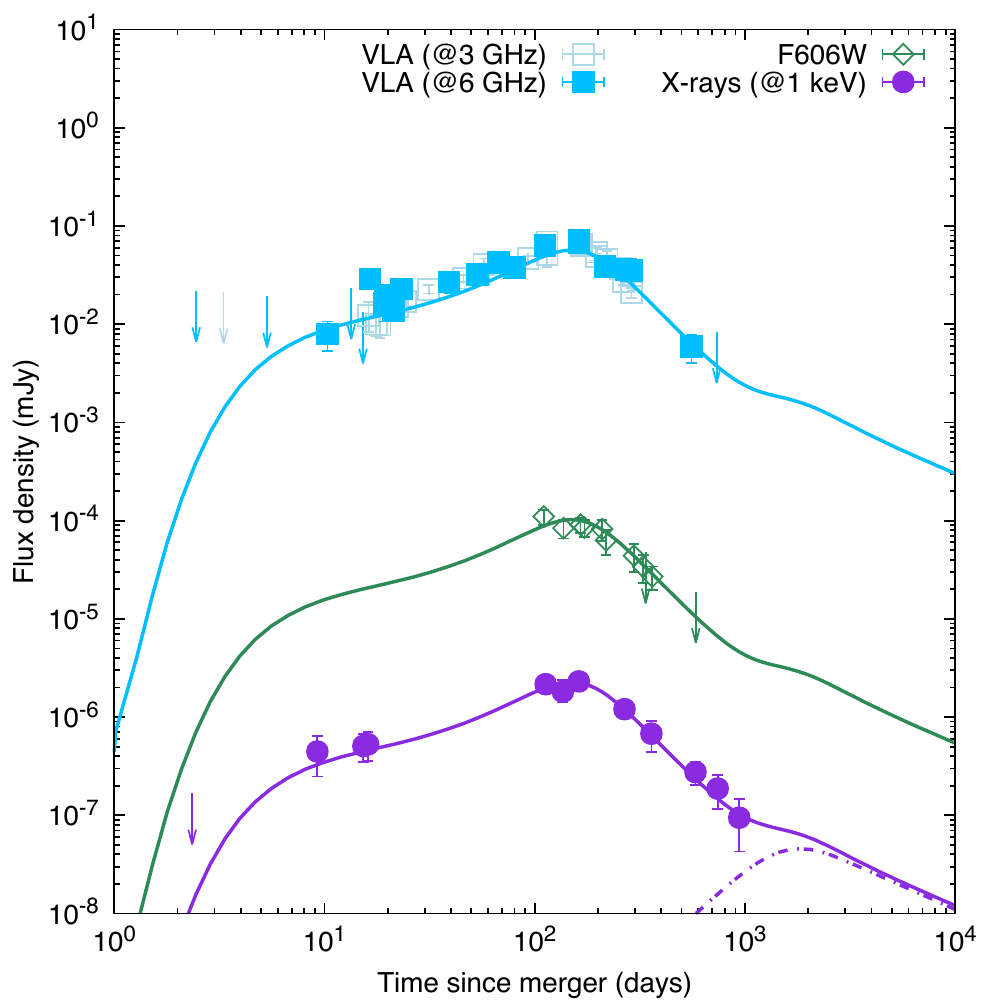}}
\resizebox*{0.5\textwidth}{0.37\textheight}
{\includegraphics{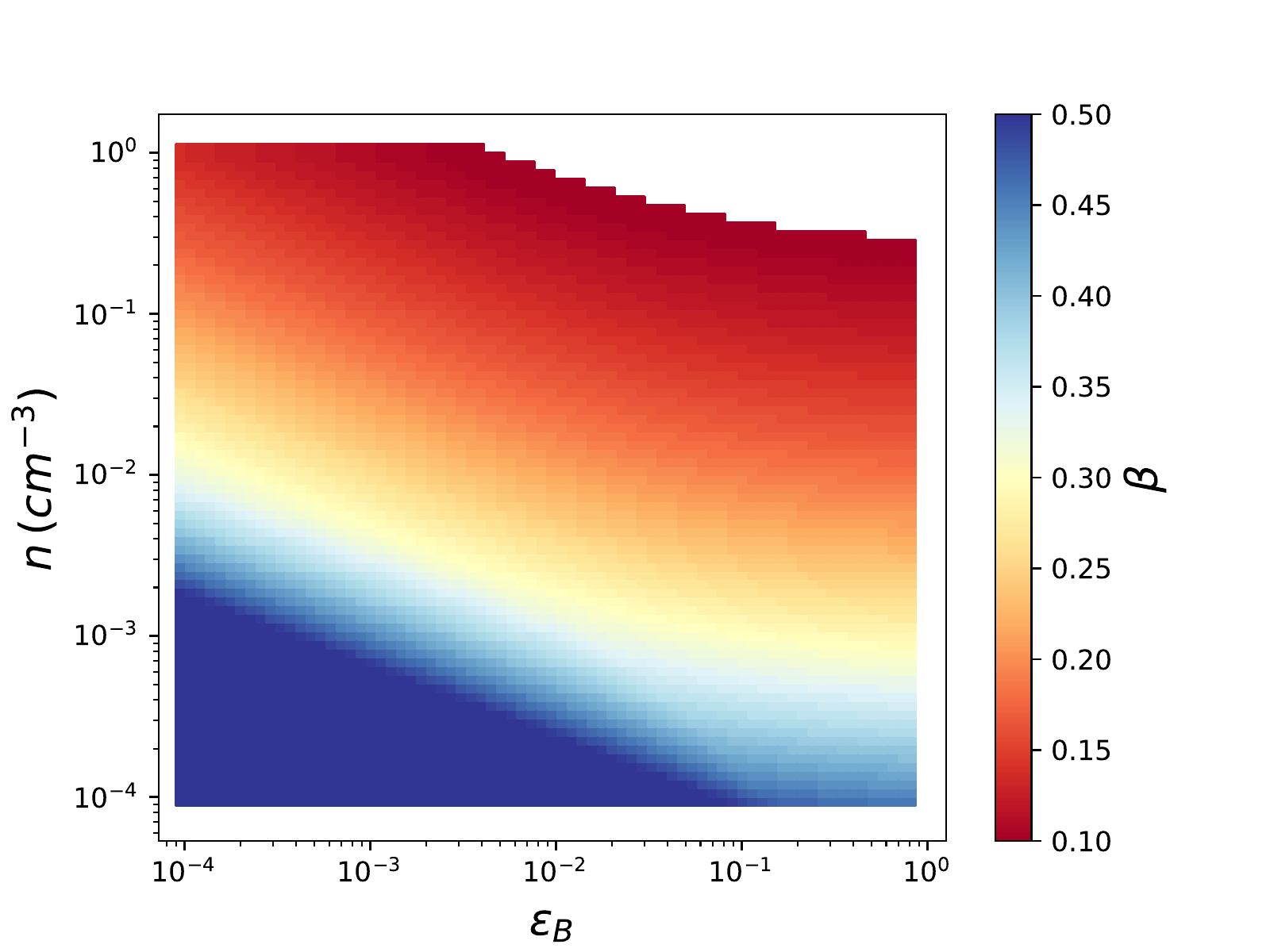}}
\resizebox*{0.5\textwidth}{0.35\textheight}
{\includegraphics{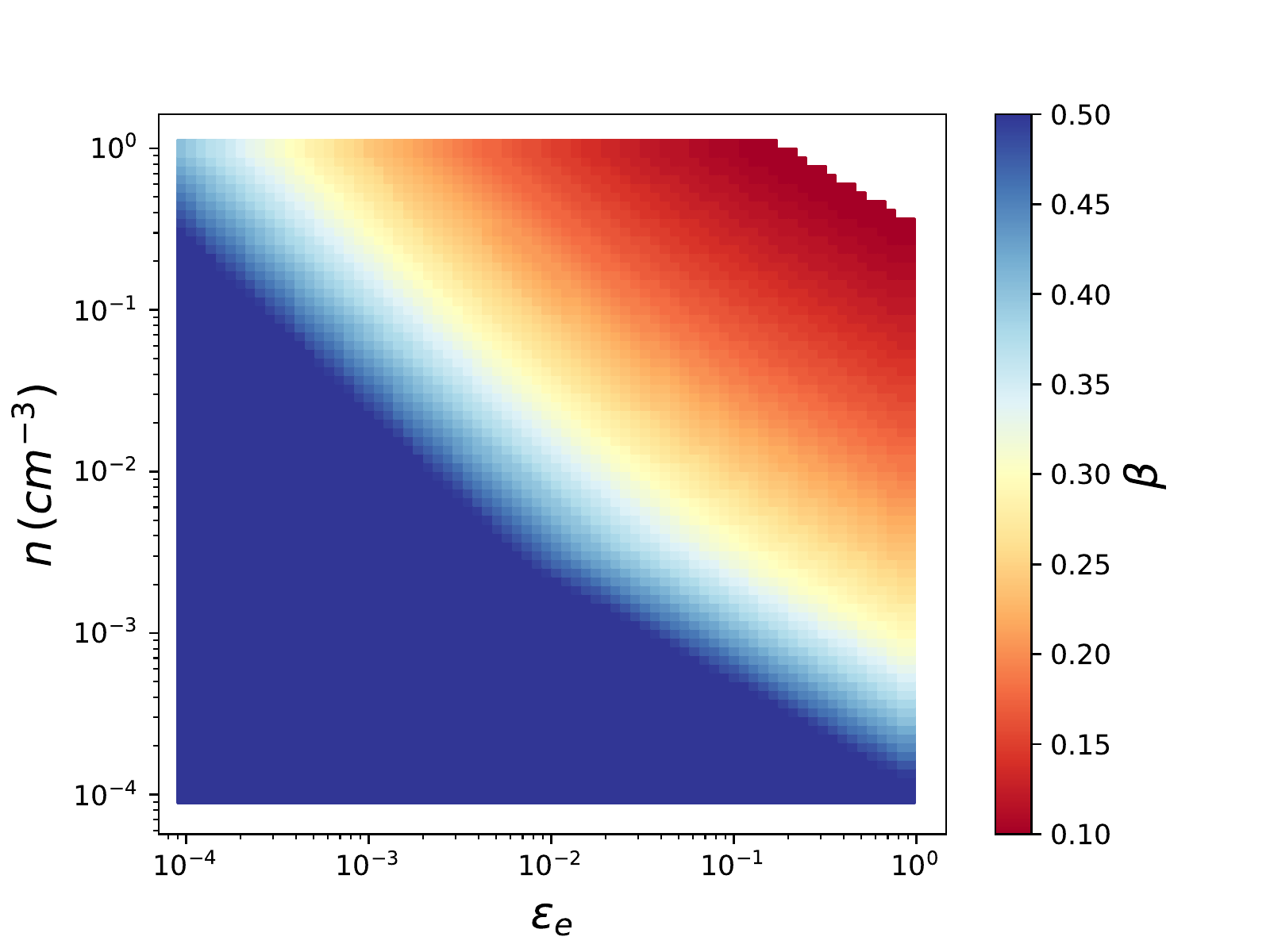}}
\resizebox*{0.5\textwidth}{0.35\textheight}
{\includegraphics{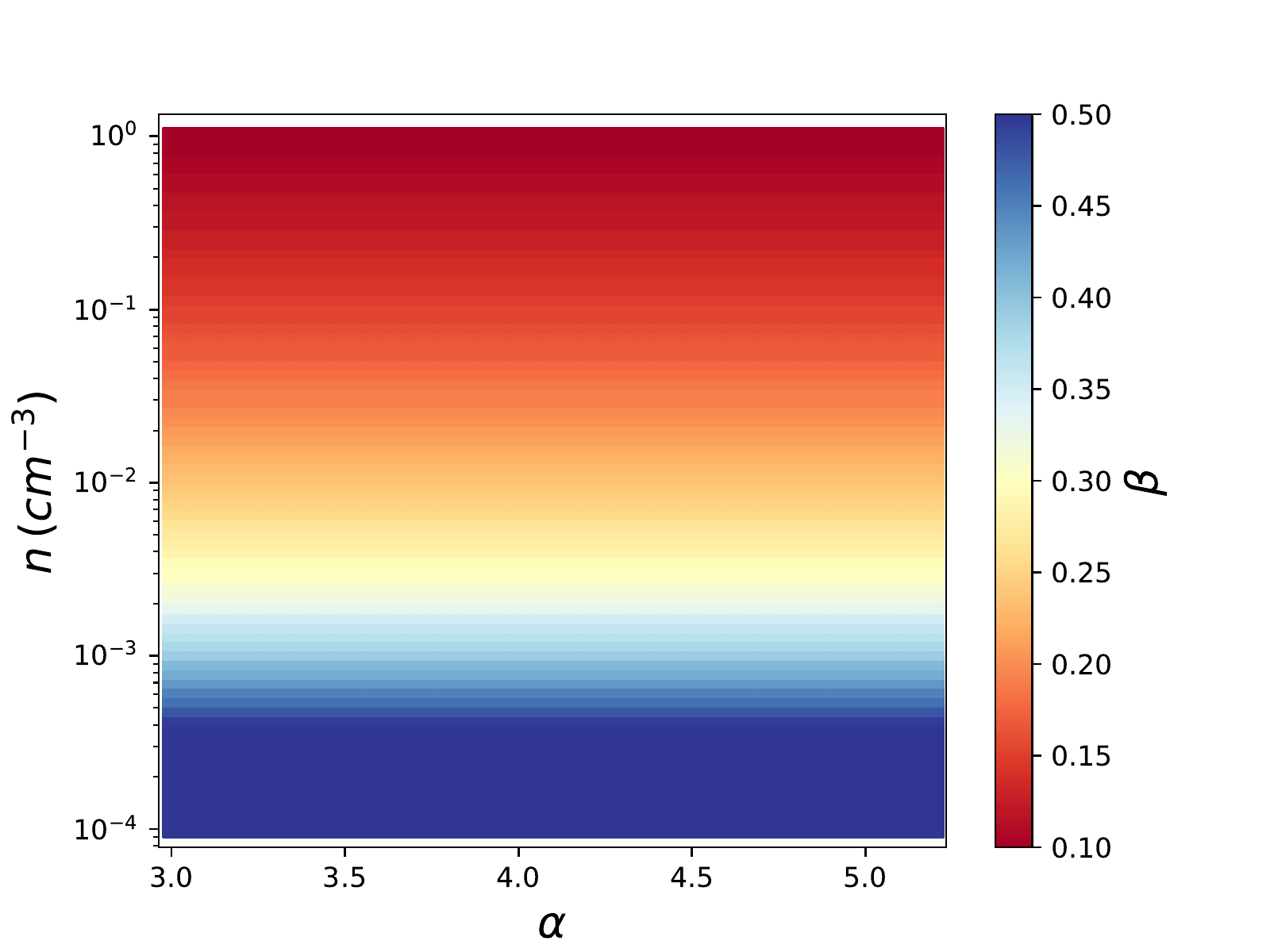}}
}  
\caption{The upper left-hand panel shows the multi-wavelength data points of GRB 170817A with the best-fit curves obtained with the structure jet model presented in \cite{2019ApJ...884...71F} and  a  possible synchrotron light curve from the deceleration of the non-relativistic ejecta.   The upper right-hand and lower panels show the allowed  parameter space of the uniform density of the circumstellar medium (${\rm n}$), the velocity of the non-relativistic ejecta ($\beta$) and the microphysical parameter ($\epsilon_{\rm e}$ and  $\epsilon_{\rm B}$) for  the fiducial energy $E=10^{49}\,{\rm erg}$,  $\alpha=3.0$ and  the spectral index $p=2.15$. The value of the microphysical parameter $\epsilon_{\rm e}=10^{-1}$ is used  in the left-hand panel, and $\epsilon_{\rm B}=10^{-3}$ in the right-hand panel.}
\label{grb170817a_lc}
\end{figure} 
\begin{figure}
{\centering
\resizebox*{0.49\textwidth}{0.35\textheight}
{\includegraphics{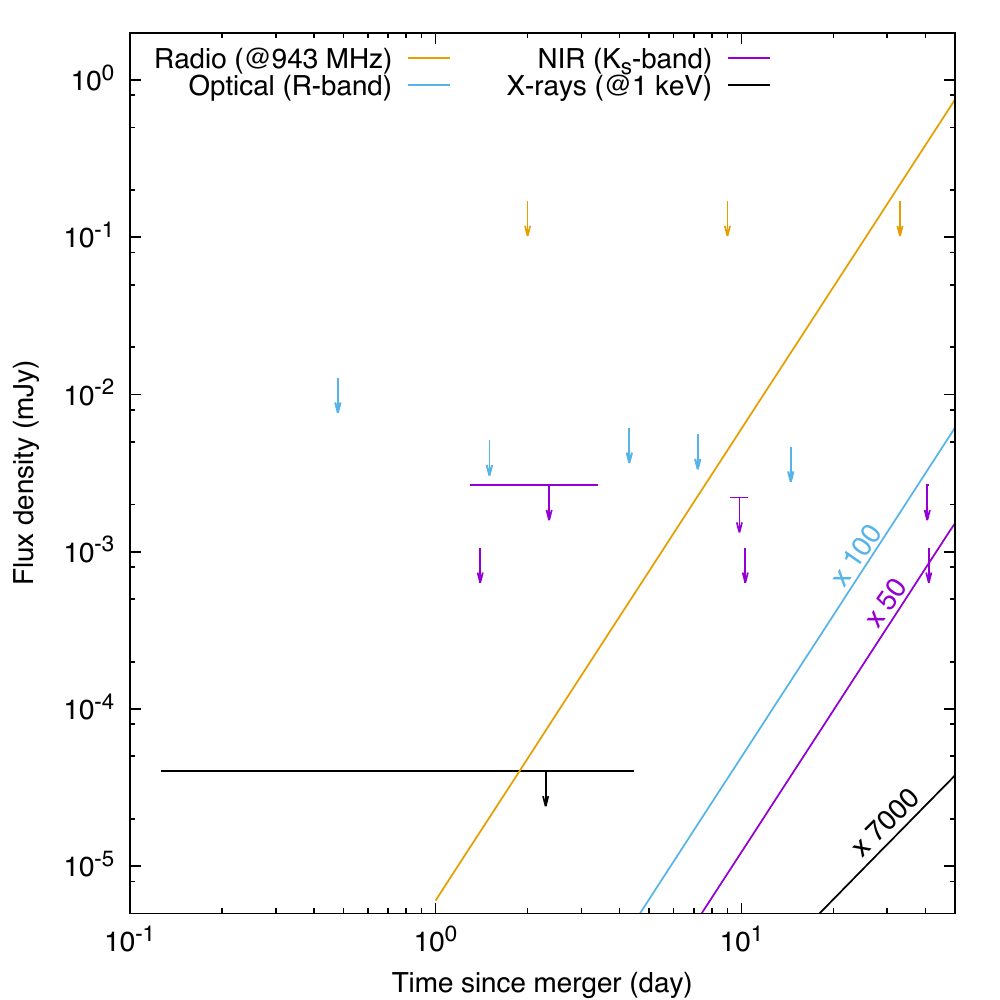}}
\resizebox*{0.49\textwidth}{0.35\textheight}
{\includegraphics{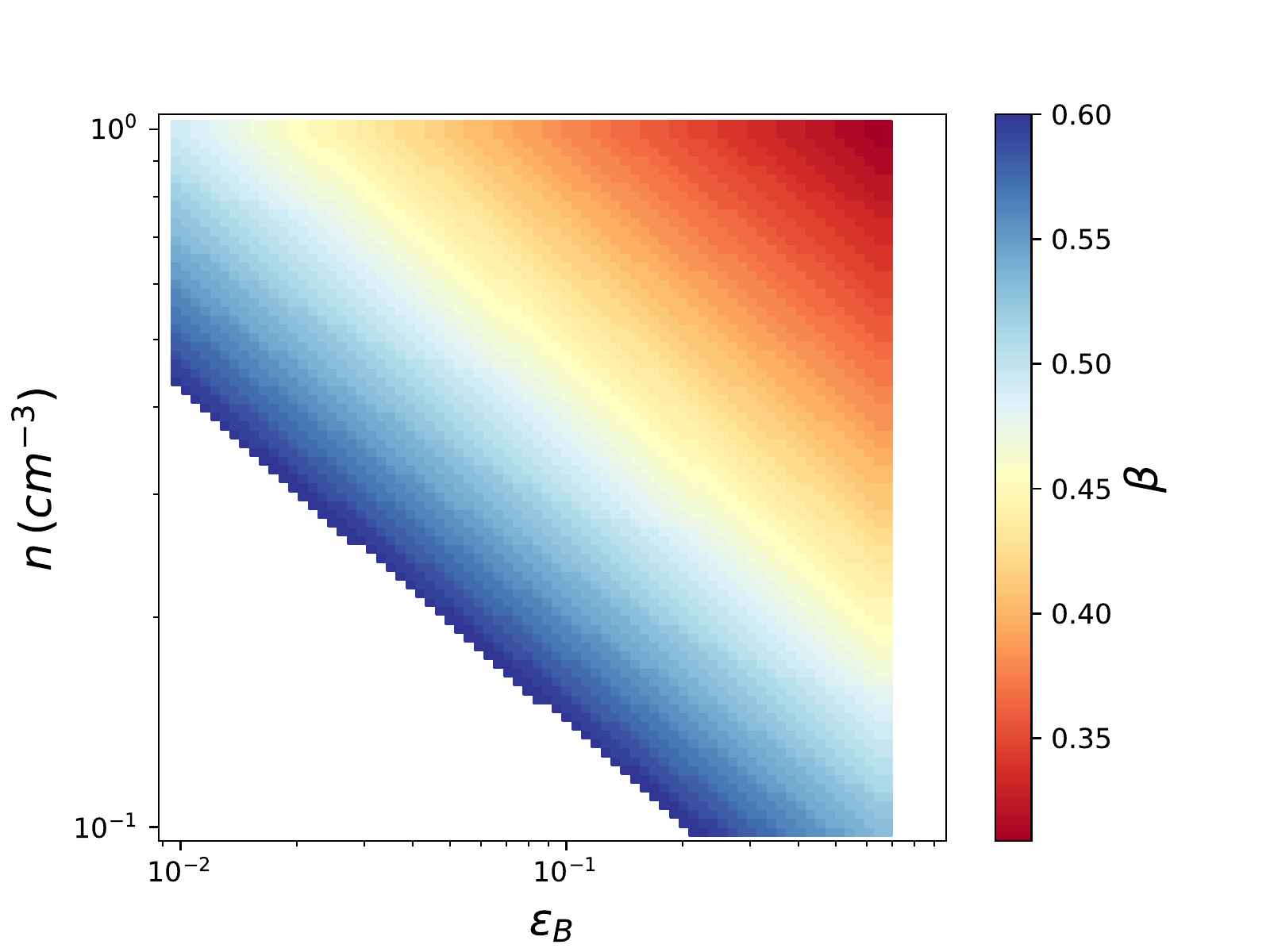}}
\resizebox*{0.49\textwidth}{0.35\textheight}
{\includegraphics{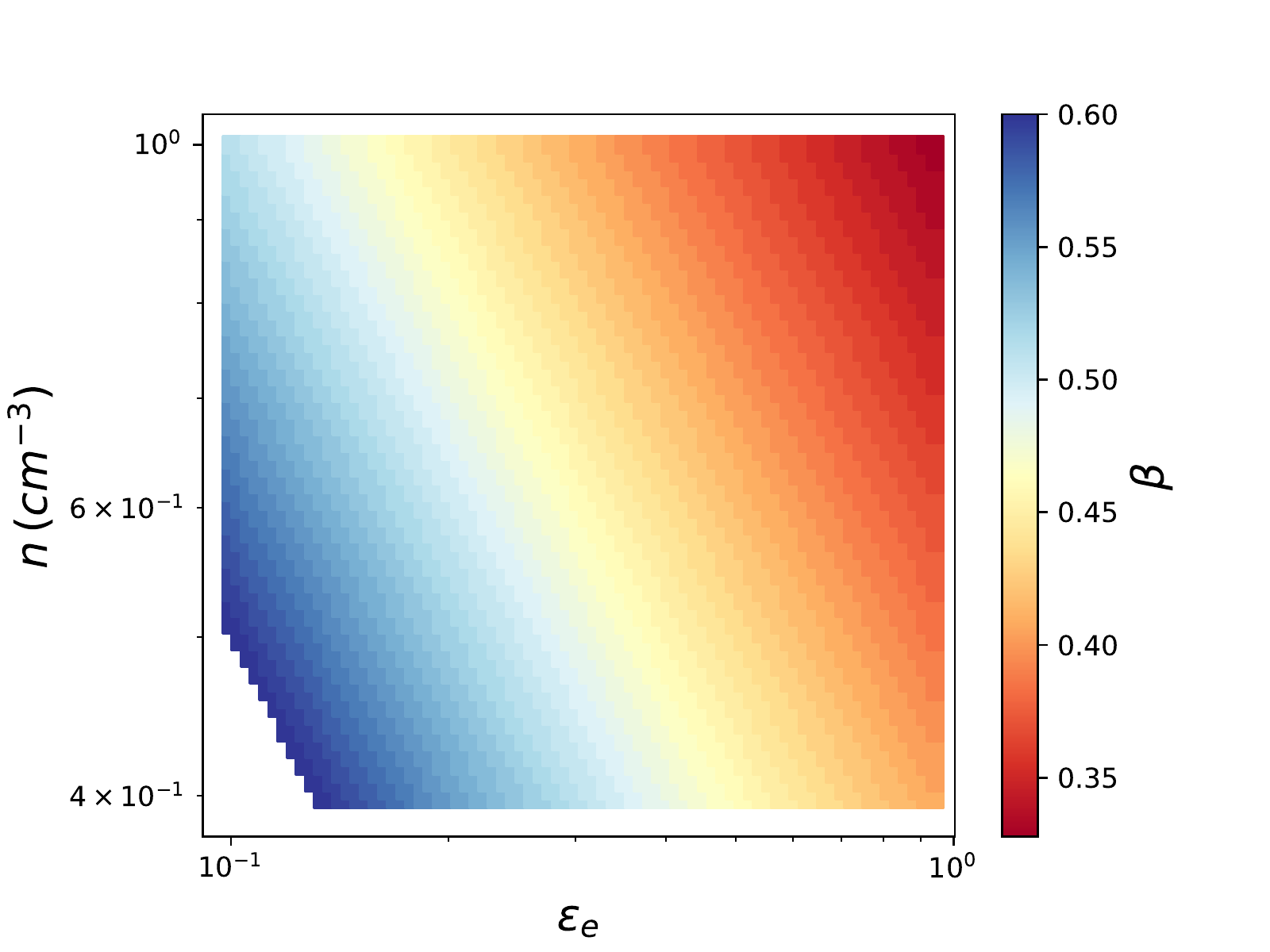}}
\resizebox*{0.49\textwidth}{0.35\textheight}
{\includegraphics{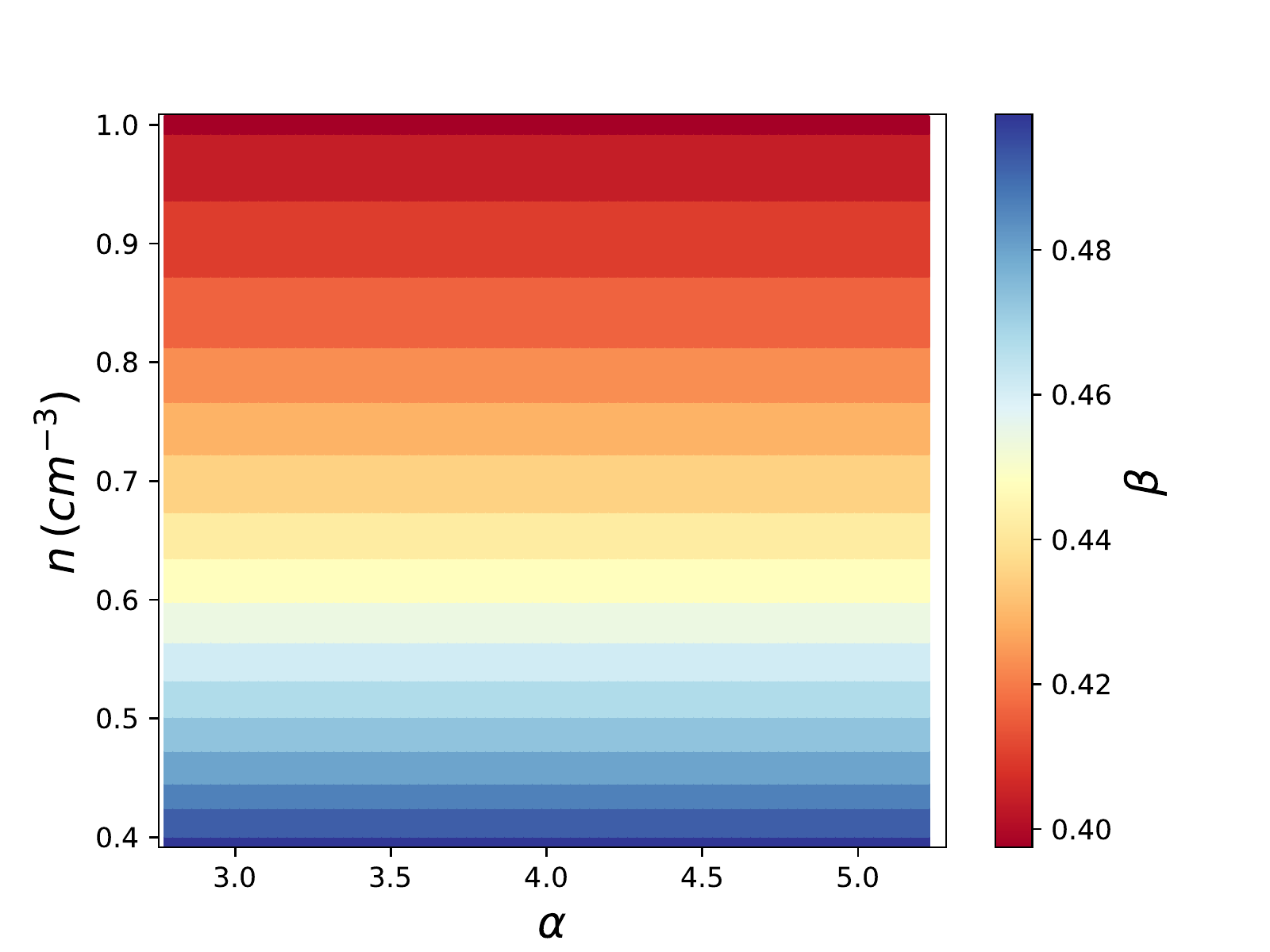}}
}   
\caption{The upper left-hand panel shows the multi-wavelength upper limits of S190814bv and the parameter space ruled out with the synchrotron model presented in this work.  The upper right-hand and lower panels show the parameter space of the uniform density of the circumstellar medium (${\rm n}$), the velocity of the non-relativistic ejecta ($\beta$) and the microphysical parameter ($\epsilon_{\rm e}$ and  $\epsilon_{\rm B}$) which is ruled out in our model for  the fiducial energy $E=10^{50}\,{\rm erg}$,  $\alpha=3.0$ and  the spectral index $p=2.6$.   The value of the microphysical parameter $\epsilon_{\rm e}=10^{-1}$ is used in the left-hand panel, and $\epsilon_{\rm B}=10^{-2}$ in the right-hand panel.}
\label{lc_gw190814}
\end{figure} 
\end{document}